\documentclass[a4paper,12pt]{article}
\usepackage[utf8x]{inputenc}
\usepackage{graphicx}
\usepackage{overcite}
\usepackage{amsmath}
\usepackage{amssymb}
\usepackage{float}
\usepackage{rotating} 
\usepackage{longtable}
\usepackage{scrextend}
\usepackage{cancel}
\usepackage[font=small,labelfont=bf]{caption}

\newcommand{\onlinecite}[1]{\hspace{-1 ex} \nocite{#1}\citenum{#1}}

\title{Reliable estimation of prediction uncertainty for physico-chemical property models}
\author{Jonny Proppe and Markus Reiher\thanks{corresponding author: markus.reiher@phys.chem.ethz.ch; Phone: +41446334308; Fax: +41446331594}}

\begin{document}

\maketitle

\vspace*{-0.9cm}\begin{center}
ETH Z\"urich, Laboratorium f\"ur Physikalische Chemie, \\ Vladimir-Prelog-Weg 2, 8093 Z\"urich, Switzerland. \\
\end{center}

\begin{abstract}

	One of the major challenges in computational science is to determine the uncertainty of a virtual measurement, that is the prediction of an observable based on calculations.
	As highly accurate first-principles calculations are in general unfeasible for most physical systems, one usually resorts to parameteric property models of observables, which require calibration by incorporating reference data.
	The resulting predictions and their uncertainties are sensitive to systematic errors such as inconsistent reference data, parametric model assumptions, or inadequate computational methods.
	Here, we discuss the calibration of property models in the light of bootstrapping, a sampling method that can be employed for identifying systematic errors and for reliable estimation of the prediction uncertainty.
	We apply bootstrapping to assess a linear property model linking the $^{57}$Fe M\"ossbauer isomer shift to the contact electron density at the iron nucleus for a diverse set of 44 molecular iron compounds.
	The contact electron density is calculated with twelve density functionals across Jacob's ladder (PWLDA, BP86, BLYP, PW91, PBE, M06-L, TPSS, B3LYP, B3PW91, PBE0, M06, TPSSh).
	We provide systematic-error diagnostics and reliable, locally resolved uncertainties for isomer-shift predictions.
	Pure and hybrid density functionals yield average prediction uncertainties of 0.06--0.08~mm~s$^{-1}$ and 0.04--0.05~mm~s$^{-1}$, respectively, the latter being close to the average experimental uncertainty of 0.02~mm~s$^{-1}$.
	Furthermore, we show that both model parameters and prediction uncertainty depend significantly on the composition and number of reference data points.
	Accordingly, we suggest that rankings of density functionals based on performance measures (e.g., the squared coefficient of correlation, $r^{2}$, or the root-mean-square error, RMSE) should not be inferred from a single data set.
	This study presents the first statistically rigorous calibration analysis for theoretical M\"ossbauer spectroscopy, which is of general applicability for physico-chemical property models and not restricted to isomer-shift predictions.
We provide the statistically meaningful reference data set MIS39 and a new calibration of the isomer shift based on the PBE0 functional.
\end{abstract}


\section{Introduction}
	
	Predicting observables from a combination of scientific (knowledge-based) and statistical (data-based) information is at the heart of any parameteric property model applied in computational science.\cite{kennedy2001}
	In chemical physics, property models are applied whenever it is unfeasible or too demanding to calculate a \textit{target} observable for a desired range of chemical systems with benchmark methods.
	To resolve this issue, the target observable is represented by a property model, which is a parametric representation of the former.
	The statistical variables of a property model (\textit{parameters}) represent its unknown part, and their optimization (\textit{calibration} of the model) requires both a reference data set and an objective, e.g., minimization of the mean squared error.
	The reference data set comprises pairs of values for (a) the target observable (obtained from measurements or benchmark results) and (b) the corresponding \textit{input} variable.
	The input variable can represent the target observable itself or another physically motivated variable.
	It is also possible that the input variable is a vectorial quantity representing, e.g., nuclear coordinates.
	
	The purpose of calibration is the estimation of parameter values that maximize the transferability of a property model to measurements or benchmark results of its target observable not included in the reference data set.
	In the field-specific literature, there are numerous applications of property models, e.g., for the calibration of force fields\cite{cailliez2011, angelikopoulos2012}, exchange--correlation density functionals\cite{mortensen2005, simm2016}, dispersion-corrected potentials for density functional theory\cite{grimme2010}, semi-empirical electronic structure methods\cite{thiel2014}, vibrational frequencies\cite{rauhut1995, neugebauer2003, irikura2005}, kinetic models\cite{xing2015, sargsyan2015, sutton2016}, ionization potentials\cite{edwards2014}, thermochemical properties \cite{ruscic2014}, properties of semi-conductors and insulators\cite{pernot2015}, linear free energy relationships\cite{wells1963}, or melting point models\cite{karthikeyan2005}, to name only a few.
	
	Concomitant with the maximization of transferability is the assessment of model prediction uncertainty (MPU), i.e., the expected random deviation of a prediction from a measurement or benchmark result.
	MPU can be estimated analytically (or at least iteratively) if certain parametric assumptions are made on the population distribution underlying the reference data set.
	For a continuous variable, the most likely of all other parametric population distributions is the normal distribution, which is parameterized by mean and variance.\cite{bishop2009}
	In that case, Bayesian inference is an efficient way to estimate MPU,\cite{bishop2009} which we will also address in this paper.
	Generally, the more input--target pairs are included in the reference data set, the more reliable it does represent the underlying population distribution, and the less ambiguous is the selection of an adequate property model.
	Consequently, MPU estimation becomes increasingly reliable for an increasing number of data points given a reasonable property model and a specific input domain (the interval of the input variable studied).
	However, in the usual case of a limited number of data points, MPU estimation is error-prone and requires a thorough analysis of parametric population assumptions.
	Moreover, the property model under consideration may be inadequate such that systematic deviations of predictions from measurements or benchmark results are observed.\cite{ohagan2013,pernot2015,pernot2016,pernot2016a}
	Another source of systematic errors are inconsistent data (e.g., outliers).\cite{lira2007,toman2009,pernot2016}
	
	Here, we study nonparametric bootstrapping introduced by Efron\cite{efron1979}, a statistical inference method that meets the challenge of unknown population distributions by sampling from available data.
	This approach produces an \textit{empirical} population distribution, where the reference data set itself represents the underlying population.
	By drawing samples from the empirical population distribution, uncertainty in the parameters can be inferred, which is an essential part in MPU estimation and, hence, for the assessment of transferability.
	
	We apply nonparametric bootstrapping to estimate the uncertainty of predictions of the $^{57}$Fe M\"ossbauer \textit{isomer shift} (target observable) inferred from a linear property model.
	The corresponding input variable is the contact electron density at the iron nucleus of a molecular iron compound, which is obtained from calculations based on density functional theory.
	For 44 molecular iron compounds, we generate input data with 12 exchange--correlation density functionals across Jacob's ladder (PWLDA, BP86, BLYP, PW91, PBE, M06-L, TPSS, B3LYP, B3PW91, PBE0, M06, TPSSh).
	We examine (a) systematic errors in both the reference data and the property model, (b) the effect of experimental uncertainty on the model parameters, (c) the reliability of different performance measures (e.g., the squared coefficient of correlation, $r^{2}$, or the root-mean-square error, RMSE), and (d) the dependency of density functional rankings on the number and composition of reference data points.

	This paper is organized as follows:
	In Section \ref{chap:sampling}, we introduce the idea of sampling methods to estimate MPU, explain the essential concepts of nonparametric bootstrapping, discuss its advantages for a specific class of property models (linear least-squares regression), and demonstrate its ability to unravel different sources of systematic errors.
	In Section \ref{chap:moessbauer}, we discuss challenges for the prediction of the $^{57}$Fe M\"ossbauer isomer shift on the basis of density functional theory, and reevaluate these challenges in the light of a statistically rigorous calibration analysis.
	

\section{Prediction uncertainty from sampling methods}
\label{chap:sampling}

	Since statistically rigorous calibration of physico-chemical property models is the objective of this paper, we first provide a brief and concise review of the relevant concepts and notation needed for this purpose.
	
	To discuss the general concepts of sampling methods, we focus on the common case of a single target observable, $y$, linked to a scalar input variable, $x$, or to a vectorial input variable, $\mathbf{x}$ (in the following, we choose the more general notation $\mathbf{x}= (x_0, x_1, ...)^\top$).

	Given a reference data set, $\mathcal{D}$, comprising $N$ data points ($\mathcal{D} \equiv \lbrace (\mathbf{x}_n,y_n) \rbrace$ with $n = 1, ..., N$) we would like to learn predictions of the target observable by calibration of the underlying property model, $f(\mathbf{x},\mathbf{w}) \approx y$, where $\mathbf{w}= (w_0, w_1, ...)^\top$ is the vector of parameters.
	
	The mean squared error, MSE, and the coefficient of determination, $R^2$, are determined with respect to a reference data set, $\mathcal{D}$, and represent common performance measures of a property model,
	\begin{equation}
	\label{eq:mse}
	\text{MSE} \equiv \text{MSE}_{\mathcal{D},\mathbf{w}} = \frac{1}{N} \sum_{n=1}^N \big(y_n - f(\mathbf{x}_n,\mathbf{w})\big)^2 \,
	\end{equation}
	and
	\begin{equation}
	\label{eq:r2}
	R^2 \equiv R_\mathcal{D}^2 = 1 - N \frac{\min_\mathbf{w}(\text{MSE}_{\mathcal{D},\mathbf{w}})}{\sum_{n=1}^N (y_n - \bar{y})^2} \, ,
	\end{equation}
	where $\bar{y}$ is the arithmetic mean of the target values, $\lbrace y_n \rbrace$, and $\mathbf{w}$ is the vector of parameters.
	Minimizing the MSE with respect to the parameters, $\partial \text{MSE}_{\mathcal{D},\mathbf{w}} / \partial w_m = 0 \ \forall \ m$, is equivalent to the method of least squares and yields $\min_\mathbf{w}(\text{MSE}_{\mathcal{D},\mathbf{w}})$.
	In the following, we refer to the corresponding parameter vector as $\mathbf{w}_\mathcal{D}$, and the shorthand notation $\text{MSE}_\mathcal{D} \equiv \min_\mathbf{w}(\text{MSE}_{\mathcal{D},\mathbf{w}})$ implies the least-squares objective.
	
	It is important to distinguish the \textit{coefficient of determination}, $R^2$, from the \textit{squared coefficient of correlation}, $r^2$, the latter being independent of the parametric model.
	Only in some cases (such as linear least-squares regression with a single input variable), both quantities are equivalent.
	
	MSE$_\mathcal{D}$ and $R^2$ are established measures of the model performance conditioned on $\mathcal{D}$, respectively.
	However, in terms of transferability we would like to know the model performance \textit{independent} of a specific data set.
	In the following thought experiment, we assume that we can generate an asymptotically large number, $B$, of new (training) samples, $\mathcal{D}_b^\ast \equiv \lbrace (\mathbf{x}_{n_b}^\ast,y_{n_b}^\ast) \rbrace$ with $n = 1, ..., N$ and $b = 1, ..., B$, where the asterisk means ``drawn from the population distribution underlying $\mathcal{D}$''.
	If the data points are randomly drawn from a smooth population distribution, we can safely assume that all samples have no data in common (the importance of independent samples for MPU estimation will be discussed in Section \ref{chap:bootstrap}).
	For every training sample, $\mathcal{D}_b^\ast$, we learn the least-squares parameters, $\mathbf{w}_b^\ast$, of the corresponding property model, $f(\mathbf{x},\mathbf{w}_b^\ast)$, and evaluate its deviation from the target values of the reference sample, $\mathcal{D}$, 
	\begin{equation}
	\label{eq:mseb}
	\text{MSE}_{\mathcal{D},\mathbf{w}_b^\ast} = \frac{1}{N} \sum_{n=1}^N \big(y_n - f(\mathbf{x}_n,\mathbf{w}_b^\ast)\big)^2 \geq \text{MSE}_\mathcal{D} \, .
\end{equation}
	In Eq.\ (\ref{eq:mseb}), the target values $y_n$ are elements of $\mathcal{D}$, whereas the predictions $f(\mathbf{x}_n,\mathbf{w}_b^\ast)$ have been learned from $\mathcal{D}_b^\ast$.
	
	To estimate the model performance independent of $\mathcal{D}$, $\mathbb{E}[\text{MSE}]$, we average over all training samples, $\mathcal{D}_b^\ast$,
	\begin{equation}
	\label{eq:emse}
	\mathbb{E}[\text{MSE}] = \frac{1}{B} \sum_{b=1}^B \text{MSE}_{\mathcal{D},\mathbf{w}_b^\ast} \geq \text{MSE}_\mathcal{D} \, ,
	\end{equation}
	where the equality only holds in the artificially ideal case of $f(\mathbf{x}_n,\mathbf{w}_b^\ast) = f(\mathbf{x}_n,\mathbf{w}_\mathcal{D}) \ \forall \ b$.
	Compared to the $\text{MSE}_\mathcal{D}$, the $\mathbb{E}[\text{MSE}]$ additionally incorporates uncertainty in the parameters.
	
	Note that under the assumption of normally distributed data with respect to a parametric model, the $\mathbb{E}[\text{MSE}]$ can be calculated analytically, i.e., sampling is not required in that case.
	Here, however, we examine the implications of such a critical assumption, which is why we explicitly refrain from considering a parameteric population distribution underlying $\mathcal{D}$.


\subsection{Nonparametric bootstrapping --- Sampling from available data}
\label{chap:bootstrap}

	The bootstrapping class of sampling methods by Efron\cite{efron1979} has been continuously developed\cite{chernick1999, davison1997}.
	In nonparametric bootstrapping, the reference sample, $\mathcal{D}$, itself acts as population and, hence, new samples are drawn from $\mathcal{D}$.
	Consequently, given $N$ input--target pairs in $\mathcal{D}$, these pairs are drawn with equal probability, $p(\mathbf{x}_n,y_n) = N^{-1}$.
	This procedure of sampling from available data is referred to as \textit{resampling}.
	The term \textit{nonparametric} refers to the exclusion of a parameteric population distribution.
	We drop this term in the following and \textit{bootstrapping}, if not otherwise mentioned, implies its (original) nonparametric variant.
	
	To generate a \textit{bootstrap sample}, $N$ elements are drawn from $\mathcal{D}$ with replacement.
	This procedure is repeated $B$ times, say $B = 1000$.
	There exist two variants of bootstrapping if the reference data set is composed of input--target pairs.\cite{chernick1999, davison1997}
	The first variant is independent of the property model under consideration and referred to as \textit{pair resampling}.
	In that case, the input--target pairs themselves are drawn with replacement, i.e., the bootstrap samples are constructed as $\mathcal{D}_b^\ast \equiv \lbrace (\mathbf{x}_n,y_n)_b^\ast \rbrace$.
	The second variant requires a pre-calibrated property model conditioned on the reference data set, $\mathcal{D}$.
	The resulting residuals, $r_n = y_n - f(\mathbf{x}_n,\mathbf{w}_\mathcal{D})$, are then subject to resampling, which is referred to as \textit{residual resampling}.
	Since independent and identically distributed residuals are assumed in that case, they have the same probability to occur anywhere along the input domain for which data is available.
	Consequently, the input values, $x_n$, are fixed in residual resampling, whereas the residuals, $r_n$, are randomly drawn (with replacement) and added to the pre-calibrated property model, $f(\mathbf{x}_n,\mathbf{w}_\mathcal{D})$.
	Hence, the corresponding bootstrap samples are constructed as $\mathcal{D}_b^\ast \equiv \lbrace (\mathbf{x}_n,f(\mathbf{x}_n,\mathbf{w}_\mathcal{D})+r_{n_b}^\ast) \rbrace$.
	Here, we will exclusively apply pair resampling since calibration prior to bootstrapping imposes further critical assumptions.
	
	Bootstrapping allows us to sample mean and variance of the parameters contained in a calibration model (and of arbitrary other statistics) without relying on parametric population assumptions.
	Mean and covariance of $\mathbf{w}$ read
	\begin{equation}
	\label{eq:mean_w}
	\bar{\mathbf{w}}^\ast = \frac{1}{B} \sum_{b=1}^B \mathbf{w}_b^\ast
	\end{equation}
	and
	\begin{equation}
	\label{eq:var_w}
	\sigma_{\mathbf{w}^\ast}^2 = \frac{1}{B-1} \sum_{b=1}^B (\mathbf{w}_b^\ast - \bar{\mathbf{w}}^\ast)(\mathbf{w}_b^\ast - \bar{\mathbf{w}}^\ast)^\top \, ,
	\end{equation}
	respectively.
	
	The model performance can be estimated from bootstrapping according to Eq.\ (\ref{eq:emse}), where the square root of $\mathbb{E}[\text{MSE}]$ represents an estimate of the MPU.
	Note that this bootstrapped variant of the $\mathbb{E}[\text{MSE}]$ is different from that derived in the thought experiment introduced in the beginning of this section.
	Here, the reference data set, $\mathcal{D}$, and the training data sets, $\mathcal{D}_b^\ast$, have many data points in common.
	This overlap promotes underestimation of MPU, and there exist straightforward corrections to the bootstrapped $\mathbb{E}[\text{MSE}]$, for instance, the \textit{.632} estimator, \cite{efron1983, hastie2016}
	\begin{equation}
	\mathbb{E}[\text{MSE}]_{.632} = 0.368 \ \text{MSE}_\mathcal{D} + 0.632 \ \frac{1}{N}\sum_{n=1}^N \mathbb{E}[\text{MSE}]_{-n} \, ,
	\end{equation}
	where $\mathbb{E}[\text{MSE}]_{-n}$ refers to the $\mathbb{E}[\text{MSE}]$ with respect to the $n$-th data point, which is \textit{not} involved in the calibration of the corresponding property models $f(\mathbf{x},\mathbf{w}_b^\ast)$,
	\begin{equation}
	  \mathbb{E}[\text{MSE}]_{-n} = \frac{1}{\vert \mathcal{B}_{-n} \vert} \sum_{b \in \mathcal{B}_{-n}} \big(y_n - f(\mathbf{x}_n,\mathbf{w}_b^\ast)\big)^2 \, .
	\end{equation}
	Here, $\mathcal{B}_{-n}$ and $\vert \mathcal{B}_{-n} \vert$ represent the set and number of bootstrap samples \textit{not} comprising the $n$-th input--target pair, respectively.
	The constant 0.632 $\approx 1 - \text{e}^{-1}$ relates to the probability of a data point to be included in a bootstrap sample.\cite{hastie2016}
	In overfitting situations (when the model is too complex such that data noise is fitted), the .632 estimator can be biased.
	In such a case, the constant 0.632 requires correction, resulting in the improved .632+ estimator.\cite{efron1997}
	
	Another resampling method that is most popular with respect to MPU estimation is referred to as \textit{cross-validation}.\cite{hastie2016}.
	In $k$-fold cross-validation, one splits the reference data set into $k$ subsets of most similar size.
	The model is trained on $k-1$ subsets and validated with respect to the remaining one.
	This procedure can be performed in $k$ distinct ways.
	\textit{Leave-one-out} cross-validation is a frequently applied variant of $k$-fold cross-validation where $k = N$.
	The corresponding measure of model performance, $\mathbb{E}[\text{MSE}]_\text{LOO}$, reads
	\begin{equation}
	\mathbb{E}[\text{MSE}]_\text{LOO} = \frac{1}{N} \sum_{n=1}^N \big( y_n - f(\mathbf{x}_n,\mathbf{w}_{\mathcal{D}_{-n}}) \big)^2 \, ,
	\end{equation}
	where $\mathbf{w}_{\mathcal{D}_{-n}}$ refers to the least-squares estimate of $\mathbf{w}$ with the $n$-th data point removed from the reference data set.
	Since cross-validation usually deals with a limited number of $k$ training samples, it is, in general, computationally more efficient than bootstrapping.
	One of the advantages of bootstrapping over cross-validation is, however, the direct assessment of variability for estimated parameters.
	Furthermore, cross-validation may suffer from a poor \textit{bias--variance tradeoff}.\cite{hastie2016}
	This issue relates to the observations that (a) for large values of $k$ (in particular for $k = N$), the variance of model parameters is overestimated, while (b) for small values of $k$ (in particular for $k = 2$), the expected value of model parameters is biased.
	Both effects may strongly misestimate the $\mathbb{E}[\text{MSE}]$.
	In bootstrapping, this issue is resolved by the .632 and .632+ estimators.
	
	We would like to highlight that the sampled parameter distributions resemble empirical variants of posterior distributions employed in Bayesian inference.\cite{rubin1981}
	The latter approach allows for a direct estimation of MPU, which bypasses the need for corrections.
	For the inference of arbitrary posterior distributions, a Bayesian approach is generally much more involved than bootstrapping.
	However, if Gaussian posterior distributions are enforced (by choosing Gaussian prior distributions and a Gaussian likelihood function), the MPU can be estimated efficiently through Bayesian inference,\cite{bishop2009} which we will discuss in the next subsection.
	In this study, we will compare the .632 bootstrap estimate of MPU to that obtained from Bayesian inference, assuming normality of the population distribution in the latter case.
	
	Bootstrapping is an appealing alternative to statistical methods implying parametric assumptions on population distributions.
	On the one hand, modern general-purpose computers can generate and analyze thousands of bootstrap samples in a few seconds given a light fitting problem, i.e., one where calibration is not the limiting step such as in linear least-squares regression.
	On the other hand, it is an objective approach as it allows for inferring statistical quantities solely from available information (the reference data set).
	Clearly, if the available data is biased in a way that it badly represents the true underlying population distribution, application of bootstrapping or any other statistical method is not sensible.
	Major sources of data bias are small sample sizes\cite{chernick1999} and gross outliers. 
	For the latter exist established detection methods\cite{hastie2016}.
	Strictly speaking, even bootstrapping builds upon a population assumption, i.e., the reference data set itself being the population.
	To express it in Chernick's words\cite{chernick1999}, bootstrapping does not mean ``getting nothing for something'', but ``getting the most from the little that is available''.
	

\subsection{Prediction uncertainty of linear regression models}
\label{chap:regression}

	Linear regression refers to a class of calibration procedures, where a target observable, $y$, is estimated to be a linear combination, $f(\mathbf{x},\mathbf{w})$, of $M$ input variables, $x_m$ ($m = 1, ..., M$), and $M+1$ parameters, $w_m$ ($m = 0, 1, ..., M$, where the zero-index refers to the intercept, which is quasi-multiplied by $x_0 = 1$),
\begin{equation}
\label{eq:regression}
f(\mathbf{x}_n,\mathbf{w}) = \sum_{m=0}^{M} w_m (x_m)_n = \mathbf{x}_n^\top \mathbf{w} \, ,
\end{equation}
	where $n = 1, ..., N$ enumerates the input--target pairs of the reference data set, $\mathcal{D}$.
	
	Given a least-squares objective, the parameter vector $\mathbf{w}_\mathcal{D}$ reads
	\begin{equation}
	\label{eq:ls}
	\mathbf{w}_\mathcal{D} = \Big( \mathbf{X}^\top \mathbf{X} \Big)^{-1} \mathbf{X}^\top \mathbf{y} \, ,
	\end{equation}
	where the so-called design matrix\cite{bishop2009} $\mathbf{X}= (\mathbf{x}_1, ..., \mathbf{x}_N)^\top$ contains all instances of the input vector $\mathbf{x}$ contained in $\mathcal{D}$, and $\mathbf{y}= (y_1, ..., y_N)^\top$ is the vector of target values.
	
	Even though linear least-squares regression is a well-established approach with eligibility for many applications, it implies certain critical assumptions\cite{chernick1999}; (a) independent and identically distributed residuals, $r_n = y_n - f(\mathbf{x}_n,\mathbf{w}_\mathcal{D})$, (b) finite variance of residuals, and (c) variance-free input variables.
	Only if these assumptions are valid, the least-squares approach yields the best linear unbiased estimate of the regression parameters.
	For instance, violation of assumption (c) is a ubiquitous phenomenon in the calibration of property models.
	It is a central topic of this paper and will be discussed in detail in Section \ref{chap:moessbauer}.
	Furthermore, if assumption (a) is violated because the variance of the residuals is correlated with one or more input variables, one should instead minimize the weighted MSE, WMSE,\cite{gentle2007} for calibration,
	\begin{equation}
	\text{WMSE} \equiv \text{WMSE}_{\mathcal{D},\mathbf{w},\mathbf{U}} = \frac{1}{N} \sum_{n=1}^N u_n^{-2}\big(y_n - f(\mathbf{x}_n,\mathbf{w_\mathbf{U}})\big)^2 \, ,
	\end{equation}
	where $\mathbf{U}$ is a diagonal weight matrix (assumption of independent residuals),
	\begin{equation}
	\label{eq:umatrix}
	\mathbf{U} =
	\begin{pmatrix}
		u_1^2 & \cdots & 0 \\
		\vdots & \ddots & \vdots \\
		0 & \cdots & u_N^2
	\end{pmatrix} \, .
	\end{equation}
	For property models, the elements $u_n^2$ usually represent experimental variances.
	The parameter vector obtained from weighted least-squares regression, $\mathbf{w}_{\mathcal{D}, \mathbf{U}}$, reads
	\begin{equation}
	\label{eq:wls}
	\mathbf{w}_{\mathcal{D},\mathbf{U}} = \Big( \mathbf{X}^\top \mathbf{U}^{-1} \mathbf{X} \Big)^{-1} \mathbf{X}^\top \mathbf{U}^{-1} \mathbf{y} \, .
	\end{equation}
	This expression is a generalization of the special case covered by Eq.\ (\ref{eq:ls}) where all elements $u_n$ are equal, which is why we drop the subscript $\mathbf{U}$ in the following (it will be evident from the context to which definition of $\mathbf{U}$ we are referring).
	For the analysis of MPU, it is sufficient to consider the average (root-mean-square) experimental uncertainty, $\langle u \rangle$, in the special case of equality,
	\begin{equation}
	\langle u \rangle \equiv \sqrt{\frac{1}{N} \sum_{n=1}^N u_n^2} \, .
	\end{equation}

	In bootstrapped linear least-squares regression, each bootstrap sample yields a parameter vector
\begin{equation}
\mathbf{w}_b^\ast = \Big( (\mathbf{X}_b^\ast)^\top (\mathbf{U}_b^\ast)^{-1} \mathbf{X}_b^\ast \Big)^{-1} (\mathbf{X}_b^\ast)^\top (\mathbf{U}_b^\ast)^{-1} \mathbf{y}_b^\ast \, ,
\end{equation}
	where the asterisk indicates that $\mathbf{X}$, $\mathbf{y}$ and $\mathbf{U}$ are affected by the resampling procedure.
	Properly speaking, when explicitly considering uncertainty on the target observable, pair resampling is replaced by triple resampling where bootstrap samples are constructed as $\mathcal{D}_b^\ast \equiv \lbrace (\mathbf{x}_n,y_n,u_n)_b^\ast \rbrace$.
	Locally resolved MPU for a prediction at the input value $\mathbf{x}_0$, $u(\mathbf{x}_0)$, can be estimated from the reduced $\text{MSE}$---by a factor of $N/(N-M-1)$---and the covariance matrix of the model parameters, $\sigma^2_{\mathbf{w}^\ast}$, introduced in Eq.\ (\ref{eq:var_w}),\cite{pernot2015}
	\begin{equation}
	\label{eq:local_error}
	u(\mathbf{x}_0) = \sqrt{\frac{N}{N-M-1} \text{MSE}_{\mathcal{D},\bar{\mathbf{w}}^\ast} + \mathbf{x}_0^\top \sigma^2_{\mathbf{w}^\ast} \mathbf{x}_0} \, .
	\end{equation}
	Note that $u^2(\mathbf{x}_0)$ estimates the prediction variance of the target observable at the input value $\mathbf{x}_0$ with respect to the average experimental variance, $\langle u \rangle^2$, which would converge to the reduced MSE if all systematic errors have been removed.
	For a new series of measurements at input value $\mathbf{x}_0$ with experimental variance $u_0^2$, it is necessary to add the difference $u_0^2 \, - \, \langle u \rangle^2$ to the reduced MSE in Eq.\ (\ref{eq:local_error}).
	The relation between the $\mathbb{E}[\text{MSE}]_{.632}$ and the locally resolved MPU is given by
	\begin{equation}
	\mathbb{E}[\text{MSE}]_{.632} \approx \frac{1}{N} \sum_{n=1}^N u^2(\mathbf{x}_n) \, ,
	\end{equation}
	where the approximation sign arises, i.a., from the assumption of normality of the parameter distributions as indicated by the second term on the right-hand side of Eq.\ (\ref{eq:local_error}).
	The entire right-hand side of Eq.\ (\ref{eq:local_error}) can be replaced by its sampled analog (bootstrapping of prediction intervals\cite{davison1997}), but for the sake of user-friendliness, we decided to employ a limited number of distinct characteristic values (in this study, two for a prediction, $\bar{\mathbf{w}}^\ast$, and three for the corresponding uncertainty, $\sigma^2_{\mathbf{w}^\ast}$) to facilitate comparisons of different computational methods (e.g., density functionals) with respect to their performance.
	
	Bayesian linear regression is an alternative to bootstrapped linear least-squares regression.
	If there is good reason to assume that the parameter distributions are Gaussian, Bayesian linear regression can be much more efficient than sampling-based regression.
	Bayesian linear regression provides analytical posterior distributions of parameters,\cite{bishop2009} the maxima of which represent the best-fit parameter vector, $\mathbf{w}_\text{MAP}$ (MAP, maximum posterior).
	The procedure is outlined in the Appendix and yields a measure of model performance referred to as mean predictive variance (MPV),
	\begin{equation}
	\label{eq:mpv}
	\text{MPV} = \frac{1}{N} \sum_{n=1}^N s^2(\mathbf{x}_n) \, ,
	\end{equation}
	where $s^2(\mathbf{x}_n)$ is the analogue of $u^2(\mathbf{x}_n)$ obtained from bootstrapping.
	Alternatively, the MPV can be obtained by summing over a dense grid of input values (with a number of grid points $\gg N$),\cite{pernot2015} which yields a smoother result in the sense that it is less dependent on the particular choice of reference input values.
	In that case, it is important to specify the bounds of the input interval and the spacing between or distribution of grid points.
	
	In the following, we will report the square roots (RMSE, R632, RMPV, RLOO) of the performance measures introduced ($\text{MSE}_\mathcal{D}$, $\mathbb{E}[\text{MSE}]_{.632}$, $\text{MPV}$, $\mathbb{E}[\text{MSE}]_\text{LOO}$) for the sake of better comparability with the experimental uncertainty.
	
	
\subsection{Jackknife-after-bootstrapping --- Data diagnostics}
\label{chap:jackknife}

	So far, the discussion of statistical inference was built on the implicit assumption of data sets representing their underlying population distributions well.
	Verifying the validity of this assumption is a tedious task, but a diagnostic referred to as \textit{jackknife-after-bootstrapping} provides a good approximation to the problem\cite{chernick1999,riu2003}.
	In the first step of that diagnostic (given a data set with $N$ input--target pairs), the sampled ensemble of parameters, $\lbrace \mathbf{w}_b^\ast \rbrace$, is decomposed into $N$ different sets, $\lbrace \mathbf{w}_b^\ast \rbrace_{-n}$ ($n = 1, ..., N$), the $n$-th of them containing only those parameters $\mathbf{w}_b^\ast$ learned from bootstrap samples in which the $n$-th input--target pair is \textit{not} included.
	For instance, one obtains a bootstrap estimate of the mean of a parameter with the $n$-th data point removed, $\bar{\mathbf{w}}_{-n}^\ast$.
	If this jackknifed mean deviates significantly from that inferred from the complete reference data set, $\bar{\mathbf{w}}^\ast$, we have an indication that the $n$-th data point biases the calibration.
	
	An appealing feature of the jackknife-after-bootstrapping method is its efficiency. Instead of running $N$ extra bootstrap simulations, one only performs a single bootstrap simulation on the complete reference data set.
	The reason is that some bootstrap samples do not contain certain data points.
	In this way, the decomposition of parameters into $N$ subsets can be performed simultaneously to the bootstrap simulation on the complete reference data set.
	Since the probability of a data point to be excluded from a bootstrap sample is roughly\cite{hastie2016} 0.368 $\approx \text{e}^{-1}$, we recommend to increase the default number of bootstrap samples by a factor of approximately 3 to preserve the intended calibration accuracy.
	

\section{Case study: Isomer-shift calibration in theoretical $^\text{57}$Fe M\"ossbauer spectroscopy}
\label{chap:moessbauer}

	Calibration has been frequently applied to predict the isomer shift observed in $^{57}$Fe M\"ossbauer spectroscopy.\cite{lovell2002, neese2002, zhang2002, liu2003, vrajmasu2003, zhang2003, sinnecker2005, han2006, nemykin2006, shoji2007, han2008, remacle2008, hopmann2009, ling2009, long2009, romelt2009, bochevarov2010, kurian2010, harris2011, sandala2011, gubler2013, papai2013, hedegard2014, long2015, casassa2016, grandjean2016, bjornsson2017}
	The corresponding theory\cite{gutlich2011} postulates a linear relationship between the measurable isomer shift, $\delta_\text{exp}$, and the difference in the contact electron density (CED), $\rho_\text{absorber} - \rho_\text{source}$, of the $^{57}$Fe isotope embedded in two different chemical environments (referred to as \textit{absorber} and \textit{source}),
	\begin{equation}
	\label{eq:mis}
	\delta_\text{exp} = g(r) \Bigg( \frac{\Delta r}{r} \Bigg) [\rho_\text{absorber} - \rho_\text{source}] \, .
	\end{equation}
	$g(r)$ is a function of the average charge radius, $r$, of an iron nucleus, and $\Delta r$ is the difference between the charge radii of the excited state and the ground state of an iron nucleus.
	In the corresponding property model, $\delta(\rho_\text{absorber},\mathbf{w})$ with $\mathbf{w} = (w_0 \ w_1)^\top$, all quantities of the right-hand side of Eq.\ (\ref{eq:mis}) except for $\rho_\text{absorber}$ are hidden in the regression parameters, $w_0$ (intercept) and $w_1$ (slope),
	\begin{equation}
	\label{eq:model}
	\delta(\rho_\text{absorber},\mathbf{w}) = w_0 + w_1 \rho_\text{absorber} \approx \delta_\text{exp} \, .
	\end{equation}

	The absorber CED (ACED), $\rho_\text{absorber}$, is determined from an electronic structure method, typically based on density functional theory.
	Since every specification employed in an electronic structure calculation (e.g., density functionals, basis sets, integration grids, convergence criteria) may affect the value of $\rho_\text{absorber}$, the property model needs to be calibrated every time a specification is changed.
	As noted in Section \ref{chap:regression}, one of the key assumptions in applying linear least-squares regression are variance-free input values.
	While electronic structure calculations yield virtually variance-free results (neglecting numerical errors and convergence threshold effects), the expectation value of an observable remains unpredictable due to method-inherent systematic errors, which are collectively referred to as method inadequacy.\cite{ohagan2013, pernot2015}
	Method inadequacy causes the average uncertainty of the reference isomer shifts to be non-reproducible, no matter which values we choose for $w_0$ and $w_1$.
	Consequently, one should have serious doubt on the validity of the free-of-variance assumption.
	
	Another frequent source of systematic errors are inconsistent data.
	On the one hand, the reference isomer shifts employed in this study have been recorded at different temperatures (4--100 K), which can lead to signed deviations of about $-$0.02~mm~s$^{-1}$ (second-order Doppler shift).\cite{gutlich2011}
	Bochevarov, Friesner, and Lippard\cite{bochevarov2010} proposed to consider only those isomer shifts recorded at liquid helium temperature (4.2 K).
	This situation is clearly desirable, but it would have limited our reference data set in terms of chemical diversity.
	On the other hand, it is difficult to ensure that the molecular-structure representations employed are sufficiently accurate for reliable ACED calculations.
	Even though crystal structures guide the search for the correct minimum on the Born--Oppenheimer surface, there is no guarantee that structure optimization yields reliable results.
	Furthermore, not only the iron-containing compound itself may be important for isomer-shift calibration, but also the closer environment of the solid sample such as adjacent iron complexes, counter-ions, or solvent molecules.
	The effect of the molecular-structure representation on the ACED remains an issue to be studied, which is beyond the scope of this work.
	
	In essentially all previous calibration studies of the isomer shift,\cite{lovell2002, neese2002, zhang2002, liu2003, vrajmasu2003, zhang2003, sinnecker2005, han2006, nemykin2006, shoji2007, han2008, remacle2008, hopmann2009, ling2009, long2009, romelt2009, bochevarov2010, kurian2010, harris2011, sandala2011, gubler2013, papai2013, hedegard2014, long2015, casassa2016, grandjean2016, bjornsson2017} the squared coefficient of correlation, $r^2$, served as a measure to assess the performance of an electronic structure method (note that in linear least-squares regression with a single input variable, as applied here, $r^2$ is identical to the coefficient of determination, $R^2$).
	However, according to Eqs.\ (\ref{eq:r2}) and (\ref{eq:emse}), there is no guarantee that $R^2$ ($= r^2$) allows for a reliable comparison of two electronic structure methods with respect to their transferability, because this performance measure does not take into account uncertainty in the model parameters.
	
	The incompleteness of $r^2$ (or the RMSE) as model performance measure, the possible existence of inconsistent reference data, and the unpredictable variability in the input variable has motivated us to reexamine isomer-shift calibration in the light of a statistically rigorous analysis.
	We assess the reliability of different performance measures and study the transferability of 12 density functionals with respect to isomer-shift predictions for 44 iron compounds of considerable chemical diversity (cf.\ Table \ref{tab:compounds}).
	
	Details on the computational protocol employed for both statistical calibration analysis and quantum chemical calculations are provided in the Appendix. Our statistical calibration program \texttt{reBoot} developed for the analysis presented in this paper
will be made available on our webpage\cite{scriptref}.
In combination with the data provided in the Supporting Information, \texttt{reBoot} allows one to reproduce all results of our paper.
        Furthermore, \texttt{reBoot} can be harnessed to apply the statistical calibration methods presented in our paper to arbitrary polynomial property models that are linear with respect to their parameters.
        Note that the statistical methods implemented in \texttt{reBoot} are not limited to this family of models.
        For instance, implementation of non-polynomial models or models being nonlinear in their parameters would be straightforward.
	
	
\subsection{Reference set of molecular iron compounds}

	In previous calibration studies of the isomer shift\cite{lovell2002, neese2002, zhang2002, liu2003, vrajmasu2003, zhang2003, sinnecker2005, han2006, nemykin2006, shoji2007, han2008, remacle2008, hopmann2009, ling2009, long2009, romelt2009, bochevarov2010, kurian2010, harris2011, sandala2011, gubler2013, papai2013, hedegard2014, long2015, casassa2016, grandjean2016, bjornsson2017}, a variety of iron complexes was considered.
	Here, we chose a diverse set of molecular iron compounds (Table \ref{tab:compounds} and Figs.\ S6--S16) representing wide ranges of formal oxidation states, spin multiplicities, total charges, and ligand environments (type, number, and spatial orientation).
	We blended parts of previous reference sets by Friesner and Lippard\cite{bochevarov2010}, Neese\cite{romelt2009}, Noodleman\cite{han2006}, and our group\cite{gubler2013} with iron(I) complexes\cite{stoian2005, hendrich2006, mock2008, lee2010, lee2011, dugan2012, macleod2014, lichtenberg2015}.
	We explicitly excluded linear and T-shaped iron compounds with a formal oxidation state of +1, since these species are known to reveal pronounced spin--orbit coupling\cite{zadrozny2013, zadrozny2013a, bill2013, samuel2014, ung2014} not considered in our computational approach.
	While the effect of strong spin--orbit coupling on iron CEDs remains a subject to be studied in more detail, scalar-relativistic effects have been found to induce only a constant shift of nonrelativistic CEDs for iron-containing molecules, which is why isomer-shift calibration is frequently based on nonrelativistic calculations.\cite{gutlich2011}
	In our previous study on $^{57}$Fe isomer-shift calibration \cite{gubler2013}, we found that scalar-relativistic calculations lead to a slightly higher correlation between the target observable and the input variable compared to nonrelativistic calculations.
	However, in preparation of the present work, we detected a specification error for iodine in the EMSL basis-set database\cite{emsl} for the def2-TZVP basis set and the {\sc Molcas} computer program 
that turned out to be the source of the slightly better performance.
	Correction of this error reveals equivalent input--target correlation for scalar-relativistic and nonrelativistic results.
	
	\begin{sidewaystable}
	\center
	\caption{Reference set of molecular iron compounds compiled for this study (to be continued).}
	\label{tab:compounds}
	\small
	\begin{tabular}{llcccccccl}
	no.\ & compound\footnote{\footnotesize Abbreviations in alphabetical order: 
	ac, CH$_3$CO$^-$;
	ArNC, 2,6-Me$_2$-C$_6$H$_3$NC;
	ArS, 2,6-Me$_2$-C$_6$H$_3$S;
	$\beta$-DKI$^\text{Me}$, bulky $\beta$-diketiminate;
	cyclam, 1,4,8,11-tetraazacyclotetradecane;
	dae, N--CH2--CH2--N;
	Dipp, 2,6-diisopropylphenyl;
	DTSQ, bis(1,2-dithiosquarato-\textit{S},\textit{S}');
	HBpz$_3$, hydrotris-1-(pyrazolyl)borate;
	MAC$^\ast$, tetraanion of 1,4,8,11-tetraaza-13,13-diethyl-2,2,5,5,7,7,10,10-octamethyl-3,6,9,12,14-pentaoxocyclotetradecane;
	MBTHx, bis(\textit{N}-methylbenzothiohydroxamato) anion;
	Me$_3$TACN, 1,4,7-trimethyl-1,4,7-triazacyclononane;
	OEC, dianion of \textit{trans}-7,8-dihydro-octaethylporphyrin;
	OEP, dianion of octaethylporphyrin;
	opda, \textit{o}-phenylenediamine;
	PhTt$^\text{\textit{t}Bu}$, phenyltris((\textit{tert}-butylthio)-methyl)borate;
	por, porphyrin;
	PyO, 6-methyl-2-pyridonate;
	PyS, 6-methyl-2-thiolate;
	pyS$_4$, dianion of 2,6-bis(2-mercaptophenylthiomethyl)pyridine;
	``S2'', 1,2-benzenedithiolato-\textit{S},\textit{S}' dianion;
	$^{\textit{t}\text{Bu}}$py, 4-\textit{tert}-butylpyridine;
	tmc, 1,4,8,11-tetramethyl-1,4,8,11-tetraazacyclotetradecane;
	trop, 5H-dibenzo[a,d]cyclo-hepten-5-yl.} 
	& $2S+1$ & ox.\ state & CN & $\delta_\text{exp}$ (mm s$^{-1}$)\footnote{\footnotesize All experimental isomer shifts reported in this study refer to bulk $\alpha$-iron at room temperature.} & $T_\text{exp}$ (K) & ref.\ $\delta_\text{exp}$ & ref.\ coords. & code\footnote{\footnotesize The six-letter codes refer to the Cambridge Structural Database identifiers. The N09-X code refers to the structure \#X in ascending order as found in the Supporting Information of the study by R\"omelt, Ye, and Neese \cite{romelt2009}. The R13-Y code refers to the structure \#Y as found in Table 1 of our previous isomer-shift calibration study \cite{gubler2013}. For two compounds, no code is available (--).} \\
	\hline
	1\footnote{\footnotesize This compound refers to an inconsistent input--target pair according to the jackknife-after-bootstrapping method.\label{fn:critical}} & Fe(NCS)$_{4}^{{2}-}$ & 5 & +2 & 4 & 0.97($-$) & 4.2 & \cite{edwards1967} & \cite{fan1989} & KEFFEG \\
	2\footref{fn:critical} & Fe(opda)$_{2}$Cl$_{2}$ & 5 & +2 & 6 & 0.88($-$) & 80 & \cite{renovitch1969} & \cite{maxcy2000} & FUJQOQ \\
	3\footnote{\footnotesize The molecular structure was truncated prior to structure optimization.\label{fn:trunc}} & Fe($^\text{\textit{t}Bu}$py)$_{2}\beta$-DKI$^\text{Me}$ & 4 & +1 & 4 & 0.79($-$) & 80 & \cite{dugan2012} & \cite{dugan2012} & YEZSAZ \\
	4\footref{fn:trunc} & Fe(PhTt$^\text{\textit{t}Bu}$)(PEt$_{3}$) & 4 & +1 & 4 & 0.76(3) & 4.5 & \cite{mock2008} & \cite{mock2008} & VIYHER \\
	5\footref{fn:trunc} & Fe($\eta^{6}$-benzene)$\beta$-DKI$^\text{Me}$ & 2 & +1 & 8 & 0.68($-$) & 80 & \cite{macleod2014} & \cite{macleod2014} & $-$ \\
	6 & Fe(DTSQ)$_{2}^{{2}-}$ & 5 & +2 & 4 & 0.67($-$) & 4.2 & \cite{coucouvanis1981} & \cite{coucouvanis1981} & PTSQFE10 \\
	7\footref{fn:critical} & Fe(por)(O$_{2}$)$^-$ & 6 & +3 & 5 & 0.67($-$) & 4.2 & \cite{burstyn1988} & \cite{romelt2009} & N09-14 \\
	8 & Fe(SPh)$_{4}^{{2}-}$ & 5 & +2 & 4 & 0.66($-$) & 4.2 & \cite{coucouvanis1981} & \cite{coucouvanis1981} & PTHPFE10 \\
	9 & Fe(OEC) & 3 & +2 & 4 & 0.63(1) & 4.2 & \cite{strauss1985} & \cite{strauss1985} & BUYKUB10 \\
	10\footref{fn:trunc} & Fe(PhTt$^\text{\textit{t}Bu}$)(PhCCPh) & 4 & +1 & 5 & 0.62(3) & 4.5 & \cite{mock2008} & \cite{mock2008} & VIYHOB \\
	11 & Fe(OEP) & 3 & +2 & 4 & 0.62(1) & 4.2 & \cite{strauss1985} & \cite{strauss1985} & DEDWUE \\
	12\footref{fn:trunc} & Fe(PMe)(PhBP$^\text{\textit{i}Pr}_{3}$) & 4 & +1 & 4 & 0.57(2) & 80 & \cite{hendrich2006} & \cite{daida2004} & QAJNUL \\
	13\footref{fn:critical} & Fe$_{2}$O(HBpz$_{3}$)$_{2}$(OAc)$_{2}$ & 1 & +3 & 6 & 0.52(3) & 4.2 & \cite{armstrong1984} & \cite{armstrong1984} & CACZIP10 \\
	14\footref{fn:trunc} & Fe(HCCPh)(HC(C$^\text{\textit{t}Bu}$N--Dipp)$_{2}$)$^-$ & 4 & +1 & 4 & 0.50(1) & 4.2 & \cite{stoian2005} & \cite{stoian2005} & MATVOT \\
	15 & Fe$_{2}$O(OAc)$_{2}$(Me$_{3}$TACN)$_{2}^{{2}+}$ & 1 & +3 & 6 & 0.47(3) & 4.2 & \cite{hartman1987} & \cite{hartman1987} & DIBXAN10 \\
	16 & Fe(SMe$_{3}$)(pyS$_{4}$) & 1 & +2 & 6 & 0.44(2) & 80 & \cite{li2002} & \cite{romelt2009} & N09-24 \\
	17 & FeCl(MBTHx)$_{2}$ & 6 & +3 & 5 & 0.43(1) & 4.2 & \cite{berry1983} & \cite{berry1983} & CELVEU \\
	18\footref{fn:trunc} & Fe(SiP$^\text{\textit{i}Pr}_{3}$)N$_{2}$ & 2 & +1 & 5 & 0.38($-$) & 77 & \cite{lee2010} & \cite{mankad2007} & NIHQIF \\
	19 & Fe(PH$_{3}$)(pyS$_{4}$) & 1 & +2 & 6 & 0.34(2) & 80 & \cite{li2002} & \cite{romelt2009} & N09-23 \\
	20 & Fe(NO)(pyS$_{4}$) & 2 & +2 & 6 & 0.33(2) & 80 & \cite{li2002} & \cite{romelt2009} & N09-22 \\
	21 & \textit{trans}-Fe(cyclam)(N$_{3}$)$_{2}^+$ & 2 & +3 & 6 & 0.28(6) & 80 & \cite{meyer1999} & \cite{romelt2009} & N09-6 \\
	22 & LiFe(trop$_{2}$dae) & 2 & +1 & 6 & 0.28(1) & 77 & \cite{lichtenberg2015} & \cite{lichtenberg2015} & RUDHAB \\
	\hline
	\end{tabular}
	\end{sidewaystable}	
	
	\begin{sidewaystable}
	\center
	\caption*{Continuation of Table \ref{tab:compounds}.}
	\small
	\begin{tabular}{llcccccccl}
	no.\ & compound\footnote{\footnotesize Abbreviations in alphabetical order: 
	ac, CH$_3$CO$^-$;
	ArNC, 2,6-Me$_2$-C$_6$H$_3$NC;
	ArS, 2,6-Me$_2$-C$_6$H$_3$S;
	$\beta$-DKI$^\text{Me}$, bulky $\beta$-diketiminate;
	cyclam, 1,4,8,11-tetraazacyclotetradecane;
	dae, N--CH2--CH2--N;
	Dipp, 2,6-diisopropylphenyl;
	DTSQ, bis(1,2-dithiosquarato-\textit{S},\textit{S}');
	HBpz$_3$, hydrotris-1-(pyrazolyl)borate;
	MAC$^\ast$, tetraanion of 1,4,8,11-tetraaza-13,13-diethyl-2,2,5,5,7,7,10,10-octamethyl-3,6,9,12,14-pentaoxocyclotetradecane;
	MBTHx, bis(\textit{N}-methylbenzothiohydroxamato) anion;
	Me$_3$TACN, 1,4,7-trimethyl-1,4,7-triazacyclononane;
	OEC, dianion of \textit{trans}-7,8-dihydro-octaethylporphyrin;
	OEP, dianion of octaethylporphyrin;
	opda, \textit{o}-phenylenediamine;
	PhTt$^\text{\textit{t}Bu}$, phenyltris((\textit{tert}-butylthio)-methyl)borate;
	por, porphyrin;
	PyO, 6-methyl-2-pyridonate;
	PyS, 6-methyl-2-thiolate;
	pyS$_4$, dianion of 2,6-bis(2-mercaptophenylthiomethyl)pyridine;
	``S2'', 1,2-benzenedithiolato-\textit{S},\textit{S}' dianion;
	$^{\textit{t}\text{Bu}}$py, 4-\textit{tert}-butylpyridine;
	tmc, 1,4,8,11-tetramethyl-1,4,8,11-tetraazacyclotetradecane;
	trop, 5H-dibenzo[a,d]cyclo-hepten-5-yl.} 
	& $2S+1$ & ox.\ state & CN & $\delta_\text{exp}$ (mm s$^{-1}$)\footnote{\footnotesize All experimental isomer shifts reported in this study refer to bulk $\alpha$-iron at room temperature.} & $T_\text{exp}$ (K) & ref.\ $\delta_\text{exp}$ & ref.\ coords. & code\footnote{\footnotesize The six-letter codes refer to the Cambridge Structural Database identifiers. The N09-X code refers to the structure \#X in ascending order as found in the Supporting Information of the study by R\"omelt, Ye, and Neese \cite{romelt2009}. The R13-Y code refers to the structure \#Y as found in Table 1 of our previous isomer-shift calibration study \cite{gubler2013}. For two compounds, no code is available (--).} \\
	\hline
	23\footref{fn:trunc} & Fe(OEP)CO & 1 & +2 & 5 & 0.27($-$) & 4.2 & \cite{silvernail2006} & \cite{silvernail2006} & YEQPOA \\
	24 & Fe(SEt)$_{4}^-$ & 6 & +3 & 4 & 0.25($-$) & 4.2 & \cite{maelia1992} & \cite{maelia1992} & CANDAW10 \\
	25\footref{fn:trunc} & Fe(SiP$^\text{\textit{i}Pr}_{3}$)CO & 2 & +1 & 5 & 0.21($-$) & 77 & \cite{lee2011} & \cite{lee2011} & UTOXOR \\
	26 & NaFe(trop$_{2}$dae) & 2 & +1 & 6 & 0.20(1) & 77 & \cite{lichtenberg2015} & \cite{lichtenberg2015} & RUDGUU \\
	27\footref{fn:trunc} & Fe(OEC)Cl & 3 & +4 & 5 & 0.19($-$) & 77 & \cite{vogel1994} & \cite{vogel1994} & SUMWUS \\	  	  
	28\footref{fn:critical} & Fe(NO)$_{2}$(S(\textit{p}-Me)Ph)$_{2}^-$ & 6 & +3 & 4 & 0.18(2) & 4.2 & \cite{harrop2008} & \cite{harrop2008} & SONMUE \\
	29 & Fe(PPh$_{3}$)$_{2}$(``S2'')$_{2}$ & 3 & +4 & 6 & 0.17(1) & 4.2 & \cite{sellmann1991} & \cite{sellmann1991} & SOCVUB \\
	30 & Fe$=$O(tmc)(NCCH$_{3}$)$^{{2}+}$ & 3 & +4 & 6 & 0.17(1) & 4.2 & \cite{rohde2003} & \cite{romelt2009} & N09-18 \\
	31 & Fe(PPh$_{3}$)(``S2'')$_{2}$ & 3 & +4 & 5 & 0.12(1) & 4.2 & \cite{sellmann1991} & \cite{sellmann1991} & SOCWAI \\
	32 & Fe($\eta^{4}$-butadiene)(CO)$_{3}$ & 1 & 0 & 7 & 0.12($-$) & 4.2 & \cite{havlin1998} & \cite{kukolich1993} & $-$ \\
	33 & Fe(PyO)I(CO)$_{2}$PPh$_{3}$ & 1 & +2 & 6 & 0.10(2) & 55 & \cite{obrist2009} & \cite{obrist2009} & PUCFID \\
	34 & Fe(PyS)I(CO)$_{2}$PPh$_{3}$ & 1 & +2 & 6 & 0.10(2) & 77 & \cite{chen2011} & \cite{gubler2013} & R13-L \\
	35 & Fe(PyO)(ArS)(CO)$_{2}$PPh$_{3}$ & 1 & +2 & 6 & 0.06(2) & 100 & \cite{chen2011} & \cite{gubler2013} & R13-K \\
	36 & Fe$_{2}$(PyS)$_{2}$(ac)$_{2}$(CO)$_{4}$ & 1 & +2 & 6 & 0.06(1) & 77 & \cite{chen2011} & \cite{chen2010} & FUMZAP \\
	37 & Fe(NO)(pyS$_{4}$)$^+$ & 1 & +2 & 6 & 0.04(2) & 80 & \cite{li2002} & \cite{romelt2009} & N09-21 \\
	38 & Fe(SC$_{5}$H$_{4}$N-CO)I(CO)$_{2}$(ArNC) & 1 & +2 & 6 & 0.01(2) & 77 & \cite{chen2011} & \cite{chen2011} & YAKZAN \\
	39 & Fe(PyS)(ac)(CO)$_{2}$(ArNC) & 1 & +2 & 6 & 0.00(2) & 100 & \cite{chen2011} & \cite{gubler2013} & R13-O \\
	40 & Fe(PyS)(ac)(CO)$_{2}$PPh$_{3}$ & 1 & +2 & 6 & 0.00(2) & 77 & \cite{chen2011} & \cite{chen2010} & FUMZET \\
	41 & Fe(PyS)(ac)(CN)(CO)$_{2}$ & 1 & +2 & 6 & $-$0.02(2) & 77 & \cite{chen2011} & \cite{gubler2013} & R13-Q \\
	42\footref{fn:trunc} & FeCl($\eta^{4}$-MAC$^\ast$)$^-$ & 5 & +4 & 5 & $-$0.04(2) & 4.2 & \cite{kostka1993} & \cite{kostka1993} & JESGUJ \\	  
	43 & Fe(CO)$_{5}$ & 1 & 0 & 5 & $-$0.09(1) & 78 & \cite{greatrex1969} & \cite{romelt2009} & N09-20 \\
	44 & Fe(OEC)C$_{6}$H$_{5}$ & 3 & +4 & 5 & $-$0.11($-$) & 77 & \cite{vogel1994} & \cite{vogel1994} & SUMXED \\
	\hline
	\end{tabular}
	\end{sidewaystable}	
	
	
\subsection{Effect of experimental uncertainty on model parameters}

\begin{figure} [!ht]
\center
\includegraphics[width=\textwidth]{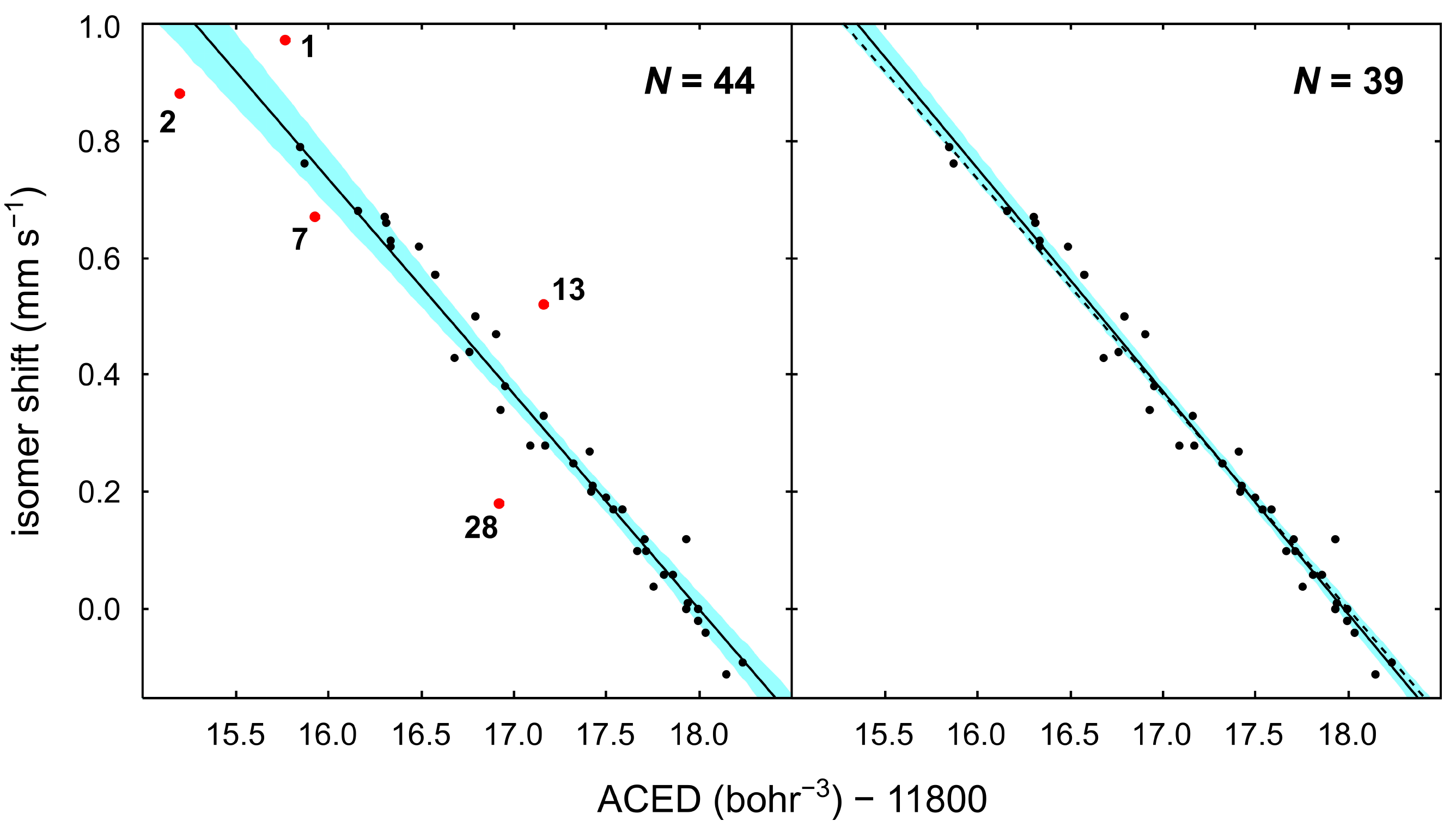}
\caption{Results of bootstrapped ($B = 10^5$) linear least-squares regressions in the presence (left) and in the absence (right) of the inconsistent data points \#1, \#2, \#7, \#13, \#28 (cf.\ Table \ref{tab:compounds}).
The width of the 95\% confidence band (turquoise area, obtained from the 0.025 and 0.975 quantiles of the bootstrapped joint distribution of model parameters) decreases by a factor of about 2 when removing the inconsistent data points.
The solid lines represent mean values of regression parameters over all bootstrap samples, $\bar{\mathbf{w}}^\ast$, and are visually indistinguishable from the least-squares solutions to the regression problem, $\mathbf{w}_\mathcal{D}$.
The dashed line (right) is a replication of the solid line shown in the left frame (identical values for $w_0$ and $w_1$).
The corresponding ACED values were obtained from B3LYP/def2-TZVP calculations.
}
\label{fig:regression}
\end{figure}

	When applying linear least-squares regression to all reference isomer shifts ($N = 44$, see Fig.\ \ref{fig:regression}, left), we find an RMSE of 0.07~mm~s$^{-1}$ for all hybrid density functionals (B3LYP, B3PW91, PBE0, M06, TPSSh).
	For the pure density functionals (PWLDA, BP86, BLYP, PW91, PBE, M06-L, TPSS), the RMSE ranges from 0.08~mm~s$^{-1}$ (M06-L, TPSS) to 0.10~mm~s$^{-1}$ (BLYP); see also Table \ref{tab:error44}.
	
\begin{table} [!ht]
\small
\caption{Performance measures (RMSE, R632, RMPV, RLOO) of density functionals regarding isomer-shift prediction.
The reference data set is the original one ($N = 44$).
RMSE, R632, and RLOO were calculated on the basis of ordinary (unweighted) least-squares regression.
The RMPV was obtained from Bayesian linear regression based on the evidence approximation.
For the calculation of the .632 estimator, R632, $B = 10^5$ bootstrap samples were generated. 
All performance measures are reported in mm~s$^{-1}$.
The corresponding ACED values were obtained with the def2-TZVP basis set for ligand atoms.}
\label{tab:error44}
\begin{center}
\begin{tabular}{lcccc}
functional & RMSE & R632 & RMPV & RLOO \\
\hline
PWLDA & 0.09 & 0.10 & 0.09 & 0.10 \\
BP86 & 0.09 & 0.09 & 0.09 & 0.09 \\
BLYP & 0.10 & 0.10 & 0.10 & 0.10 \\
PW91 & 0.09 & 0.09 & 0.09 & 0.09 \\
PBE & 0.09 & 0.09 & 0.09 & 0.09 \\
M06-L & 0.08 & 0.08 & 0.08 & 0.08\\
TPSS & 0.08 & 0.09 & 0.09 & 0.09 \\
B3LYP & 0.07 & 0.07 & 0.07 & 0.07 \\
B3PW91 & 0.07 & 0.07 & 0.07 & 0.07 \\
PBE0 & 0.07 & 0.07 &0.07 & 0.07 \\
M06 & 0.07 & 0.07 & 0.07 & 0.07 \\
TPSSh & 0.07 & 0.08 & 0.08 & 0.08 \\
\hline
\end{tabular}
\end{center}
\end{table}	

	Even though the RMSE is a lower bound to the MPU inferred from this specific set of reference isomer shifts, it is already significantly larger than the average experimental uncertainty of $\langle u \rangle = 0.02$~mm~s$^{-1}$ found for molecular iron compounds.\cite{li2002}
	However, it is possible that $\langle u \rangle$ is larger than 0.02~mm~s$^{-1}$ for the compounds studied here, since several isomer-shift measurements were reported without uncertainty (cf.\ Table \ref{tab:compounds}).
	Selecting only those measurements for which uncertainty has been reported ($N = 30$), we also find an average experimental uncertainty of $\langle u \rangle = 0.02$~mm~s$^{-1}$.
	The RMSE still ranges from 0.06~mm~s$^{-1}$ (B3LYP, B3PW91, PBE0, M06) to 0.10~mm~s$^{-1}$ (BLYP).
	This discrepancy between the average experimental uncertainty and the RMSE indicates that explicit consideration of experimental uncertainty (through weighted least-squares regression) may have a minor effect on the model parameters.
	This hypothesis can be examined by iteratively reweighted linear least-squares regression \cite{pernot2015,gentle2007} (see the Appendix for more details).
	Table \ref{tab:weight} summarizes the results for B3LYP, which was repeatedly found to perform superior to other density functionals in isomer-shift calibration.\cite{neese2002, zhang2002, nemykin2006, romelt2009, bochevarov2010, papai2013}
	
	\begin{table} [!ht]
	\center
	\caption{Parameters of the linear isomer-shift model, $\delta(\rho_\text{absorber},\mathbf{w})$, optimized with different calibration procedures based on linear least-squares regression.
	Only those reference isomer shifts with reported experimental uncertainty ($N = 30$) were employed.
	The corresponding ACED values were obtained from B3LYP/def2-TZVP calculations.
	Standard deviations in parentheses were obtained from bootstrapping ($B = 10^4$).}
	\label{tab:weight}
	\begin{tabular}{lcc}
	calibration procedure & $w_0$ (mm s$^{-1}$) & $w_1$ (mm s$^{-1}$ bohr$^{3}$) \\
	\hline
	ordinary (unweighted) & 4454.80(15303) & $-$0.38(1) \\
	weighted & 4397.39(14260) & $-$0.37(1) \\
	iteratively reweighted & 4451.15(14963) & $-$0.38(1) \\
	\hline
	\end{tabular}
	\end{table}
	
	While the slope, $w_1$, is almost invariant to the approaches applied, the intercept, $w_0$, changes by 57.41~mm~s$^{-1}$ when switching from ordinary (unweighted) least-squares regression to weighted least-squares regression, whereas least-squares regression and iteratively reweighted least-squares regression yield almost identical results (deviation of 3.65~mm~s$^{-1}$).
	In the first instance, these results suggest that experimental uncertainty does not need to be taken into account since method inadequacy and/or data inconsistency appear to play a dominant role.
	However, the bootstrapped standard deviation of the intercept, which is larger than 100~mm~s$^{-1}$ in each case, indicates that the differences between all calibration procedures are insignificant.
	Qualitatively analogous results were obtained for the remaining density functionals.
	As a consequence, we assume that calibration of our property model, $\delta(\rho_\text{absorber},\mathbf{w})$, is not perturbed by the inclusion of reference isomer shifts for which experimental uncertainty was not reported.
	If not mentioned otherwise, the application of bootstrapping will imply ordinary (unweighted) least-squares regression, which is equivalent to assigning the average experimental uncertainty of $\langle u \rangle = 0.02$~mm~s$^{-1}$ to all reference isomer shifts.
	
	In the next two subsections, we will explore the possibilities to unravel the effects of method inadequacy and data inconsistency on the model residuals.

	
\subsection{Selection of the property model with Occam's razor}

	First of all, we examine whether systematic method inadequacy is present.
	Systematic method inadequacy would result in residuals which show a trend with respect to the underlying property model instead of random scatter.
	While Gaussian process regression\cite{rasmussen2006} is a reliable approach to infer the model complexity with the highest transferability, we will apply a simple alternative based on Occam's razor.
	In addition to the linear model, which is based on a physical theory\cite{gutlich2011}, we choose quadratic, cubic, and quartic models,
	\begin{equation}
	\label{eq:quadratic}
	\delta_M(\rho_\text{absorber},\mathbf{w}) = \sum_{m=0}^{M} w_m \rho^m_\text{absorber} \, ,
	\end{equation}
	with $M = 2, 3, 4$.
	In bootstrapping procedures, the model parameters are calibrated with respect to a bootstrap sample and then validated at a reference sample.
	Therefore, if the model is too rigid or too flexible (i.e., when it features too few or too many parameters, respectively), we will observe decreased transferability compared to a balanced model through an increase of the MPU.
	Table \ref{tab:occam} summarizes the results for B3LYP.
	
	\begin{table} [!ht]
	\center
	\caption{Estimated MPU (R632) for four property models of increasing polynomial degree, $M$, on the basis of $B = 10^4$ bootstrap samples ($N = 44$), respectively.
	The corresponding ACED results were obtained from B3LYP/def2-TZVP calculations.
	The linear and quadratic models reveal the lowest MPU.}
	\label{tab:occam}
	\begin{tabular}{lcc}
	property model & $M$ & R632 (mm s$^{-1}$) \\
	\hline
	linear & 1& 0.07 \\
	quadratic & 2 & 0.07 \\
	cubic & 3 & 0.08 \\
	quartic & 4 & 0.10  \\
	\hline
	\end{tabular}
	\end{table}
	
	The linear and quadratic models reveal the lowest MPU as measured by the .632 estimator.
	According to Occam's razor, we choose the simpler of both models, which is also the only one built on physical grounds\cite{gutlich2011}.
	By inspection of Fig.\ \ref{fig:regression}, the residuals of the linear model appear randomly distributed.
	Qualitatively identical results were obtained for the remaining density functionals.
	Consequently, we may assume that the discrepancy between RMSE and $\langle u \rangle$ is rooted in inconsistent data and/or random method inadequacy.
	The latter effect would be a consequence of non-constant systematic errors introduced by the electronic structure method under consideration, which cause an increase in data noise.\cite{pernot2015}
		
	
\subsection{Assessment of data inconsistency based on jackknife-after-bootstrapping}		
		
	In Fig.\ \ref{fig:regression} (left), uncertainty in the regression parameters is represented by the 95\% confidence band (turquoise area, see the Supporting Information for details) obtained from bootstrapping ($B = 10^5$).
	The corresponding ACED values were obtained from B3LYP calculations.
	The black line represents the mean of regression parameters over all bootstrap samples, $\bar{\mathbf{w}}^\ast$, as defined in Eq.\ (\ref{eq:mean_w}).
	R632, RMPV, and RLOO equal the RMSE (0.07~mm~s$^{-1}$); see also Table \ref{tab:error44}.
	Note that the experimental resolution is limited to 0.01~mm~s$^{-1}$.
	If instead it would be artificially increased to 0.001~mm~s$^{-1}$, which can be approximated by adding a trailing zero to the reference isomer shifts, we obtain RMSE $=$ 0.066~mm~s$^{-1}$, R632 $=$ 0.070~mm~s$^{-1}$, RMPV $=$ 0.069~mm~s$^{-1}$, and RMPV $=$ 0.070~mm~s$^{-1}$.
	For details on a statistically valid increase of the experimental resolution, see the Appendix.
	Hence, the low experimental resolution masks the effect of parameter uncertainty on the MPU as measured by R632, RMPV, and RLOO, at least for this specific composition and number of isomer shifts.
	Moreover, the RLOO appears to be an efficient alternative to the R632 for isomer-shift calibration.
	Qualitatively similar results were also obtained for the remaining density functionals (Table \ref{tab:error44}).
	
	\begin{figure} [!ht]
\center
\includegraphics[scale=0.75]{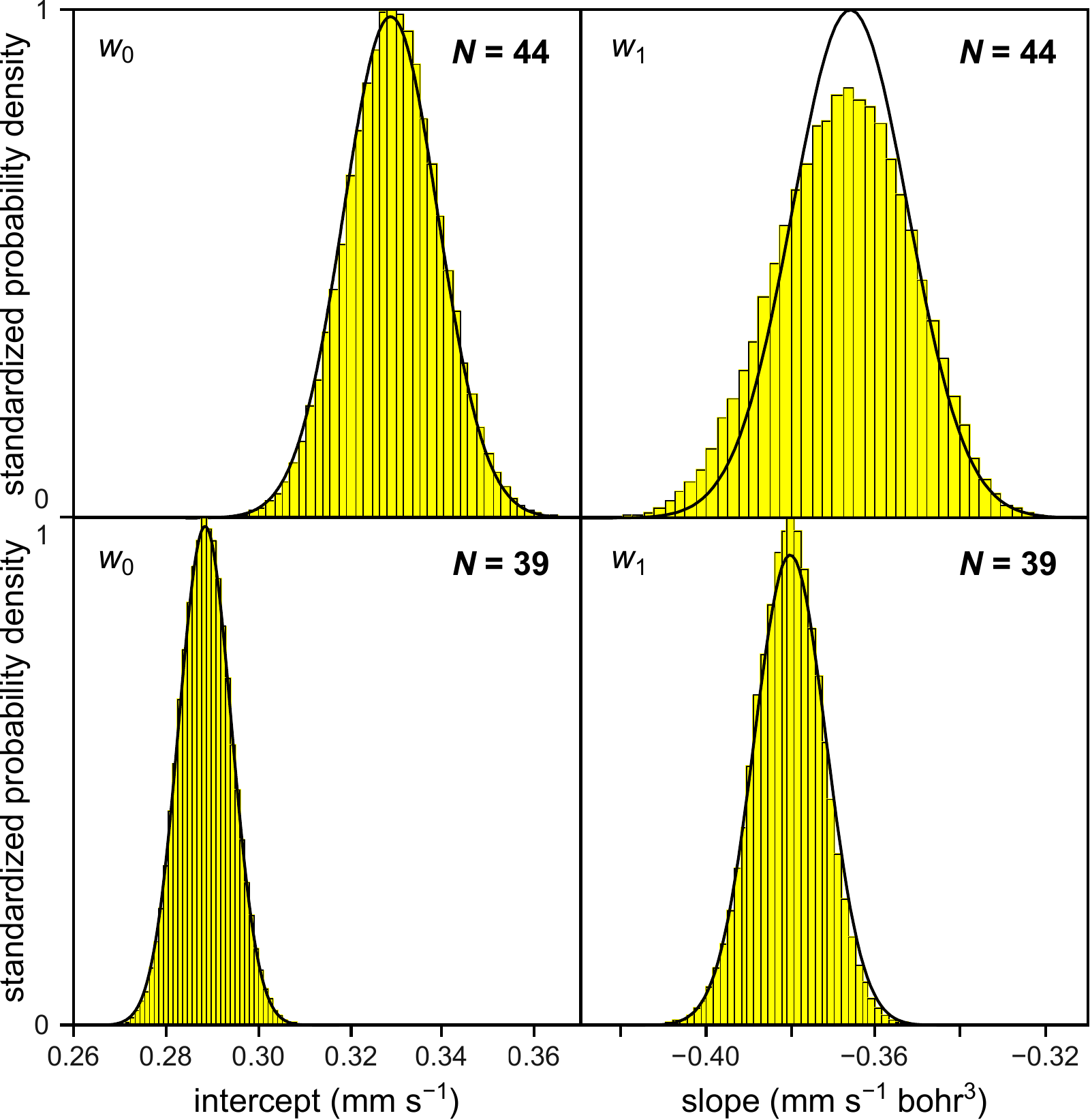}
\caption{Bootstrapped parameter distributions ($B = 10^5$, histograms with 50 bars each) and Gaussian posterior parameter distributions (solid curves) obtained in the presence (top) and in the absence (bottom) of the inconsistent data points \#1, \#2, \#7, \#13, \#28 (cf.\ Table \ref{tab:compounds} and Fig.\ \ref{fig:regression}, left).
After their removal, the standard deviation of both intercept, $w_0$, and slope, $w_1$, introduced in Eq.\ (\ref{eq:model}) decreases by a factor of about 2, and the corresponding mean values are obviously shifted.
Furthermore, the similarity between the two types of parameter distributions increases.
Here, we employed an isomer-shift model with a centered ACED, i.e., $\delta(\rho) = w_0 + w_1(\rho - \bar{\rho})$, where we dropped the subscript ``absorber''. $\bar{\rho}$ refers to the mean ACED of the reference data set considered.
The corresponding statistics of the model parameters are summarized in Table \ref{tab:stats}.
All ACED values were obtained from B3LYP/def2-TZVP calculations.
}
\label{fig:distribution}
\end{figure}
	
	In Fig.\ \ref{fig:distribution} (top left and top right), histograms of the bootstrapped parameter distributions ($B = 10^5$) are shown.
	The solid curves are Gaussian posterior distributions obtained from Bayesian linear regression.
	We find systematic deviation of the posterior slope distribution from its bootstrapped variant, in particular with respect to higher moments (skewness), which may explain the slight difference between R632 and RMPV when artificially increasing the experimental resolution.
	The skewness of the bootstrapped slope distributions may arise from data points with an above-average effect on the objective (here, sum of least squares) such as those at the boundaries of the input domain or apparent outliers.
	By contrast, the agreement between the posterior intercept distribution and its bootstrapped variant appears to be sufficiently high.
	Q--Q plots of bootstrapped and Gaussian parameter distributions (Fig.\ S1) support these findings.

	\begin{table} [!ht]
	\center
	\caption{Statistics of model parameters corresponding to the distributions shown in Fig.\ \ref{fig:distribution}.
	$\bar{w}_m$ and $\sigma_{w_m}$ ($m = \lbrace 0,1 \rbrace$) correspond to mean and standard deviation of the inferred parameter distributions, respectively. 
	$\text{corr}_{w_0,w_1}$ refers to the correlation between $w_0$ and $w_1$ as measured by the square root of $r^2$.
	For better comparability, the results have not been rounded to the experimental resolution (0.01~mm~s$^{-1}$).}
	\label{tab:stats}
	\begin{tabular}{lcccccc}
	inference method & $N$ & $\bar{w}_0$ & $\bar{w}_1$ & $\sigma_{w_0}$ & $\sigma_{w_1}$ & $\text{corr}_{w_0,w_1}$ \\
	\hline
	\hline
	bootstrapping & 44 & 0.3295 & $-$0.3674 & 0.0101 & 0.0157 & $-$0.4720 \\
	& 39 & 0.2888 & $-$0.3805 & 0.0056 & 0.0082 & $-$0.0776 \\
	\hline
	Bayesian & 44 & 0.3288 & $-$0.3658 & 0.0102 & 0.0137 & 0.0000 \\
	& 39 & 0.2886 & $-$0.3801 &  0.0057 & 0.0085 & 0.0000 \\
	\hline
	\end{tabular}
	\end{table}	
	
\begin{figure} [!ht]
\center
\includegraphics[width=\textwidth]{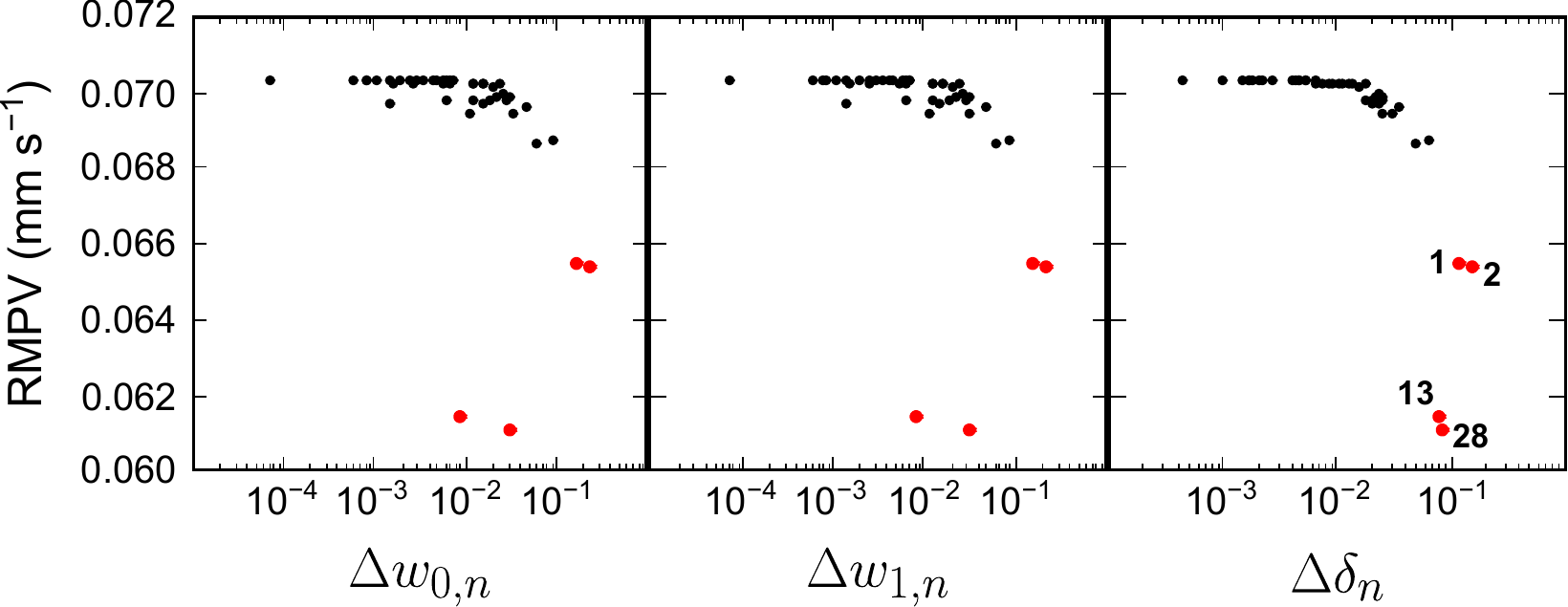}
\caption{RMPV of $N$ jackknife data sets, $\mathcal{D}_{-n}$, versus three measures of the difference between various calibrations of $\delta(\rho_\text{absorber},\mathbf{w})$.
The reference calibration ($B = 10^4$) is performed with respect to the complete reference data set, $\mathcal{D}$, yielding $\bar{\mathbf{w}}^\ast$, which is compared to $N$ calibrations based on the jackknife data sets, $\mathcal{D}_{-n}$, yielding $\bar{\mathbf{w}}^\ast_{-n}$.
Left: $\Delta w_{0,n}$, normalized absolute deviation of $\bar{w}_{0,-n}^\ast$ from $\bar{w}_0^\ast$ (intercept). 
Center: $\Delta w_{1,n}$, normalized absolute deviation of $\bar{w}_{1,-n}^\ast$ from $\bar{w}_1^\ast$ (slope). 
Right: $\Delta \delta_n$, normalized root-mean-square deviation of $\delta(\rho_\text{absorber},\bar{\mathbf{w}}^\ast_{-n})$ from $\delta(\rho_\text{absorber},\bar{\mathbf{w}}^\ast)$.
The data points highlighted (\#1, \#2, \#13, \#28, cf.\ Table \ref{tab:compounds} and Fig.\ \ref{fig:regression}, left) are potentially inconsistent as their removal leads to significant changes in the MPU as measured by the RMPV.
For better legibility, the results of the RMPV have not been rounded to the experimental resolution (0.01~mm~s$^{-1}$).
The corresponding ACED values were obtained from B3LYP/def2-TZVP calculations.}
\label{fig:jackknife}
\end{figure}

	To examine the reliability of the sampled parameter distributions and the MPU associated with them, we apply the jackknife-after-bootstrapping method.
	An overview in Fig.\ \ref{fig:jackknife} (left) of the RMPV versus the normalized unsigned deviation in the intercept, $\Delta w_{0,n}$, 
	\begin{equation}
	\Delta w_{0,n} \equiv \frac{\vert \bar{w}_{0,-n}^\ast - \bar{w}_0^\ast \vert}{\sum_{n=1}^N \vert \bar{w}_{0,-n}^\ast - \bar{w}_0^\ast \vert} \, ,
	\end{equation}	
optimized with respect to both $\mathcal{D}$ and $\mathcal{D}_{-n}$,
	 reveals no clear correlation between the quantities (here, $\mathcal{D}_{-n}$ refers to the reference data set with the $n$-th data point removed, and the RMPV is given with respect to $\mathcal{D}_{-n}$).
	The same holds true for the correlation between the RMPV and the  normalized unsigned deviation in the slope (Fig.\ \ref{fig:jackknife}, center),
	\begin{equation}
	\Delta w_{1,n} \equiv \frac{\vert \bar{w}_{1,-n}^\ast - \bar{w}_1^\ast \vert}{\sum_{n=1}^N \vert \bar{w}_{1,-n}^\ast - \bar{w}_1^\ast \vert} \, .
	\end{equation}	
	As the parameters are correlated in the property model, we also calculated the normalized root-mean-square deviation of the property model, $\Delta \delta_n$, calibrated with respect to both $\mathcal{D}$ and $\mathcal{D}_{-n}$,
\begin{equation}
\Delta \delta_n \equiv \frac{\sqrt{\sum_{i=1}^N \Big(\delta(\rho_{i,\text{absorber}},\bar{\mathbf{w}}^\ast_{-n}) - \delta(\rho_{i,\text{absorber}},\bar{\mathbf{w}}^\ast) \Big)^2}}{\sum_{n=1}^N\sqrt{\sum_{i=1}^N \Big(\delta(\rho_{i,\text{absorber}},\bar{\mathbf{w}}^\ast_{-n}) - \delta(\rho_{i,\text{absorber}},\bar{\mathbf{w}}^\ast) \Big)^2}} \, .
\end{equation}  
	A plot (Fig.\ \ref{fig:jackknife}, right) of the RMPV versus $\Delta \delta_n$ reveals a more distinct correlation.
	At smaller deviations, the RMPV is almost constant, whereas it decreases overlinearly at larger deviations.
	Hence, if the removal of a data point significantly changes the functional form of the property model, it also has a significant effect on the MPU.
	
\begin{figure} [!ht]
\center
\includegraphics[width=\textwidth]{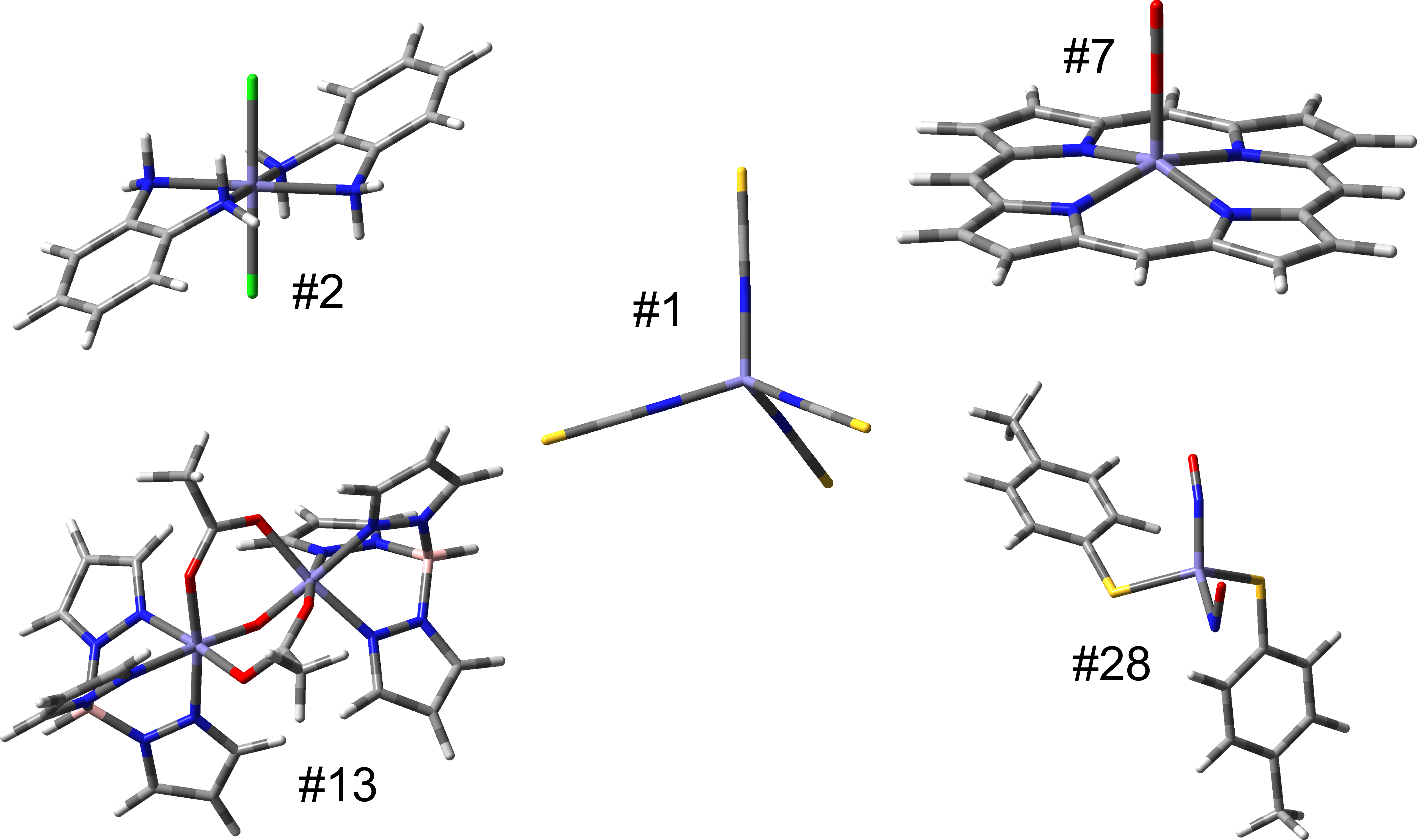}
\caption{Molecular structures (tube models) corresponding to the inconsistent input--target pairs \#1, \#2, \#7, \#13, \#28 (cf.\ Table \ref{tab:compounds} and Fig.\ \ref{fig:regression}, left).
All complexes exhibit at least one iron--nitrogen bond.
Color code: magenta, iron; light gray, carbon; white, hydrogen; green, chlorine; blue, nitrogen; red, oxygen; yellow, sulfur; dark gray, chemical bond.
}
\label{fig:critical}
\end{figure}

	The plots of Fig.\ \ref{fig:jackknife} reveal four data points (\#1, \#2, \#13, \#28, cf.\ Table \ref{tab:compounds} and Fig.\ \ref{fig:regression}, left) that lead to a distinct decrease of the MPU and high values of $\Delta \delta_n$, respectively.
	The question arises whether these data points are inconsistent in the sense that if they are present, the underlying population distribution may not be well-represented by the data set under consideration.
	Since we are studying several density functionals, we can examine whether we find the same potentially inconsistent data points in the remaining cases.
	We find that three of the four potentially inconsistent data points identified with B3LYP are also identified as most likely inconsistent with all other density functionals (we considered the four highest values of $\Delta \delta_n$ for each density functional).
	This finding indicates that these data points (\#1, \#2, \#28) are not affected by inconsistent ACED calculations as they have been identified to be inconsistent irrespective of the density functional employed.
	Rather, systematic measurement errors or deficiencies in the molecular-structure representation may be responsible for their inconsistent status.
	Moreover, the fourth potentially inconsistent data point identified with B3LYP (\#13) has also been identified with three other hybrid density functionals (B3PW91, PBE0, M06) and is ranked fifth with respect to $\Delta \delta_n$ for the remaining hybrid density functional (TPSSh).
	In this case, we interpret data point \#13 as affected by inconsistent ACED calculations (method inadequacy) as it was only identified inconsistent by a particular category of density functionals.
	Likewise, data point \#7 (cf.\ Table \ref{tab:compounds} and Fig.\ \ref{fig:regression}, left) is one of the first four data points identified potentially inconsistent by all pure density functionals (PWLDA, BP86, BLYP, PW91, PBE, M06-L, TPSS) and one hybrid density functional (TPSSh), and is ranked fifth with respect to $\Delta \delta_n$ for all remaining hybrid density functionals but M06.
	We do not find obvious similarities in the corresponding compounds regarding the different categories in Table \ref{tab:compounds}, but in all five cases, iron is coordinated to nitrogen (Fig.\ \ref{fig:critical}).
	We also compared the calculated versus ideal expectation value of the $\langle S^2 \rangle$ operator for all open-shell complexes (Table S5), but find no anomalies for the inconsistent data points.
	Likewise, we cannot confirm that the root-mean-square deviation of atomic positions is particularly high for those iron complexes corresponding to inconsistent input--target pairs (Table S1).
	
	To obtain reliable estimates of the MPU, we decided to remove all inconsistent data points (\#1, \#2, \#7, \#13, \#28, cf.\ Table \ref{tab:compounds} and Fig.\ \ref{fig:regression}, left).
	Repeating bootstrapped ($B = 10^5$) linear least-squares regression (Fig.\ \ref{fig:regression}, right), the updated property model (solid line) clearly deviates from that one calibrated with respect to the reference data set including the inconsistent input--target pairs (dashed line).
	Fig.\ \ref{fig:distribution} (bottom left and bottom right) reveals that mean and standard deviation of the bootstrapped parameter distributions (histograms) change significantly, the latter decreases by a factor of 2 (cf.\ Table \ref{tab:stats}).
	The corresponding Gaussian posterior distributions obtained from Bayesian linear regression (solid curves) are now both quite similar to their sampled counterparts.
	This finding is also supported by Q--Q plots of bootstrapped versus Gaussian parameter distributions (Fig.\ S1).
	Even though the 95\% confidence band (Fig.\ \ref{fig:regression}, right) is narrower than before, we now find a difference between the RMSE (0.03~mm~s$^{-1}$) and R632 as well as RMPV and RLOO (0.04~mm~s$^{-1}$, respectively).
	This finding is an artifact resulting from the low experimental resolution of 0.01~mm~s$^{-1}$.
	Increasing the experimental resolution artificially from 0.01~mm~s$^{-1}$ to 0.001~mm~s$^{-1}$, we obtain RMSE $=$ 0.035~mm~s$^{-1}$, R632 $=$ 0.036~mm~s$^{-1}$, RMPV $=$ 0.036~mm~s$^{-1}$, and RLOO $=$ 0.036~mm~s$^{-1}$.
	In this case, we see that the difference between RMSE and R632 decreases from 0.004~mm~s$^{-1}$ to 0.001~mm~s$^{-1}$, which we would expect for a narrowing of the confidence band.
	Furthermore, due to the higher similarity of the bootstrapped and Gaussian posterior distributions after removal of the inconsistent data points, we also find that the difference between R632 and RMPV vanishes (before: 0.001~mm~s$^{-1}$) for the increased experimental resolution.
	This result indicates that the normal-population assumption appears to be reasonable for isomer-shift calibration after removal of inconsistent data points.
	Again, the RLOO appears to be an efficient alternative to the R632 for isomer-shift calibration.
	Note that the updated RMSE (0.03~mm~s$^{-1}$ instead of 0.07~mm~s$^{-1}$) is now much closer to the average experimental uncertainty ($\langle u \rangle = 0.02$~mm~s$^{-1}$).
	For the remaining density functionals, the RMSE ranges from 0.03~mm~s$^{-1}$ (B3PW91, PBE0) to 0.08~mm~s$^{-1}$ (BLYP), while R632, RMPV, and RLOO range from 0.04~mm~s$^{-1}$ (B3PW91, PBE0, TPSSh) to 0.08~mm~s$^{-1}$ (BLYP); see also Table \ref{tab:error39}.
	
	\begin{table} [!ht]
\small
\caption{Performance measures (RMSE, R632, RMPV, RLOO) of density functionals regarding isomer-shift prediction.
The reference data set is the original one with the inconsistent data points (\#1, \#2, \#7, \#13, \#28, cf.\ Table \ref{tab:compounds} and Fig.\ \ref{fig:regression}, left) removed ($N = 39$).
RMSE, R632, and RLOO were calculated on the basis of ordinary (unweighted) least-squares regression.
The RMPV was obtained from Bayesian linear regression based on the evidence approximation.
For the calculation of the .632 estimator, R632, $B = 10^5$ bootstrap samples were generated. 
All performance measures are reported in mm~s$^{-1}$.
The corresponding ACED values were obtained with the def2-TZVP basis set for ligand atoms.}
\label{tab:error39}
\begin{center}
\begin{tabular}{lcccc}
functional & RMSE & R632 & RMPV & RLOO \\
\hline
PWLDA & 0.07 & 0.07 & 0.07 & 0.07 \\
BP86 & 0.07 & 0.07 & 0.07 & 0.07 \\
BLYP & 0.08 & 0.08 & 0.08 & 0.08 \\
PW91 & 0.07 & 0.07 & 0.07 & 0.07 \\
PBE & 0.07 & 0.07 &0.07 & 0.07 \\
M06-L & 0.06 & 0.06 & 0.06 & 0.06 \\
TPSS & 0.06 & 0.06 & 0.06 & 0.06 \\
B3LYP & 0.03 & 0.04 & 0.04 & 0.04 \\
B3PW91 & 0.03 & 0.04 & 0.04 & 0.04 \\
PBE0 & 0.03 & 0.04 & 0.04 & 0.04 \\
M06 & 0.05 & 0.05 & 0.05 & 0.05 \\
TPSSh & 0.04 & 0.04 & 0.04 & 0.04 \\
\hline
\end{tabular}
\end{center}
\end{table}

\begin{figure} [!ht]
\center
\includegraphics[width=\textwidth]{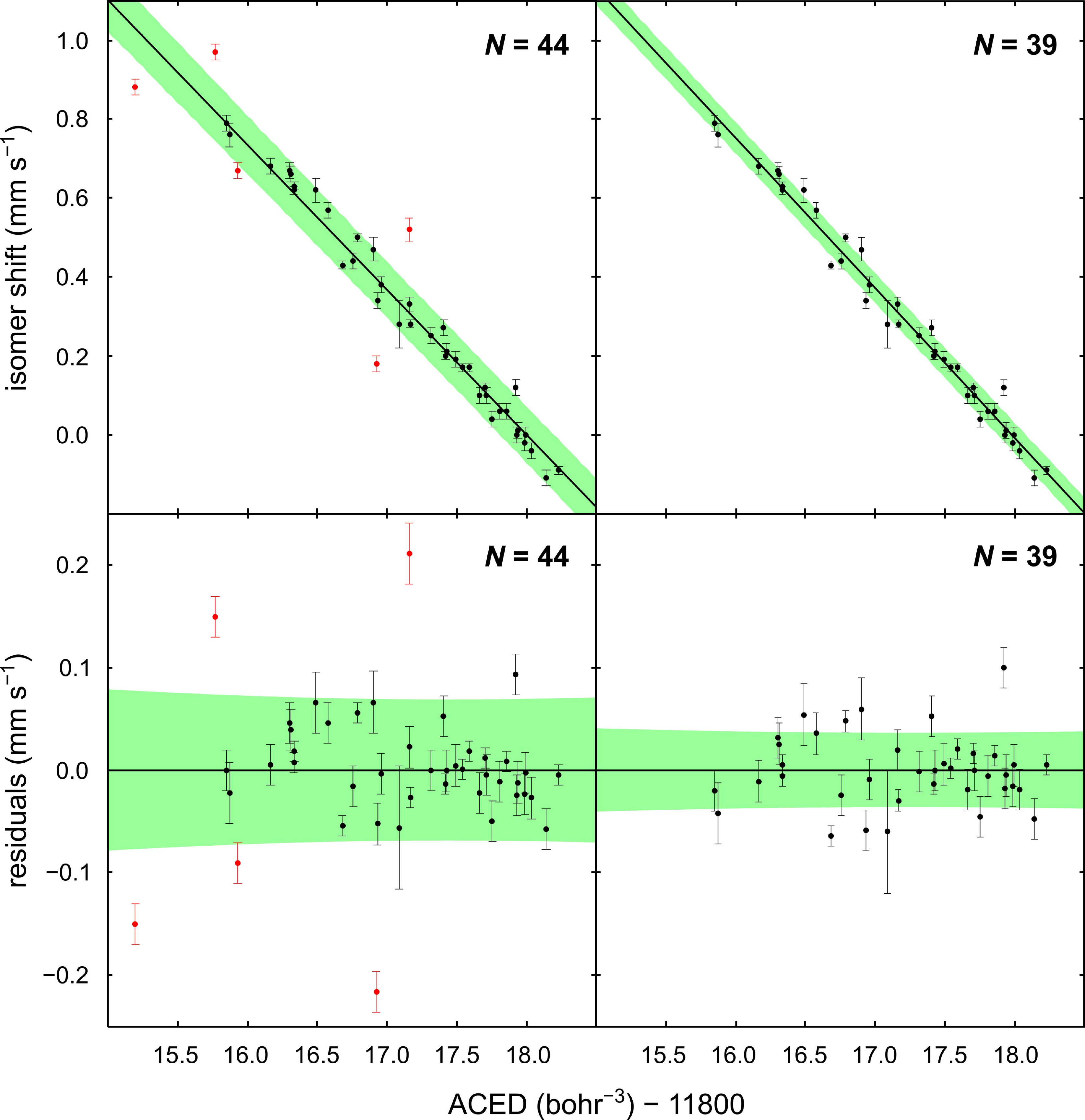}
\caption{Results of bootstrapped ($B = 10^5$) linear least-squares regressions in the presence (top left) and in the absence (top right) of the inconsistent data points \#1, \#2, \#7, \#13, \#28 (cf.\ Table \ref{tab:compounds} and Fig.\ \ref{fig:regression}, left).
The corresponding diagrams of residuals, $\delta_\text{exp} - \delta(\rho_\text{absorber},\bar{\mathbf{w}}^\ast)$, are shown at the bottom.
The 68\% prediction band (green area) comprises 86\% (left) and 72\% (right) of the data points.
Consequently, jackknife-after-bootstrapping significantly improved MPU estimation in this case.
The solid lines are equivalent to those shown in Fig.\ \ref{fig:regression}.
The corresponding ACED values were obtained from B3LYP/def2-TZVP calculations.
}
\label{fig:prediction}
\end{figure}

	Noteworthy, while inclusion of the inconsistent data points in the reference data set ($N = 44$) leads to overestimation of the MPU as measured by the R632 (86\% of the data points lie within the 68\% prediction band obtained from $u(\mathbf{x})$ defined in Eq.\ (\ref{eq:local_error})), their exclusion ($N = 39$) results in a significant improvement of MPU estimation (72\% of the data points lie within the 68\% prediction band); see also Fig.\ \ref{fig:prediction}.
	Note that the prediction bands shown in Fig.\ \ref{fig:prediction} (bottom) can be directly related to $u(\mathbf{x})$ defined in Eq.\ (\ref{eq:local_error}).
	
	Because of the decrease of the RMSE, we examined once again the importance of explicitly considering experimental uncertainty.
	The results are summarized in Table \ref{tab:weight2} and lead us again to the conclusion that calibration of $\delta(\rho_\text{absorber},\mathbf{w})$ is not perturbed by the inclusion of reference isomer shifts for which experimental uncertainty has not been reported.
	Furthermore, we applied Occam's razor to our consistent reference data set ($N = 39$).
	In all cases ($M = 1,2,3,4$), we obtained an R632 of 0.04~mm~s$^{-1}$, which confirms once again the validity of the linear (physical) model.

	\begin{table} [!ht]
	\center
	\caption{Parameters of the linear isomer-shift model, $\delta(\rho_\text{absorber},\mathbf{w})$, optimized with different calibration procedures based on linear least-squares regression.
	Only those reference isomer shifts with reported experimental uncertainty \textit{and} consistent status ($N = 28$) were employed.
	The corresponding ACED values were obtained from B3LYP/def2-TZVP calculations.
	Standard deviations in parentheses were obtained from bootstrapping ($B = 10^4$).}
	\label{tab:weight2}
	\begin{tabular}{lcc}
	calibration procedure & $w_0$ (mm s$^{-1}$) & $w_1$ (mm s$^{-1}$ bohr$^{3}$) \\
	\hline
	ordinary (unweighted) & 4510.05(11754) & $-$0.38(1) \\
	weighted & 4434.25(12639) & $-$0.38(1) \\
	iteratively reweighted & 4495.55(10932) & $-$0.38(1) \\
	\hline
	\end{tabular}
	\end{table}	
	
	Combining all these findings, we are confident that the updated, consistent reference data set represents the underlying population distribution sufficiently well if hybrid density functionals generate the input data.
Hence, our reference set of $N = 39$, already pruned by statistically critical outliers, 
provides a well-defined starting ground for further parametrization studies or for systematic extensions of the reference data.
Hence, we may call this special set of data the MIS39 data set (for 39 reference data points of M\"ossbauer isomer shifts).


	\subsection{How reliable are density functional rankings based on a specific data set?}
	
	In the previous subsection, we found for the particular composition and number of data points that the effect of parameter uncertainty on the MPU is very small ($<$0.01~mm~s$^{-1}$ for all density functionals).
	Hence, given this specific reference data set, we are confident that the RMSE (and with it the squared coefficient of correlation, $r^2$) is a good approximation to performance measures such as R632, RMPV, and RLOO, and therefore, suited to set up a ranking of density functionals.
	In the following, we only distinguish between first and other places in a ranking as only those density functionals will be considered for actual applications that reveal highest transferability.
	Regarding our reference set of isomer shifts ($N = 39$), B3LYP, B3PW91, and PBE0 are placed first in all rankings studied (RMSE, R632, RMPV, RLOO), with TPSSh being placed first in the rankings based on R632, RMPV, and RLOO (cf.\ Table \ref{tab:error39}).
	This finding is consistent with those of other calibration studies of the isomer shift.\cite{neese2002, zhang2002, nemykin2006, romelt2009, bochevarov2010, papai2013}
	
	However, in practice, it is relevant to know which density functional yields the most accurate predictions \textit{independent} of the reference data employed for calibration of the property model.
	So far, we considered rankings of density functionals conditioned on a specific reference data set, but what we aim at is an unconditional ranking of density functionals.
	Otherwise, we cannot assess the transferability of a property model trained on a specific density functional to data not involved in its calibration.
	Clearly, the dependency of any statistical measure on a specific choice of reference data cannot entirely be removed, but bootstrapping can yield a good approximation to the problem given the data under consideration is representative of the underlying population distribution, which we have assessed in detail by the jackknife-after-bootstrapping method.
	
\begin{figure} [!ht]
\center
\includegraphics[width=\textwidth]{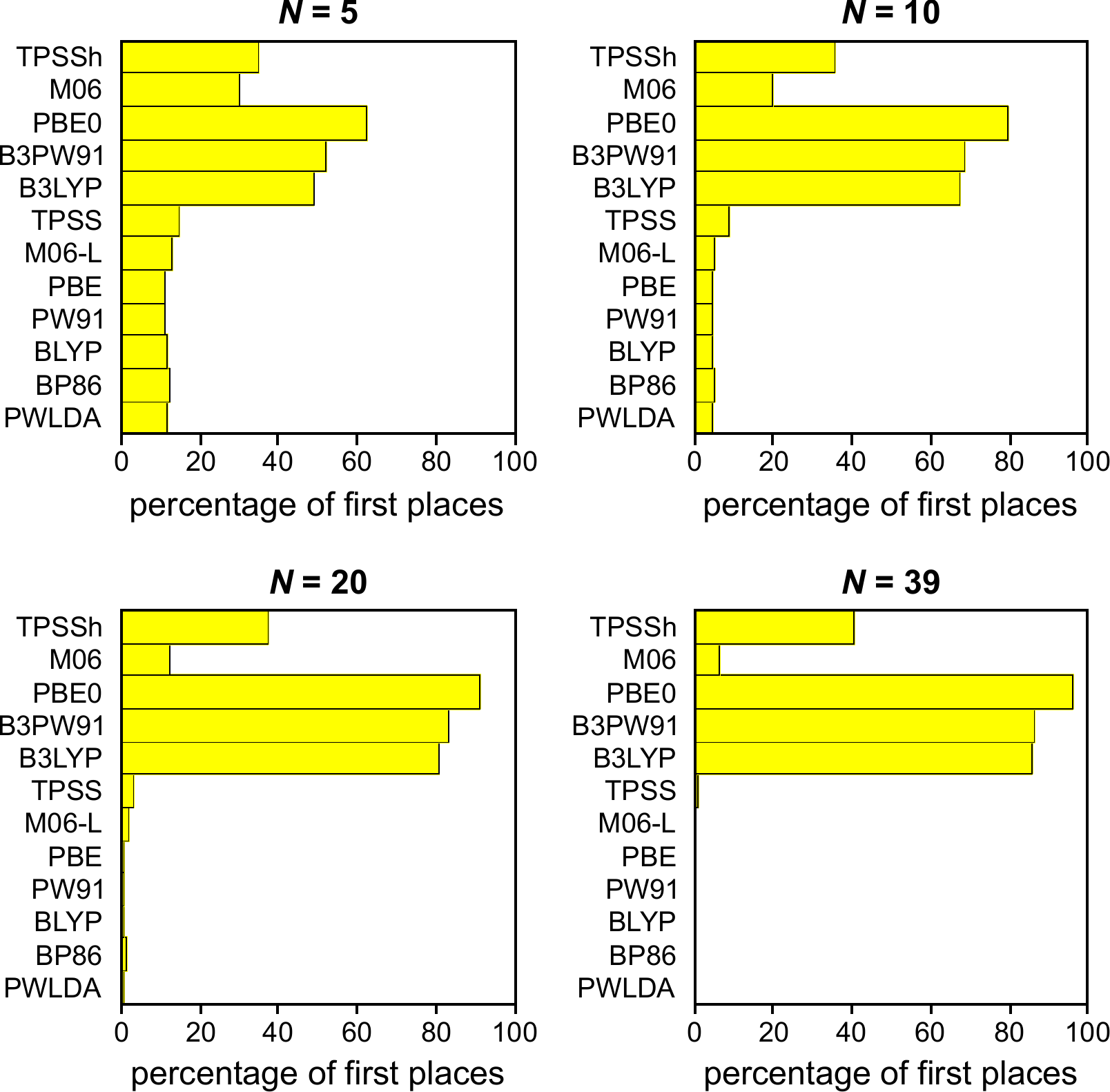}
\caption{Percentage of first places a density functional reached for $B = 2500$ synthetic data sets of different size ($N = 5, 10, 20, 39$), which were generated by drawing from the empirical population distribution of the consistent reference data set ($N = 39$) with replacement.
The rankings were determined on the basis of the RMSE, which was rounded to the experimental resolution of 0.01 mm s$^{-1}$.
Hence, more than one density functional can be placed first for a given synthetic data set, which is why the bars sum up to $>$100\%.
For every bootstrap sample, the isomer shifts have been perturbed randomly according to their experimental uncertainty (for details, see the Appendix).
While false conclusions about the transferability of a density functional are likely for 5 data points (no density functional is placed first in more than about 60\% of the cases), the ``best'' density functional, PBE0, is identified first in more than 95\% of the cases for $N = 39$, the size of the reference data set employed in this study.
The corresponding ACED values were obtained with the def2-TZVP basis set for ligand atoms.
}
\label{fig:ranking}
\end{figure}

	To study the reliability of density functional rankings with respect to the composition and number of data points, we have drawn samples of different size ($N = 5, 10, 20, 39$) with replacement ($B = 2.5 \times 10^3$, respectively) from the empirical population distribution of the consistent reference data set ($N = 39$).
	For every bootstrap sample, the isomer shifts have been perturbed randomly according to their experimental uncertainty (for details, see the Appendix) to allow for statistically justifiable variation between the synthetic data sets.
	In Fig.\ \ref{fig:ranking}, we show the percentage of first places that a density functional has reached for 2500 different data sets and four different data-set sizes.
	The rankings are based on the RMSE.
	
	For all different data sets considered, B3LYP, B3PW91, and PBE0 are still most frequently placed first, but the respective percentage varies significantly.
	For $N = 5$ data points, PBE0 reveals the highest number of first places, but only in about 60\% of the cases.
	Even all pure density functionals are placed first in at least 10\% of the cases, although the different performance measures are clearly in favor of hybrid density functionals regarding the reference data set (cf.\ Table \ref{tab:error39}).
	Hence, calibration studies based on 5 data points are highly susceptible to random conclusions about the transferability of density functionals.
	With an increasing number of data points, the percentage of first places continuously decreases for all pure density functionals and for M06, while TPSSh reveals relatively constant results for all data points considered (between 30\% and 40\% first places).
	By contrast, the percentage of first places for B3LYP, B3PW91, and PBE0 increases continuously, with PBE0 being ranked first in $> 95\%$ of the cases.
	Qualitatively identical and quantitatively similar results were obtained for density functional rankings based on the RMPV (cf.\ Fig.\ S2).
	This finding indicates that for the given experimental resolution, parameter uncertainty plays no significant role, even for 5 data points where the standard deviation in the parameters is about 4 times larger compared to 39 data points, the number of reference data points considered in this study (Fig.\ S5).
	Consequently, both RMSE and $r^2$ can be considered stable performance measures for density functional rankings applied in isomer-shift predictions of molecular iron compounds.
	Note that we did not employ the R632 as performance measure as it is very costly to sample an error for every bootstrap sample, which is equivalent to bootstrapping bootstrap samples (double bootstrapping\cite{davison1997}).
	However, as shown above, the RMPV can be expected a good approximation to the R632 (as well as RLOO).
	
\begin{figure} [!ht]
\center
\includegraphics[width=\textwidth]{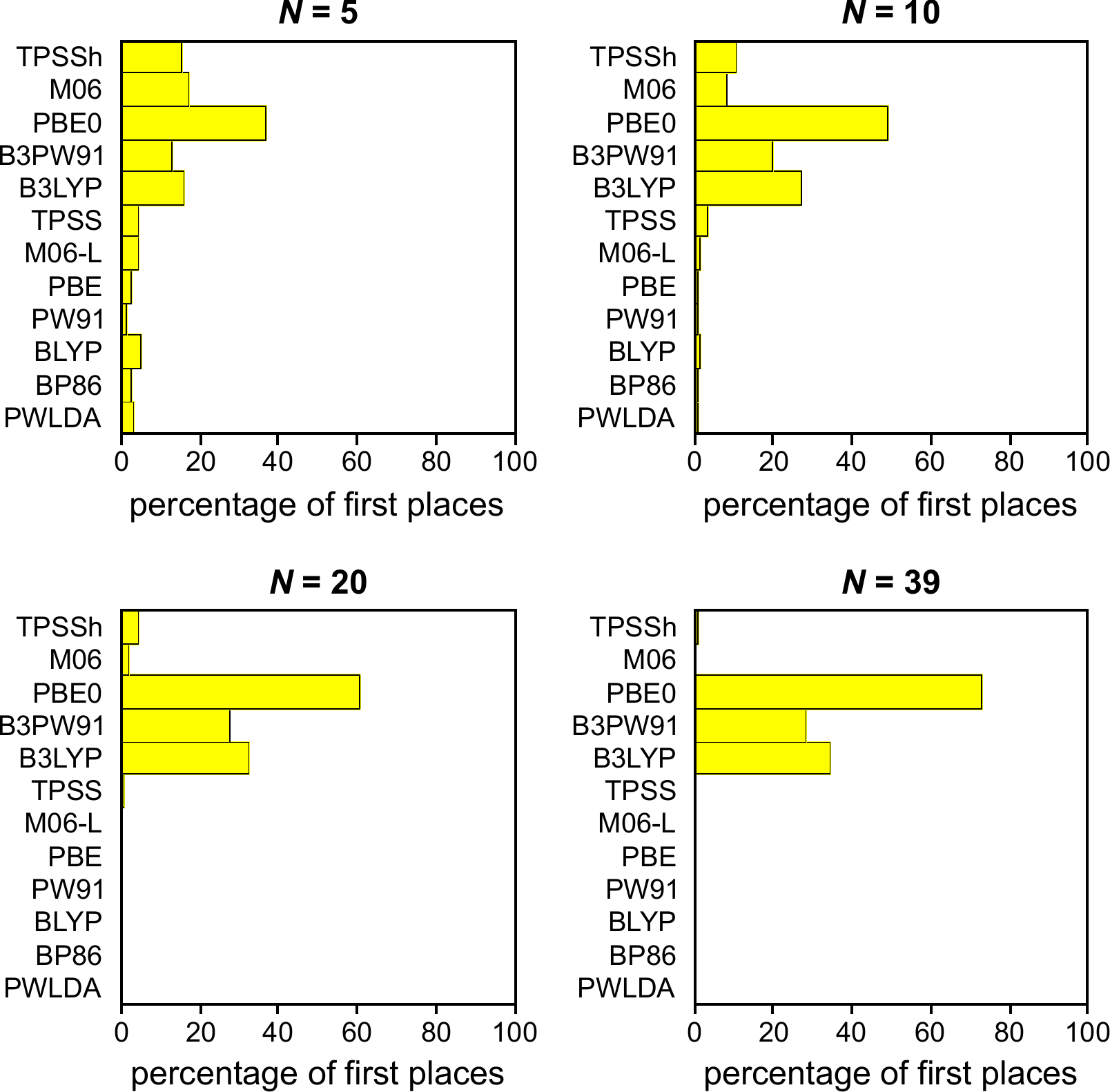}
\caption{Percentage of first places a density functional reached for $B = 2500$ synthetic data sets of different size ($N = 5, 10, 20, 39$), which were generated by drawing from the empirical population distribution of the consistent reference data set ($N = 39$) with replacement.
The rankings were determined on the basis of the RMSE, which was rounded to an artifically increased experimental resolution of 0.001 mm s$^{-1}$ (for details, see the Appendix).
Hence, more than one density functional can be placed first for a given synthetic data set, which is why the bars sum up to $>$100\%.
For every bootstrap sample, the reference isomer shifts have been perturbed randomly according to their experimental uncertainty (for details, see the Appendix).
The dispersion of the rankings increases compared to the actual experimental resolution, and even for 39 data points, the ``best'' density functional, PBE0, is identified as such in only 70--75\% of the cases.
The corresponding ACED values were obtained with the def2-TZVP basis set for ligand atoms.
}
\label{fig:highres}
\end{figure}

	The fact that more than one density functional is placed first on average is clearly an indicator of the low experimental resolution.
	When increasing the experimental resolution from 0.01~mm~s$^{-1}$ to 0.001~mm~s$^{-1}$ in a statistically sound way (for details, see the Appendix), the results based on the RMSE change significantly (Fig.\ \ref{fig:highres}).
	The dispersion of the rankings is much more pronounced for 5 data points (all density functionals are placed first in less than 40\% of the cases), and a clear preference for a density functional (more than 95\% first places) can be expected only for a number of data points significantly larger than that employed in this study ($N = 39$).
	Similar results were obtained for density functional rankings based on the RMPV (cf.\ Fig.\ S3). 

	Reducing the data-set dependency of the MPU by the application of bootstrapping and increasing the experimental resolution in a statistically sound way, we find a clear preference for PBE0 as the density functional with the highest transferability.
	This preference even remains (a) when resubstituting the 5 inconsistent input--target pairs (\#1, \#2, \#7, \#13, \#28, cf.\ Table \ref{tab:compounds} and Fig.\ \ref{fig:regression}, left) into our consistent reference data set ($N = 44$), and (b) when selecting only those reference isomer shifts for which experimental uncertainties were reported (without resubstitution of inconsistent data points, $N = 28$); cf.\ Fig.\ S4.
	Therefore, we suggest that PBE0 is the density functional of choice for applications (with the CP(PPP) basis for Fe and def2-TZVP for all other elements; see Appendix), which leads to the following linear isomer-shift model obtained from bootstrapped ($B = 10^6$) linear least-squares regression,
	\begin{eqnarray}
\label{eq:prediction}
	\delta(\rho_0,\bar{\mathbf{w}}_\text{PBE0}^\ast) &=& 2.888(1) \times 10^{-1} \ \text{mm s}^{-1} \\ &&- \big( \rho_0 - \bar{\rho} \big) \times 3.619(1) \times 10^{-1} \ \text{mm s}^{-1} \ \text{bohr}^{3} \nonumber \, ,\\
	\bar{\rho} &=& \frac{1}{N}\sum_{n=1}^{N=39} \rho_n = 11819.0531906 \ \text{bohr}^{-3} \, ,
	\end{eqnarray}
the corresponding covariance matrix,
        \begin{equation}
        \label{eq:mpu}
        \sigma^2_{{\mathbf{w}}_\text{PBE0}^\ast} = \begin{pmatrix}
        2.959(4) \times 10^{-5} & 5.994(48) \times 10^{-6} \text{ bohr}^3 \\
        5.994(48) \times 10^{-6} \text{ bohr}^3 & 6.641(9) \times 10^{-5} \text{ bohr}^6
\end{pmatrix}    \text{mm}^2 \ \text{s}^{-2} \, ,
        \end{equation}
        and the corresponding reduced MSE,
        \begin{equation}
        \frac{N}{N - M - 1} \text{MSE}_{\mathcal{D},\bar{\mathbf{w}}_\text{PBE0}^\ast} = 1.185(1) \times 10^{-3} \text{ mm}^2 \text{ s}^{-2} \, .
        \end{equation}
        We omitted the subscript ``absorber'' for the ACED, $\rho$.
        For reasons of reproducibility, we specify four significant figures for the characteristic values contained in $\bar{\mathbf{w}}_\text{PBE0}^\ast$ and $\sigma_{{\mathbf{w}}_\text{PBE0}^\ast}$, and for the reduced MSE.
        For the mean ACED, $\bar{\rho}$, we employed the raw-data precision provided in Tables S2 and S3.
        The standard deviations in parentheses were obtained from bootstrapping ($B = 10^3$) the $10^6$ parameter estimates.

        Given a single new ACED, $\rho_0$, the corresponding prediction uncertainty, $u(\boldsymbol{\rho}_0)$ with $\boldsymbol{\rho}_0 = (1,\rho_0 - \bar{\rho})^\top$, would, according to Eq.\ (\ref{eq:local_error}), read
        \begin{equation}
        u(\boldsymbol{\rho}_0) = \sqrt{\frac{N}{N-M-1} \text{MSE}_{\mathcal{D},\bar{\mathbf{w}}_\text{PBE0}^\ast} + \boldsymbol{\rho}_0^\top \sigma^2_{\mathbf{w}_\text{PBE0}^\ast} \boldsymbol{\rho}_0} \, .
        \end{equation}


\subsection{Effect of exact exchange on the MPU}

In this last section, we briefly discuss our observation why hybrid density functionals yielded significantly lower MPU estimates 
compared to pure density functionals. A systematic (even linear) behavior of relative energies of states of different spin multiplicity
on the admixture of exact exchange (i.e., Hartree--Fock-type exchange) 
in an energy density functional has already been observed more than 15 years ago\cite{reiher2001,reiher2002,salomon2002}.
Since then, the exact exchange admixture, measured by the linear parameter $c_3$ following the notation in Ref.\ \onlinecite{reiher2001}, 
has been well recognized as one of the most crucial parameters determining the accuracy of (hybrid) density functionals
(see Refs.\ \onlinecite{ioannidis2015,gani2016,janet2017,ioannidis2017} for recent systematic investigations). The linear dependence of such relative energies in many (but not all)
cases is remarkable and results from a linear dependence of the absolute electronic energies 
on $c_3$, in which the effect of a self-consistently
optimized (and hence, changing) electron density plays a negligible role.

\begin{figure} [!ht]
\center
\includegraphics[width=\textwidth]{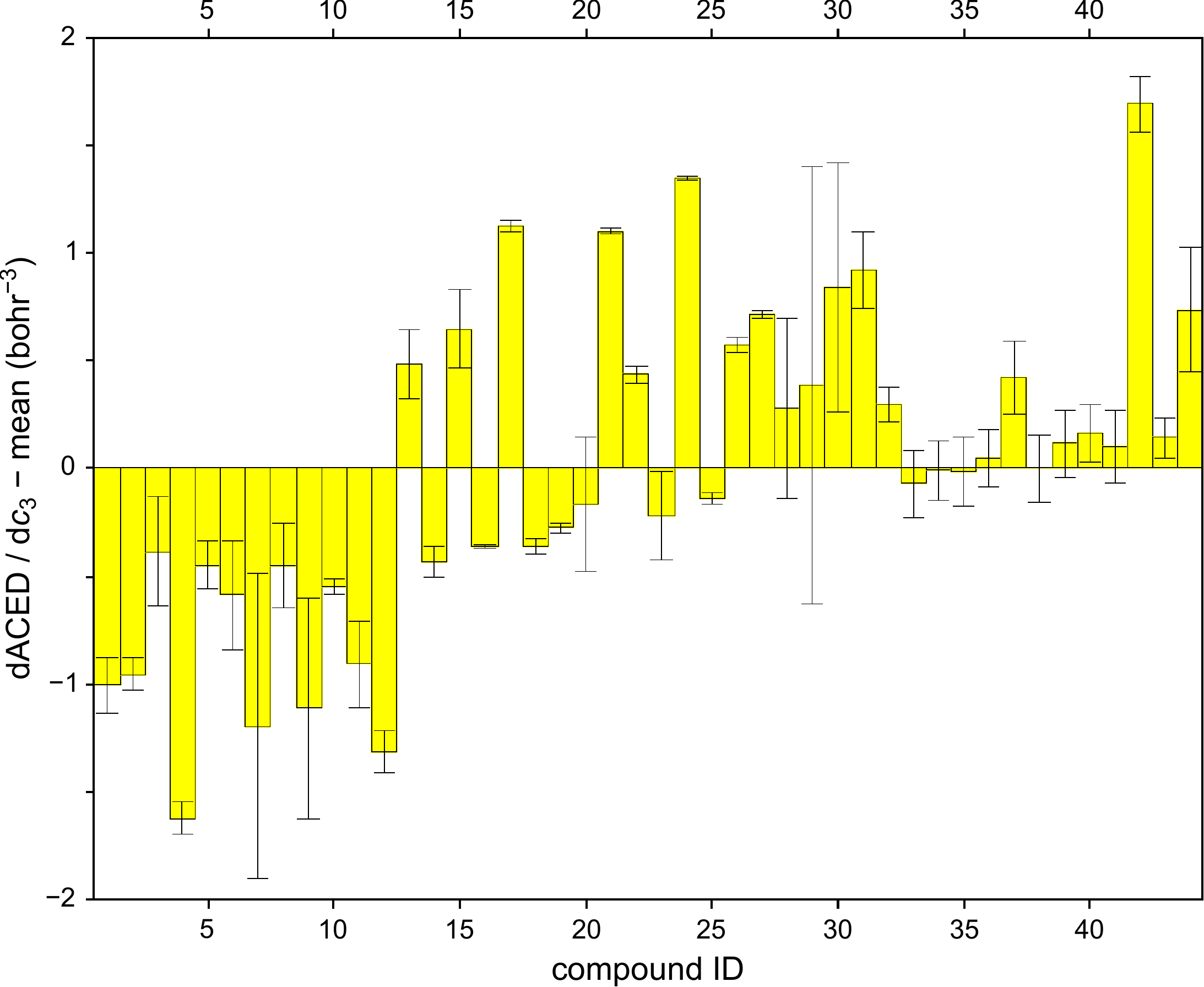}
\caption{Change of the B3LYP/def2-SVP ACED with respect to the exact exchange admixture, dACED / d$c_3$, for all 44 complexes (compound ID on the abscissa; cf.\ Table \ref{tab:compounds} and Figs.\ S6--S16). The constant first derivative indicates the linear behavior of the ACED with respect
to the exact exchange parameter $c_3$.
For better comparability, we subtracted the mean derivative 'mean' obtained for all complexes.
}
\label{fig:slope}
\end{figure}

Here, we calculated the ACED of all 44 reference compounds (cf.\ Table \ref{tab:compounds}) for varying amounts of exact exchange 
from $c_3=0.00$ to $c_3=0.50$ in steps of 0.05 in the B3LYP density functional (for which $c_3=0.20$ was originally set).
Interestingly, also the contact density features a linear dependence on $c_3$ ($r^2 > 0.9995$ in all compounds), although
with different slopes. In other words,
the first derivative of the ACED with respect to $c_3$ is constant for a given compound (Fig. \ref{fig:slope}),
but different for all complexes. 
This complex-specificity is also confirmed by bootstrapping ($B = 10^3$) the uncertainty of the first derivative of the ACED (see the error bars in Fig.\ \ref{fig:slope},
which represent three standard deviations).

\begin{figure} [!ht]
\center
\includegraphics[width=\textwidth]{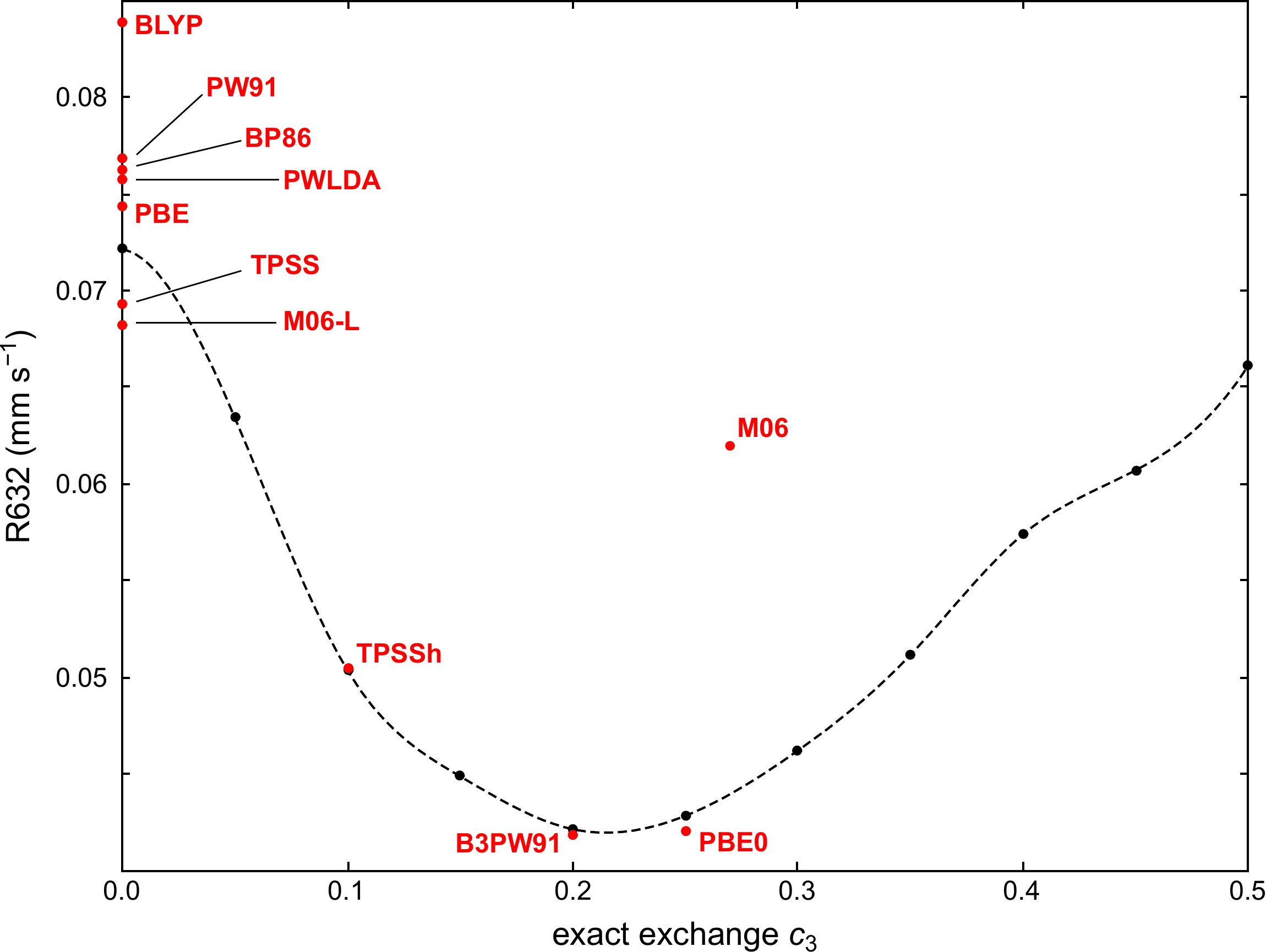}
\caption{R632 versus exact exchange admixture parameter $c_3$ for isomer-shift models obtained from B3LYP (black dots) 
with varying values of $c_3$, and for the other density functionals investigated in this study (without further modification: red dots).
The dashed curve was obtained from a cubic spline. 
For clarity, we did not round the results to the experimental resolution of 0.01 mm s$^{-1}$.
We employed our consistent reference data set ($N = 39$) for calculating the R632 (def2-SVP basis set).
}
\label{fig:hfx}
\end{figure}

Consequently, a change in exact exchange admixture does not lead to a unique ACED shift for all compounds, which is why the 
MPU is a nonlinear function of $c_3$.
Plotting the R632 against the exact exchange admixture reveals that this performance measure is minimized 
for $c_3=0.20$--$0.25$ (cf.\ Fig.\ \ref{fig:hfx} where $N = 39$), which indicates why B3LYP, B3PW91 (both featuring  $c_3=0.20$), and 
PBE0 ($c_3=0.25$) are the most transferable density functionals with respect to isomer-shift calibration.

The observation of linear dependence of the ACED is not straightforward to explain. It is clear that only basis functions
with angular momentum quantum number $l=0$ ($s$-type functions) can contribute to the nonrelativistic contact density in
an atomic system (note the
short-range behavior $\propto r^l$ of the radial function in a one-electron atom\cite{schwabl2007}).
This short-range behavior is (also in the spherically averaged case of an atom in a molecule\cite{toth2010}),
determined by the lowest-order contributions 
of the potential to a Taylor series expansion 
in terms of the radial distance in the short-range quantum mechanical differential equations for a single electron.
The lowest-order constributions are the centrifugal and the nuclear point-charge potentials, but not the electron--electron
interaction potentials, which provide a constant contribution at the nuclear position\cite{neugebauer2002}. Hence, for a spherically
symmetric atom, this is also true for the Hartree--Fock exchange interaction, which contributes a constant term to the potential
at the nuclear origin\cite{neugebauer2002} and would therefore not, in this case, affect the radial function and hence not the
electron density at the nucleus.

For a (nonspherical) molecular system, additional contributions to the contact density need to be considered.
Following the partitioning scheme of the ACED by Neese,\cite{neese2002} we find for compound \#43 (iron pentacarbonyl) that the 
iron $s$-functions in the CP(PPP) basis set contribute dominantly to the ACED ($>99.9\%$).
Hence, their contribution to the ACED alone already reveals the linear trend of the ACED with respect to exact exchange admixture.
However, the three $s$-functions in the CP(PPP) basis set with the largest exponent are constant with respect to a change of 
$c_3$, whereas the smallest 11 $s$-functions change linearly with $c_3$, even though their slope is sometimes positive 
(for larger exponents) and sometimes negative (for smaller exponents), which makes a detailed analysis difficult.
Interestingly, considering not only the charge at one specific point, i.e., the ACED, but the integral of the electron density
over some space, i.e., a partial charge of the iron atom in iron pentacarbonyl, also reveals a linear behavior with $c_3$ 
although with negative slope (see also Ref.\ \onlinecite{ioannidis2015}).
Hence, Mulliken and L\"owdin partial charges decrease with increasing $c_3$, while the ACED increases.


\section{Conclusions and outlook}

	To reliably estimate the prediction uncertainty of a property model, it is important to identify and correct for systematic errors due to, e.g., inconsistent measurements, parametric population assumptions, or inadequate computational methods.
	Here, we studied this issue at the example of $^{57}$Fe M\"ossbauer isomer-shift predictions based on a linear model.
	Twelve density functionals across Jacob's ladder were considered for the calculation of the ACED for 44 chemically diverse molecular iron compounds (formal oxidation states: 0, +1, +2, +3, +4), whereas the corresponding target data refer to measured isomer shifts.
	We explicitly considered uncertainty in the model parameters, which may be a crucial ingredient for the estimation of MPU.
	For this purpose, we employed both bootstrapping\cite{efron1979, davison1997, chernick1999, hastie2016} and Bayesian linear regression based on the evidence approximation.\cite{bishop2009}.
	
	First of all, we found that the RMSE, which measures the standard deviation of the model residuals, is significantly larger than the average experimental uncertainty (0.07--0.10~mm~s$^{-1}$ versus 0.02~mm~s$^{-1}$).
	This discrepancy cannot be explained by the simplicity of the linear model as more complex property models (quadratic, cubic, quartic) were found to yield either equal or larger MPU estimates when applying bootstrapping.
	However, with the jackknife-after-bootstrapping approach, which probes the sensitivity in the MPU with respect to small changes in the reference data set, we could identify 5 inconsistent data points.
	The reliability of the jacknife-after-bootstrapping approach can be assessed by determining the fraction of data points lying in a certain prediction interval.
	When including the inconsistent data points, 86\% of the reference isomer shifts lie in the 68\% prediction band (representing $u(\mathbf{x})$ as defined in Eq.\ (\ref{eq:local_error})), whereas excluding them leads to only 72\% of the reference isomer shifts contained in the 68\% prediction band, which reveals a significant improvement of the statistical reliability.
	Furthermore, the high agreement of bootstrapped and Gaussian posterior parameter distributions (the latter were obtained from Bayesian linear regression) suggests that the normal-population assumption is reasonable for isomer-shift calibration given the reference data set is carefully selected.
	
	The new (consistent) reference data set still leads to overestimation of the average experimental uncertainty (RMSE of 0.03--0.08~mm~s$^{-1}$), which can be assigned to non-constant systematic errors of the density functionals leading to random shifts of the (unknown) true input values.
	This random method inadequacy suggests that weighted least-squares regression, where experimental uncertainties are explicitly introduced as a weight, may bias the calibration of the property model.
	However, on the basis of bootstrapped regression we did not find a significant difference between the ordinary (unweighted) and the weighted setup due to pronounced parameter uncertainty, whereas Grandjean and Long\cite{grandjean2016} recommend the application of weighted least-squares regression for the calibration of isomer-shift models.
		
	Comparison of the RMSE with more reliable performance measures, which consider uncertainty in the parameters (such as R632, RMPV, or RLOO), reveals nearly equal results in most of the cases.
	This finding can be explained by the low experimental resolution of 0.01~mm~s$^{-1}$ compared to the width of the input domain (here, 1.08~mm~s$^{-1}$), which almost completely masks parameter uncertainty.
	Therefore, simple performance measures such as the RMSE and the squared coefficient of correlation, $r^2$, are expected to be reliable for the construction of density functional rankings, where relative rather than absolute MPU estimates are required.
	The same holds for the root-mean-square deviation (RMSD) and the mean unsigned error (MUE) as defined and discussed in a calibration study on physical properties of crystals by Pernot et al.\cite{pernot2015}

	Next, we examined the sensitivity of density functional rankings with respect to the composition and number of data points.
	For this purpose, we generated 10000 synthetic data sets (based on bootstrapping) with 5, 10, 20, and 39 (the size of our reference data set) data points.
	For 5 data points, false conclusions about the performance of a density functional are likely.
	For instance, the ``best'' density functional, PBE0, is placed first in only about 60\% of the cases.
	In the case of 39 data points, B3LYP and B3PW91 are placed first in 85--90\% of the cases, whereas PBE0 is placed first in more than 95\% of the cases.
	Still, there remains a serious chance of favoring a subprime density functional for usual data-set sizes ($N < 40$)\cite{lovell2002, neese2002, zhang2002, liu2003, vrajmasu2003, zhang2003, sinnecker2005, han2006, nemykin2006, shoji2007, han2008, remacle2008, hopmann2009, ling2009, long2009, romelt2009, bochevarov2010, kurian2010, harris2011, sandala2011, gubler2013, hedegard2014, long2015, casassa2016, grandjean2016} employed in calibration studies of the isomer shift.
	RMSE and RMPV produced equivalent density functional rankings.
	However, due to the low experimental resolution, important effects stemming from parameter uncertainty may be hidden.
	We examined this hypothesis by artificially increasing the experimental resolution from 0.01~mm~s$^{-1}$ to 0.001~mm~s$^{-1}$.
	While RMSE and RMPV still yielded similar rankings, their dispersion increased significantly.
	For instance, the ``best'' density functional, PBE0, is placed first in less than 40\% of the cases for 5 data points.
	Even for 39 data points, the identification of PBE0 as ``best'' density functional is only successful in 70--75\% of the cases.
	
	Finally, we discussed our observation that hybrid density functionals yielded significantly lower MPU estimates compared to pure density functionals.
	When varying the exact exchange admixture of the B3LYP density functional (for which the linear exact exchange parameter $c_3$\cite{reiher2001} was originally set to $c_3 = 0.20$), we found a linear dependence of the ACED on $c_3$ for all 44 iron compounds studied ($r^2 > 0.9995$).
	However, the first derivative of the ACED with respect to $c_3$ is complex-specific, which is why the MPU is a nonlinear function of $c_3$.
	It is minimized for $c_3 = 0.20$--$0.25$, matching with our observation that B3LYP, B3PW91 ($c_3 = 0.20$), and PBE0 ($c_3 = 0.25$) yielded isomer-shift models with the lowest prediction uncertainty.
	
	When selecting a density functional for actual applications, one is interested in a reliable uncertainty estimation of an isomer-shift prediction for a given ACED.
	For this purpose, the prediction bands employed in this study represent locally resolved MPU based on the .632 estimator.
	The validity of the MPU is highest at the mean ACED for a given density functional and decreases continuously from there, which is why extrapolations outside the input domain studied are not recommended.
	This limitation motivated us to cover a wide range of possible ACED values such that chemically diverse iron complexes can be investigated on the basis of our calibration analysis.
	
	Another possibility to approach accurate isomer-shift predictions is to calculate the isomer shift directly from first principles.
		Filatov derived the corresponding computational scheme\cite{Filatov2007} and applied it to examine the isomer shift of molecular iron compounds \cite{kurian2008}.
	Post-calibration of the resulting isomer-shift pairs (measured versus calculated) would allow for an assessment of the more adequate input variable (isomer shift versus ACED).
	
	We discussed several issues in this paper ranging from error diagnostics, MPU estimation, and the role of experimental resolution to the reliability of conclusions drawn on the basis of a specific data set.
	Bootstrapping clearly shows that performance assessment of density functionals is error-prone for a small number of data points, and for univariate linear regression models, one should consider at least about 40 data points.
	While the main message of previous calibration studies of the isomer shift remains (hybrid density functionals perform superior to pure density functionals\cite{neese2002, zhang2002, nemykin2006, romelt2009, bochevarov2010, papai2013}), we now provide a solid statistical framework to examine the certainty of such findings.
	This framework allowed us to suggest a new, statistically well-justified property model for the isomer shift, given in Eqs.\ (\ref{eq:prediction}) and (\ref{eq:mpu}), based on PBE0 calculations. Moreover, we introduced a new reference data set for such parametrizations, MIS39, which may be employed and extended for future work.
	In particular, this calibration study presents the first statistically rigorous analysis for theoretical M\"ossbauer spectroscopy providing the practitioner with reliable, locally resolved uncertainties for isomer-shift predictions.
	Moreover, our calibration analysis is of general applicability and not restricted to property models applied in M\"ossbauer spectroscopy.
	We will provide the scripts which were written for this work on our webpage so that the methodology can be applied in other parametrization studies of physical and chemical property models.
	

\section*{Acknowledgments}

	Financial support from the Swiss National Science Foundation (project no.\ 200020\_169120) is gratefully acknowledged.		
	

\section*{Appendix}

\subsection*{Statistical calibration analysis}
All statistical results presented in this work were produced with our suite of scripts \texttt{reBoot}\cite{scriptref} 
        developed in the {\sc GNU Octave}\cite{eaton2009} programming language that is mostly compatible with {\sc Matlab}.
	The basic functionalities of the calibration methods employed are already described in Section \ref{chap:sampling}.
In this subsection, we discuss some further details.

\subsubsection*{Bayesian linear regression}
	Bayesian linear regression\cite{bishop2009} is an efficient calibration procedure for normally distributed parameters.
	The best-fit parameter vector, $\mathbf{w}_\text{MAP}$, is obtained from
	\begin{equation}
	\label{eq:map}
	\mathbf{w}_\text{MAP} = \beta \mathbf{S} \mathbf{X}^\top \mathbf{y} \, ,
	\end{equation}
	where $\mathbf{S}$ is the covariance matrix of the parameters,
	\begin{equation}
	\mathbf{S}^{-1} = \alpha \mathbf{I} + \beta \mathbf{X}^\top \mathbf{X} \, ,
	\end{equation}
	$\mathbf{I}$ is the $(M + 1) \times (M + 1)$ unit matrix ($M$ $+$ $1$ is the number of parameters contained in the property model), and $\alpha$ and $\beta$ are so-called hyperparameters,
	\begin{equation}
	\alpha = \frac{\gamma}{\mathbf{w}_\text{MAP}^\top \mathbf{w}_\text{MAP}} \, ,
	\end{equation}
	\begin{equation}
	\beta^{-1} = \frac{N}{N - \gamma} \text{MSE}_{\mathcal{D},\mathbf{w}_\text{MAP}} \, ,
	\end{equation}
	which need to be iteratively refined, a procedure referred to as \textit{evidence approximation} or \textit{generalized maximum likelihood}.\cite{bishop2009}
	For this purpose, the effective number of parameters, $\gamma$, is calculated according to
	\begin{equation}
	\gamma = \sum_{m={0}}^M \frac{\lambda_m \beta}{\alpha + \lambda_m \beta} \, ,
	\end{equation}
	where the $\lambda_m$ are eigenvalues of the $(\mathbf{X}^\top \mathbf{X})$ matrix of dimension $(M + 1) \times (M + 1)$,
	\begin{equation}
	(\mathbf{X}^\top \mathbf{X}) \mathbf{v}_m = \lambda_m \mathbf{v}_m \, ,
	\end{equation}
	and the $\mathbf{v}_m$ are the corresponding eigenvectors.
	For the initialization of $\alpha$ and $\beta$, one can simply choose $\gamma = M + 1$ and $\mathbf{w}_\text{MAP} = \mathbf{w}_\mathcal{D}$.
	Finally, the locally resolved MPU at input value $\mathbf{x}_0$, $s(\mathbf{x}_0)$, is estimated as
	\begin{equation}
	s(\mathbf{x}_0) = \sqrt{\beta^{-1} + \mathbf{x}_0^\top \mathbf{S} \mathbf{x}_0} \, .
	\end{equation}
	
	The hyperparameter $\beta$ represents the inverse noise variance, whereas $\alpha$ represents the belief on the parameter distributions prior to considering the actual data.
	Choosing $\alpha = 0$, one assumes prior parameter distributions with infinite variance, which is equivalent to assigning each value on the real line the same probability density. 
	In that case, Eq.\ (\ref{eq:map}) reduces to Eq.\ (\ref{eq:ls}) resulting in $\mathbf{w}_\text{MAP} = \mathbf{w}_\mathcal{D}$, i.e., one simply performs linear least-squares regression and additionally obtains the covariance matrix of the model parameters.
	Here, we observe that $\alpha$ optimized with the evidence approximation and $\alpha$ constrained to zero lead to values for the RMPV deviating by less than 0.05\% (i.e., $\mathbf{w}_{\mathcal{D},\text{MAP}} \approx \mathbf{w}_\mathcal{D}$).
	Hence, we can directly compare the RMPV with the R632 obtained from bootstrapped linear least-squares regression where $\alpha = 0$ (the same holds for comparisons between RMPV and RLOO).
	
	The evidence approximation is appealing as it provides, for a given initial model complexity (here, the polynomial degree, $M$), a maximum-transferable set of model parameters, $\mathbf{w}_\text{MAP}$, by optimizing the hyperparameters $\alpha$ and $\beta$.
	Hence, the evidence approximation is an approach suited for model selection.\cite{bishop2009}
	With this approach, one also obtains the model parameters that minimize the regularized MSE,\cite{bishop2009}
	\begin{equation}
	\text{regMSE} \equiv \text{regMSE}_{\mathcal{D},\mathbf{w},\varepsilon} = \frac{1}{N} \sum_{n=1}^N \big(y_n - f(\mathbf{x}_n,\mathbf{w}_\varepsilon)\big)^2 + \varepsilon \mathbf{w}_\varepsilon^\top \mathbf{w}_\varepsilon \, ,
	\end{equation}
	where $\varepsilon = \alpha / \beta$ is a penalty factor.
	Given the normal-population assumption basically holds, the evidence approximation might yield a good initial value for the penalty factor to be learned in bootstrapped regularized regression.

	In all cases examined in this study, the RMPV (in Eq.\ (\ref{eq:mpv}), the RMPV is defined as its squared variant, MPV) did not change by more than 0.002\% after the first iteration of the evidence approximation.
	This deviation is completely masked by the low experimental resolution of 0.01~mm~s$^{-1}$ for measurements of the $^{57}$Fe M\"ossbauer isomer shift.
	Therefore, we conclude that a single iteration is sufficient to obtain a converged RMPV.
	
\subsubsection*{Iteratively reweighted linear least-squares regression}
	In the variant of iteratively reweighted linear least-squares regression\cite{gentle2007} by Pernot et al.\ \cite{pernot2015}, one starts with a weighted linear least-squares regression with respect to a reference data set, $\mathcal{D}$.
	The elements of the weight matrix $\mathbf{U}$ introduced in Eq.\ (\ref{eq:umatrix}) are updated,
	\begin{equation}
	\mathbf{U} =
	\begin{pmatrix}
		u_1^2 + d^2 & \cdots & 0 \\
		\vdots & \ddots & \vdots \\
		0 & \cdots & u_N^2 + d^2
	\end{pmatrix} \, ,
	\end{equation}
	where the term $d^2$ accounts for the discrepancy between the reduced MSE and the average experimental variance, $\langle u \rangle^2$,
	\begin{equation}
	d^2 = \frac{N}{N-M-1} \text{MSE}_\mathcal{D} - \langle u \rangle^2 \, .
	\end{equation}
	Subsequently, one performs another weighted linear least-squares regression with the updated weight matrix.
	This procedure is repeated until the change in $d^2$ becomes negligible between two iterations.
	Here, we chose a threshold of 0.001\%.
	
	Note that the approach outlined here is only of limited applicability as the constant discrepancy factor comes at the expense of two critical assumptions, i.e., a normal-population distribution and homogeneous residual variance (homoscedasticity).
	For our case study however, both assumptions can be well justified after removal of the inconsistent data points (\#1, \#2, \#7, \#13, \#28; cf.\ Table \ref{tab:compounds} and Fig.\ \ref{fig:regression}).
	On the one hand, the bootstrapped parameter distributions (Fig.\ \ref{fig:distribution}, bottom) resemble normal distribution, which is indicative of a normal-population distribution.
	On the other hand, we do not observe a trend of residual variance with respect to the input variable.
	In general, however, one should critically assess these assumptions by bootstrapping prediction intervals.\cite{davison1997}
	
\subsubsection*{Consideration of experimental uncertainty in bootstrapping}
	Bootstrap samples were drawn by pair resampling (cf.\ Section \ref{chap:bootstrap}).
	Each target value, $\delta_{\text{exp},n}$, in a bootstrap sample was randomly perturbed according to its experimental uncertainty (assumed to be 0.02~mm~s$^{-1}$ for compounds without reported experimental uncertainty).
	For instance, if a data point considered refers to the $n$-th input--target pair in the reference data set, a random value was drawn from a normal distribution with zero-mean and a standard deviation of $u_n$.
	This random value was rounded to the second decimal place to account for the experimental resolution of 0.01~mm~s$^{-1}$, and subsequently added to $\delta_{\text{exp},n}$.
	
\subsubsection*{Statistically valid increase of experimental resolution}		
	Bootstrap samples were drawn by pair resampling (cf.\ Section \ref{chap:bootstrap}).
	To increase the experimental resolution from its actual value of 0.01~mm~s$^{-1}$ to 0.001~mm~s$^{-1}$, one must respect that any value of the isomer shift with three decimal places is equiprobable as long as the result rounded to two decimal places equals the reported isomer shift.
	For this purpose, random values were drawn from a uniform distribution with boundaries of $-$0.005~mm~s$^{-1}$ and $+$0.004~mm~s$^{-1}$, and subsequently added to the actual isomer shifts, $\delta_{\text{exp},n}$, assuming the third decimal place equals zero prior to addition.
	If experimental uncertainty, $u_n$, was explicitly considered in bootstrapping (see above), the same procedure was repeated for $u_n$.
		
	
\subsection*{Quantum chemical calculations}
	References to the original crystal structures are listed in Table \ref{tab:compounds}.
First of all, solute molecules and counterions were removed from all molecular structures.
The only exceptions are structures 22 and 26 (Table \ref{tab:compounds}), where we kept the very small counterions due to their proximity to the iron nucleus (lithium and sodium, respectively).
	All molecular structures were fully optimized with the TPSS density functional \cite{tao2003}, Ahlrichs' def2-TZVP basis set on all atoms \cite{weigend2005, weigend2006}, Grimme's DFT-D3 dispersion correction \cite{grimme2010}, and the conductor-like screening model (COSMO)\cite{klamt1993} for electrostatic screening with a dielectric constant of $\varepsilon = 78$ (water).
	For iodine, the only group-5 element in the reference set of iron compounds, the effective core potential def2-ecp was employed.\cite{peterson2003}
	Structure optimizations were performed with {\sc Turbomole} 6.4.0\cite{ahlrichs1989} applying the resolution-of-the-identity approximation, which invokes an auxiliary basis set.
	The convergence thresholds were set to $10^{-7}$ hartree for the electronic energy difference and $10^{-4}$ hartree/bohr for the length of the gradient of the electronic energy with respect to the nuclear coordinates.
	Some molecular structures were further truncated prior to structure optimization (see Table \ref{tab:compounds}).
	For instance, larger alkyl groups at aromatic rings not directly attached to iron were replaced by methyl groups.

	All subsequent ACED calculations were performed with {\sc Orca} 3.0.3\cite{neese2012} (see Table S2).
	For every molecular iron compound contained in our reference set, we determined the ACED with 12 different density functionals; one LDA functional (PWLDA\cite{perdew1992}), four GGA functionals (BP86\cite{perdew1986, becke1988}, BLYP\cite{becke1988, lee1988, miehlich1989}, PW91\cite{perdew1992a}, PBE\cite{perdew1996}), two meta-GGA functionals (M06-L\cite{zhao2006a}, TPSS), three hybrid-GGA functionals (B3LYP\cite{lee1988, miehlich1989, becke1993}, B3PW91\cite{becke1993, perdew1996}, PBE0\cite{adamo1999}), and two meta-hybrid-GGA functionals (M06\cite{zhao2008}, TPSSh\cite{staroverov2003}).
	Neese's CP(PPP) basis set \cite{neese2002} was employed for iron, whereas Ahlrichs' def2-TZVP basis set was employed for all other elements.
	All calculations were based on electrostatic screening (COSMO) with a dielectric constant of $\varepsilon = 80$.
	The convergence threshold for the electronic energy difference was set to $10^{-6}$ hartree (default).
	In case of error code 16384 (the only type of error that occurred in our ACED calculations), we added either the keyword \texttt{slowconv} to the first line (starting with \texttt{!}) or \texttt{\%scf maxiter <no.> end} in a separate line (\texttt{<no.>} $=$ 500 always worked in the cases studied).
	
	We repeated all ACED calculations with def2-TZVP replaced by def2-SVP\cite{weigend2005} (see Table S3), resulting in slightly higher MPU for the latter (cf.\ Tables S6 and S7).
	We did not perform further statistical analyses based on the def2-SVP results, but exploited this smaller basis set to probe the sensitivity of the ACED with respect to the convergence threshold for the electronic energy difference (see Table S4).
	We decreased the convergence threshold for the electronic energy difference to $10^{-8}$ hartree (B3LYP/def2-SVP).
	For both thresholds, mean and maximum of the unsigned difference in the ACED are only 0.1\% and 3.7\% as large as the standard deviation of the ACED itself, respectively.
	We may conclude that default self-consistent field convergence criteria lead to stable results, especially taking into account the low experimental resolution of 0.01~mm~s$^{-{1}}$.	

\section*{Supporting Information}
All Cartesian coordinates, contact density data, and further statistical analysis
can be found in the supporting information.
This information is available free of charge via the Internet at http://pubs.acs.org/.


\begin{thebibliography}{100}

\bibitem{kennedy2001}
Kennedy,~M.~C.;\ \ O'Hagan,~A.  Bayesian Calibration of Computer Models,
  \textit{J. R. Stat. Soc. B} \textbf{2001,} \textsl{63,} 425--464.

\bibitem{cailliez2011}
Cailliez,~F.;\ \ Pernot,~P.  Statistical Approaches to Forcefield Calibration
  and Prediction Uncertainty in Molecular Simulation,  \textit{J. Chem. Phys.}
  \textbf{2011,} \textsl{134,} 054124.

\bibitem{angelikopoulos2012}
Angelikopoulos,~P.;\ \ Papadimitriou,~C.;\ \ Koumoutsakos,~P.  Bayesian
  Uncertainty Quantification and Propagation in Molecular Dynamics Simulations:
  {{A}} High Performance Computing Framework,  \textit{J. Chem. Phys.}
  \textbf{2012,} \textsl{137,} 144103.

\bibitem{mortensen2005}
Mortensen,~J.~J.;\ \ Kaasbjerg,~K.;\ \ Frederiksen,~S.~L.;\ \
  N{\o}rskov,~J.~K.;\ \ Sethna,~J.~P.;\ \ Jacobsen,~K.~W.  Bayesian {{Error
  Estimation}} in {{Density}}-{{Functional Theory}},  \textit{Phys. Rev. Lett.}
  \textbf{2005,} \textsl{95,} 216401.

\bibitem{simm2016}
Simm,~G.~N.;\ \ Reiher,~M.  Systematic {{Error Estimation}} for {{Chemical
  Reaction Energies}},  \textit{J. Chem. Theory Comput.} \textbf{2016,}
  \textsl{12,} 2762--2773.

\bibitem{grimme2010}
Grimme,~S.;\ \ Antony,~J.;\ \ Ehrlich,~S.;\ \ Krieg,~H.  A Consistent and
  Accurate Ab Initio Parametrization of Density Functional Dispersion
  Correction ({{DFT}}-{{D}}) for the 94 Elements {{H}}-{{Pu}},  \textit{J.
  Chem. Phys.} \textbf{2010,} \textsl{132,} 154104.

\bibitem{thiel2014}
Thiel,~W.  Semiempirical Quantum\textendash{}chemical Methods,  \textit{WIREs
  Comput. Mol. Sci.} \textbf{2014,} \textsl{4,} 145--157.

\bibitem{rauhut1995}
Rauhut,~G.;\ \ Pulay,~P.  Transferable {{Scaling Factors}} for {{Density
  Functional Derived Vibrational Force Fields}},  \textit{J. Phys. Chem.}
  \textbf{1995,} \textsl{99,} 3093--3100.

\bibitem{neugebauer2003}
Neugebauer,~J.;\ \ Hess,~B.~A.  Fundamental Vibrational Frequencies of Small
  Polyatomic Molecules from Density-Functional Calculations and Vibrational
  Perturbation Theory,  \textit{J. Chem. Phys.} \textbf{2003,} \textsl{118,}
  7215--7225.

\bibitem{irikura2005}
Irikura,~K.~K.;\ \ Johnson,~R.~D.;\ \ Kacker,~R.~N.  Uncertainties in {{Scaling
  Factors}} for Ab {{Initio Vibrational Frequencies}},  \textit{J. Phys. Chem.
  A} \textbf{2005,} \textsl{109,} 8430--8437.

\bibitem{xing2015}
Xing,~L.;\ \ Li,~S.;\ \ Wang,~Z.;\ \ Yang,~B.;\ \ Klippenstein,~S.~J.;\ \
  Zhang,~F.  Global Uncertainty Analysis for {{RRKM}}/Master Equation Based
  Kinetic Predictions: {{A}} Case Study of Ethanol Decomposition,
  \textit{Combust. Flame} \textbf{2015,} \textsl{162,} 3427--3436.

\bibitem{sargsyan2015}
Sargsyan,~K.;\ \ Najm,~H.~N.;\ \ Ghanem,~R.  On the {{Statistical Calibration}}
  of {{Physical Models}},  \textit{Int. J. Chem. Kinet.} \textbf{2015,}
  \textsl{47,} 246--276.

\bibitem{sutton2016}
Sutton,~J.~E.;\ \ Guo,~W.;\ \ Katsoulakis,~M.~A.;\ \ Vlachos,~D.~G.  Effects of
  Correlated Parameters and Uncertainty in Electronic-Structure-Based Chemical
  Kinetic Modelling,  \textit{Nat. Chem.} \textbf{2016,} \textsl{8,} 331--337.

\bibitem{edwards2014}
Edwards,~D.~E.;\ \ Zubarev,~D.~Y.;\ \ Packard,~A.;\ \ Lester,~W.~A.;\ \
  Frenklach,~M.  Interval {{Prediction}} of {{Molecular Properties}} in
  {{Parametrized Quantum Chemistry}},  \textit{Phys. Rev. Lett.} \textbf{2014,}
  \textsl{112,} 253003.

\bibitem{ruscic2014}
Ruscic,~B.  Uncertainty Quantification in Thermochemistry, Benchmarking
  Electronic Structure Computations, and {{Active Thermochemical Tables}},
  \textit{Int. J. Quantum Chem.} \textbf{2014,} \textsl{114,} 1097--1101.

\bibitem{pernot2015}
Pernot,~P.;\ \ Civalleri,~B.;\ \ Presti,~D.;\ \ Savin,~A.  Prediction
  {{Uncertainty}} of {{Density Functional Approximations}} for {{Properties}}
  of {{Crystals}} with {{Cubic Symmetry}},  \textit{J. Phys. Chem. A}
  \textbf{2015,} \textsl{119,} 5288--5304.

\bibitem{wells1963}
Wells,~P.~R.  Linear {{Free Energy Relationships}}.,  \textit{Chem. Rev.}
  \textbf{1963,} \textsl{63,} 171--219.

\bibitem{karthikeyan2005}
Karthikeyan,~M.;\ \ Glen,~R.~C.;\ \ Bender,~A.  General {{Melting Point
  Prediction Based}} on a {{Diverse Compound Data Set}} and {{Artificial Neural
  Networks}},  \textit{J. Chem. Inf. Model.} \textbf{2005,} \textsl{45,}
  581--590.

\bibitem{bishop2009}
Bishop,~C.~M. \textit{Pattern Recognition and Machine Learning;} Information
  Science and Statistics {Springer}: New York, NY, USA, Corrected at 8th
  printing ed.; 2009.

\bibitem{ohagan2013}
O'Hagan,~A.  Bayesian Inference with Misspecified Models: {{Inference}} about
  What?,  \textit{J. Stat. Plan. Inference} \textbf{2013,} \textsl{143,}
  1643--1648.

\bibitem{pernot2016}
Pernot,~P.;\ \ Cailliez,~F.  A Critical Review of Statistical
  Calibration/Prediction Models Handling Data Inconsistency and Model
  Inadequacy,   \textbf{2016,}  arXiv:1611.04376.

\bibitem{pernot2016a}
Pernot,~P.  The Parameters Uncertainty Inflation Fallacy,   \textbf{2016,}
  arXiv:1611.04295.

\bibitem{lira2007}
Lira,~I.  Combining Inconsistent Data from Interlaboratory Comparisons,
  \textit{Metrologia} \textbf{2007,} \textsl{44,} 415.

\bibitem{toman2009}
Toman,~B.;\ \ Possolo,~A.  Laboratory Effects Models for Interlaboratory
  Comparisons,  \textit{Accredit. Qual. Assur.} \textbf{2009,} \textsl{14,}
  553--563.

\bibitem{efron1979}
Efron,~B.  Bootstrap {{Methods}}: {{Another Look}} at the {{Jackknife}},
  \textit{Ann. Statist.} \textbf{1979,} \textsl{7,} 1--26.

\bibitem{chernick1999}
Chernick,~M.~R. \textit{Bootstrap {{Methods}}: {{A Practitioner}}'s {{Guide}};}
  {Wiley-Interscience}: New York, NY, USA, 1 edition ed.; 1999.

\bibitem{davison1997}
Davison,~A.~C.;\ \ Hinkley,~D.~V. \textit{Bootstrap {{Methods}} and Their
  {{Application}};} {Cambridge University Press}: Cambridge, UK; New York, NY,
  USA, 1 edition ed.; 1997.

\bibitem{efron1983}
Efron,~B.  Estimating the {{Error Rate}} of a {{Prediction Rule}}:
  {{Improvement}} on {{Cross}}-{{Validation}},  \textit{J. Am. Stat. Assoc.}
  \textbf{1983,} \textsl{78,} 316--331.

\bibitem{hastie2016}
Hastie,~T.;\ \ Tibshirani,~R.~J.;\ \ Friedman,~J. \textit{The Elements of
  Statistical Learning: Data Mining, Inference, and Prediction;} Springer
  series in statistics {Springer}: New York, NY, USA, Second edition, corrected
  at 11th printing 2016 ed.; 2016.

\bibitem{efron1997}
Efron,~B.;\ \ Tibshirani,~R.  Improvements on {{Cross}}-{{Validation}}: {{The}}
  .632+ {{Bootstrap Method}},  \textit{Journal of the American Statistical
  Association} \textbf{1997,} \textsl{92,} 548--560.

\bibitem{rubin1981}
Rubin,~D.~B.  The {{Bayesian Bootstrap}},  \textit{Ann. Statist.}
  \textbf{1981,} \textsl{9,} 130--134.

\bibitem{gentle2007}
Gentle,~J.~E.  Solution of {{Linear Systems}}.   In  \textit{Matrix
  {{Algebra}}}; Springer Texts in Statistics {Springer}: New York, NY, USA,
  2007.

\bibitem{riu2003}
Riu,~J.;\ \ Bro,~R.  Jack-Knife Technique for Outlier Detection and Estimation
  of Standard Errors in {{PARAFAC}} Models,  \textit{Chemometr. Intell. Lab.}
  \textbf{2003,} \textsl{65,} 35--49.

\bibitem{lovell2002}
Lovell,~T.;\ \ Han,~W.-G.;\ \ Liu,~T.;\ \ Noodleman,~L.  A {{Structural Model}}
  for the {{High}}-{{Valent Intermediate Q}} of {{Methane Monooxygenase}} from
  {{Broken}}-{{Symmetry Density Functional}} and {{Electrostatics
  Calculations}},  \textit{J. Am. Chem. Soc.} \textbf{2002,} \textsl{124,}
  5890--5894.

\bibitem{neese2002}
Neese,~F.  Prediction and Interpretation of the {{$^{57}$Fe}} Isomer Shift in
  {{M{\"o}ssbauer}} Spectra by Density Functional Theory,  \textit{Inorg. Chim.
  Acta} \textbf{2002,} \textsl{337,} 181--192.

\bibitem{zhang2002}
Zhang,~Y.;\ \ Mao,~J.;\ \ Oldfield,~E.  {{$^{57}$Fe M{\"o}ssbauer Isomer
  Shifts}} of {{Heme Protein Model Systems}}:\, {{Electronic Structure
  Calculations}},  \textit{J. Am. Chem. Soc.} \textbf{2002,} \textsl{124,}
  7829--7839.

\bibitem{liu2003}
Liu,~T.;\ \ Lovell,~T.;\ \ Han,~W.-G.;\ \ Noodleman,~L.  {{DFT Calculations}}
  of {{Isomer Shifts}} and {{Quadrupole Splitting Parameters}} in {{Synthetic
  Iron}}-{{Oxo Complexes}}:\, {{Applications}} to {{Methane Monooxygenase}} and
  {{Ribonucleotide Reductase}},  \textit{Inorg. Chem.} \textbf{2003,}
  \textsl{42,} 5244--5251.

\bibitem{vrajmasu2003}
Vrajmasu,~V.;\ \ M{\"u}nck,~E.;\ \ Bominaar,~E.~L.  Density {{Functional
  Study}} of the {{Electric Hyperfine Interactions}} and the
  {{Redox}}-{{Structural Correlations}} in the {{Cofactor}} of {{Nitrogenase}}.
  {{Analysis}} of {{General Trends}} in {{$^{57}$Fe Isomer Shifts}},
  \textit{Inorg. Chem.} \textbf{2003,} \textsl{42,} 5974--5988.

\bibitem{zhang2003}
Zhang,~Y.;\ \ Oldfield,~E.  An {{Investigation}} of the {{Unusual $^{57}$Fe
  M{\"o}ssbauer Quadrupole Splittings}} and {{Isomer Shifts}} in 2 and
  3-{{Coordinate Fe}}({{II}}) {{Complexes}},  \textit{J. Phys. Chem. B}
  \textbf{2003,} \textsl{107,} 7180--7188.

\bibitem{sinnecker2005}
Sinnecker,~S.;\ \ Slep,~L.~D.;\ \ Bill,~E.;\ \ Neese,~F.  Performance of
  {{Nonrelativistic}} and {{Quasi}}-{{Relativistic Hybrid DFT}} for the
  {{Prediction}} of {{Electric}} and {{Magnetic Hyperfine Parameters}} in
  {{$^{57}$Fe M{\"o}ssbauer Spectra}},  \textit{Inorg. Chem.} \textbf{2005,}
  \textsl{44,} 2245--2254.

\bibitem{han2006}
Han,~W.-G.;\ \ Liu,~T.;\ \ Lovell,~T.;\ \ Noodleman,~L.  {{DFT}} Calculations
  of {{$^{57}$Fe M{\"o}ssbauer}} Isomer Shifts and Quadrupole Splittings for
  Iron Complexes in Polar Dielectric Media: {{Applications}} to Methane
  Monooxygenase and Ribonucleotide Reductase,  \textit{J. Comput. Chem.}
  \textbf{2006,} \textsl{27,} 1292--1306.

\bibitem{nemykin2006}
Nemykin,~V.~N.;\ \ Hadt,~R.~G.  Influence of {{Hartree}}-{{Fock Exchange}} on
  the {{Calculated M{\"o}ssbauer Isomer Shifts}} and {{Quadrupole Splittings}}
  in {{Ferrocene Derivatives Using Density Functional Theory}},  \textit{Inorg.
  Chem.} \textbf{2006,} \textsl{45,} 8297--8307.

\bibitem{shoji2007}
Shoji,~M.;\ \ Saito,~T.;\ \ Takeda,~R.;\ \ Kitagawa,~Y.;\ \ Kawakami,~T.;\ \
  Yamanaka,~S.;\ \ Okumura,~M.;\ \ Yamaguchi,~K.  Assignments of the
  {{M{\"o}ssbauer}} Spectra of an Inorganic [{{8Fe}}\textendash{}{{7S}}]
  Complex Based on the First-Principle Calculations,  \textit{Chem. Phys.
  Lett.} \textbf{2007,} \textsl{446,} 228--232.

\bibitem{han2008}
Han,~W.-G.;\ \ Noodleman,~L.  Structural Model Studies for the High-Valent
  Intermediate {{Q}} of Methane Monooxygenase from Broken-Symmetry Density
  Functional Calculations,  \textit{Inorg. Chim. Acta} \textbf{2008,}
  \textsl{361,} 973--986.

\bibitem{remacle2008}
Remacle,~F.;\ \ Grandjean,~F.;\ \ Long,~G.~J.  A {{Density Functional Theory
  Calculation}} of the {{Electronic Properties}} of {{Several High}}-{{Spin}}
  and {{Low}}-{{Spin Iron}}({{II}}) {{Pyrazolylborate Complexes}},
  \textit{Inorg. Chem.} \textbf{2008,} \textsl{47,} 4005--4014.

\bibitem{hopmann2009}
Hopmann,~K.~H.;\ \ Ghosh,~A.;\ \ Noodleman,~L.  Density {{Functional Theory
  Calculations}} on {{M{\"o}ssbauer Parameters}} of {{Nonheme Iron Nitrosyls}},
   \textit{Inorg. Chem.} \textbf{2009,} \textsl{48,} 9155--9165.

\bibitem{ling2009}
Ling,~Y.;\ \ Zhang,~Y.  M{\"o}ssbauer, {{NMR}}, {{Geometric}}, and {{Electronic
  Properties}} in {{S}} = 3/2 {{Iron Porphyrins}},  \textit{J. Am. Chem. Soc.}
  \textbf{2009,} \textsl{131,} 6386--6388.

\bibitem{long2009}
Long,~G.~J.;\ \ Tanase,~S.;\ \ Remacle,~F.;\ \ Periyasamy,~G.;\ \ Grandjean,~F.
   Combined {{M{\"o}ssbauer Spectral}} and {{Density Functional Theory
  Determination}} of the {{Magnetic Easy}}-{{Axis}} in {{Two High}}-{{Spin
  Iron}}({{II}}) 2-{{Pyrazinecarboxylate Complexes}},  \textit{Inorg. Chem.}
  \textbf{2009,} \textsl{48,} 8173--8179.

\bibitem{romelt2009}
R{\"o}melt,~M.;\ \ Ye,~S.;\ \ Neese,~F.  Calibration of {{Modern Density
  Functional Theory Methods}} for the {{Prediction}} of {{$^{57}$Fe
  M{\"o}ssbauer Isomer Shifts}}: {{Meta}}-{{GGA}} and {{Double}}-{{Hybrid
  Functionals}},  \textit{Inorg. Chem.} \textbf{2009,} \textsl{48,} 784--785.

\bibitem{bochevarov2010}
Bochevarov,~A.~D.;\ \ Friesner,~R.~A.;\ \ Lippard,~S.~J.  Prediction of
  {{$^{57}$Fe M{\"o}ssbauer Parameters}} by {{Density Functional Theory}}: {{A
  Benchmark Study}},  \textit{J. Chem. Theory Comput.} \textbf{2010,}
  \textsl{6,} 3735--3749.

\bibitem{kurian2010}
Kurian,~R.;\ \ Filatov,~M.  Calibration of {{$^{57}$Fe}} Isomer Shift from Ab
  Initio Calculations: Can Theory and Experiment Reach an Agreement?,
  \textit{Phys. Chem. Chem. Phys.} \textbf{2010,} \textsl{12,} 2758--2762.

\bibitem{harris2011}
Harris,~T.~V.;\ \ Szilagyi,~R.~K.  Comparative {{Assessment}} of the
  {{Composition}} and {{Charge State}} of {{Nitrogenase FeMo}}-{{Cofactor}},
  \textit{Inorg. Chem.} \textbf{2011,} \textsl{50,} 4811--4824.

\bibitem{sandala2011}
Sandala,~G.~M.;\ \ Hopmann,~K.~H.;\ \ Ghosh,~A.;\ \ Noodleman,~L.  Calibration
  of {{DFT Functionals}} for the {{Prediction}} of {{$^{57}$Fe M{\"o}ssbauer
  Spectral Parameters}} in {{Iron}}\textendash{}{{Nitrosyl}} and
  {{Iron}}\textendash{}{{Sulfur Complexes}}: {{Accurate Geometries Prove
  Essential}},  \textit{J. Chem. Theory Comput.} \textbf{2011,} \textsl{7,}
  3232--3247.

\bibitem{gubler2013}
Gubler,~J.;\ \ Finkelmann,~A.~R.;\ \ Reiher,~M.  Theoretical {{$^{57}$Fe
  M{\"o}ssbauer Spectroscopy}} for {{Structure Elucidation}} of [{{Fe}}]
  {{Hydrogenase Active Site Intermediates}},  \textit{Inorg. Chem.}
  \textbf{2013,} \textsl{52,} 14205--14215.

\bibitem{papai2013}
P{\'a}pai,~M.;\ \ Vank{\'o},~G.  On {{Predicting M{\"o}ssbauer Parameters}} of
  {{Iron}}-{{Containing Molecules}} with {{Density}}-{{Functional Theory}},
  \textit{J. Chem. Theory Comput.} \textbf{2013,} \textsl{9,} 5004--5020.

\bibitem{hedegard2014}
Hedeg{\aa}rd,~E.~D.;\ \ Knecht,~S.;\ \ Ryde,~U.;\ \ Kongsted,~J.;\ \ Saue,~T.
  Theoretical {{$^{57}$Fe M{\"o}ssbauer}} Spectroscopy: Isomer Shifts of
  [{{Fe}}]-Hydrogenase Intermediates,  \textit{Phys. Chem. Chem. Phys.}
  \textbf{2014,} \textsl{16,} 4853--4863.

\bibitem{long2015}
Long,~G.~J.;\ \ Grandjean,~F.;\ \ Harrop,~T.~C.;\ \ Petroccia,~H.~M.;\ \
  Papaefthymiou,~G.~C.  Combined {{M{\"o}ssbauer Spectral}} and {{Density
  Functional Study}} of an {{Eight}}-{{Coordinate Iron}}({{II}}) {{Complex}},
  \textit{Inorg. Chem.} \textbf{2015,} \textsl{54,} 8415--8422.

\bibitem{casassa2016}
Casassa,~S.;\ \ Ferrari,~A.~M.  Calibration of {{$^{57}$Fe M{\"o}ssbauer}}
  Constants by First Principles,  \textit{Phys. Chem. Chem. Phys.}
  \textbf{2016,} \textsl{18,} 10201--10206.

\bibitem{grandjean2016}
Grandjean,~F.;\ \ Long,~G.~J.  Comment on ``{{Calibration}} of {{$^{57}$Fe
  M{\"o}ssbauer}} Constants by First Principles'' {{Phys}}. {{Chem}}. {{Chem}}.
  {{Phys}}., 2016, 18, 10201\textendash{}10206,  \textit{Phys. Chem. Chem.
  Phys.} \textbf{2016,} \textsl{18,} 26306--26309.

\bibitem{bjornsson2017}
Bjornsson,~R.;\ \ Neese,~F.;\ \ DeBeer,~S.  Revisiting the {{M{\"o}ssbauer
  Isomer Shifts}} of the {{FeMoco Cluster}} of {{Nitrogenase}} and the
  {{Cofactor Charge}},  \textit{Inorg. Chem.} \textbf{2017,} \textsl{56,}
  1470--1477.

\bibitem{gutlich2011}
G{\"u}tlich,~P.;\ \ Bill,~E.;\ \ Trautwein,~A.~X. \textit{M{\"o}ssbauer
  Spectroscopy and Transition Metal Chemistry: Fundamentals and Application;}
  {Springer}: Heidelberg, Germany, 2011.

\bibitem{scriptref}
Proppe,~J.;\ \ Reiher,~M. ``{{reBoot}}: {{A}} Program for Statistical
  Calibration of Property Models'',
  http://www.reiher.ethz.ch/software/reboot.html, 2017.

\bibitem{stoian2005}
Stoian,~S.~A.;\ \ Yu,~Y.;\ \ Smith,~J.~M.;\ \ Holland,~P.~L.;\ \
  Bominaar,~E.~L.;\ \ M{\"u}nck,~E.  M{\"o}ssbauer, {{Electron Paramagnetic
  Resonance}}, and {{Crystallographic Characterization}} of a {{High}}-{{Spin
  Fe}}({{I}}) {{Diketiminate Complex}} with {{Orbital Degeneracy}},
  \textit{Inorg. Chem.} \textbf{2005,} \textsl{44,} 4915--4922.

\bibitem{hendrich2006}
Hendrich,~M.~P.;\ \ Gunderson,~W.;\ \ Behan,~R.~K.;\ \ Green,~M.~T.;\ \
  Mehn,~M.~P.;\ \ Betley,~T.~A.;\ \ Lu,~C.~C.;\ \ Peters,~J.~C.  On the
  Feasibility of {{N2}} Fixation via a Single-Site {{FeI}}/{{FeIV}} Cycle:
  {{Spectroscopic}} Studies of {{FeI}}({{N2}}){{FeI}}, {{FeIV N}}, and Related
  Species,  \textit{Proc. Natl. Acad. Sci. U.S.A.} \textbf{2006,} \textsl{103,}
  17107--17112.

\bibitem{mock2008}
Mock,~M.~T.;\ \ Popescu,~C.~V.;\ \ Yap,~G. P.~A.;\ \ Dougherty,~W.~G.;\ \
  Riordan,~C.~G.  Monovalent {{Iron}} in a {{Sulfur}}-{{Rich Environment}},
  \textit{Inorg. Chem.} \textbf{2008,} \textsl{47,} 1889--1891.

\bibitem{lee2010}
Lee,~Y.;\ \ Mankad,~N.~P.;\ \ Peters,~J.~C.  Triggering {{N2}} Uptake via
  Redox-Induced Expulsion of Coordinated {{NH3}} and {{N2}} Silylation at
  Trigonal Bipyramidal Iron,  \textit{Nat. Chem.} \textbf{2010,} \textsl{2,}
  558--565.

\bibitem{lee2011}
Lee,~Y.;\ \ Peters,~J.~C.  Silylation of {{Iron}}-{{Bound Carbon Monoxide
  Affords}} a {{Terminal Fe Carbyne}},  \textit{J. Am. Chem. Soc.}
  \textbf{2011,} \textsl{133,} 4438--4446.

\bibitem{dugan2012}
Dugan,~T.~R.;\ \ Bill,~E.;\ \ MacLeod,~K.~C.;\ \ Christian,~G.~J.;\ \
  Cowley,~R.~E.;\ \ Brennessel,~W.~W.;\ \ Ye,~S.;\ \ Neese,~F.;\ \
  Holland,~P.~L.  Reversible {{C}}\textendash{}{{C Bond Formation}} between
  {{Redox}}-{{Active Pyridine Ligands}} in {{Iron Complexes}},  \textit{J. Am.
  Chem. Soc.} \textbf{2012,} \textsl{134,} 20352--20364.

\bibitem{macleod2014}
MacLeod,~K.~C.;\ \ Vinyard,~D.~J.;\ \ Holland,~P.~L.  A {{Multi}}-Iron {{System
  Capable}} of {{Rapid N2 Formation}} and {{N2 Cleavage}},  \textit{J. Am.
  Chem. Soc.} \textbf{2014,} \textsl{136,} 10226--10229.

\bibitem{lichtenberg2015}
Lichtenberg,~C.;\ \ Viciu,~L.;\ \ Adelhardt,~M.;\ \ Sutter,~J.;\ \ Meyer,~K.;\
  \ {de\hspace{0.25em}Bruin},~B.;\ \ Gr{\"u}tzmacher,~H.  Low-{{Valent
  Iron}}({{I}}) {{Amido Olefin Complexes}} as {{Promotors}} for
  {{Dehydrogenation Reactions}},  \textit{Angew. Chem. Int. Ed.} \textbf{2015,}
  \textsl{54,} 5766--5771.

\bibitem{zadrozny2013}
Zadrozny,~J.~M.;\ \ Xiao,~D.~J.;\ \ Atanasov,~M.;\ \ Long,~G.~J.;\ \
  Grandjean,~F.;\ \ Neese,~F.;\ \ Long,~J.~R.  Magnetic Blocking in a Linear
  Iron({{I}}) Complex,  \textit{Nat. Chem.} \textbf{2013,} \textsl{5,}
  577--581.

\bibitem{zadrozny2013a}
Zadrozny,~J.~M.;\ \ Xiao,~D.~J.;\ \ Long,~J.~R.;\ \ Atanasov,~M.;\ \
  Neese,~F.;\ \ Grandjean,~F.;\ \ Long,~G.~J.  M{\"o}ssbauer {{Spectroscopy}}
  as a {{Probe}} of {{Magnetization Dynamics}} in the {{Linear Iron}}({{I}})
  and {{Iron}}({{II}}) {{Complexes}}
  [{{Fe}}({{C}}({{SiMe3}})3)2]1\textendash/0,  \textit{Inorg. Chem.}
  \textbf{2013,} \textsl{52,} 13123--13131.

\bibitem{bill2013}
Bill,~E.  Single-Molecule Magnets: {{Iron}} Lines Up,  \textit{Nat. Chem.}
  \textbf{2013,} \textsl{5,} 556--557.

\bibitem{samuel2014}
Samuel,~P.~P.;\ \ Mondal,~K.~C.;\ \ Amin~Sk,~N.;\ \ Roesky,~H.~W.;\ \
  Carl,~E.;\ \ Neufeld,~R.;\ \ Stalke,~D.;\ \ Demeshko,~S.;\ \ Meyer,~F.;\ \
  Ungur,~L.;\ \ Chibotaru,~L.~F.;\ \ Christian,~J.;\ \ Ramachandran,~V.;\ \
  {van Tol},~J.;\ \ Dalal,~N.~S.  Electronic {{Structure}} and {{Slow Magnetic
  Relaxation}} of {{Low}}-{{Coordinate Cyclic Alkyl}}(Amino) {{Carbene
  Stabilized Iron}}({{I}}) {{Complexes}},  \textit{J. Am. Chem. Soc.}
  \textbf{2014,} \textsl{136,} 11964--11971.

\bibitem{ung2014}
Ung,~G.;\ \ Rittle,~J.;\ \ Soleilhavoup,~M.;\ \ Bertrand,~G.;\ \ Peters,~J.~C.
  Two-{{Coordinate Fe0}} and {{Co0 Complexes Supported}} by {{Cyclic}}
  (Alkyl)(Amino)Carbenes,  \textit{Angew. Chem. Int. Ed.} \textbf{2014,}
  \textsl{53,} 8427--8431.

\bibitem{emsl}
Schuchardt,~K.~L.;\ \ Didier,~B.~T.;\ \ Elsethagen,~T.;\ \ Sun,~L.;\ \
  Gurumoorthi,~V.;\ \ Chase,~J.;\ \ Li,~J.;\ \ Windus,~T.~L.  Basis {{Set
  Exchange}}: {{A Community Database}} for {{Computational Sciences}},
  \textit{Journal of Chemical Information and Modeling} \textbf{2007,}
  \textsl{47,} 1045--1052.

\bibitem{edwards1967}
Edwards,~P.~R.;\ \ Johnson,~C.~E.;\ \ Williams,~R. J.~P.  M{\"o}ssbauer
  {{Spectra}} of {{Some Tetrahedral Iron}} ({{II}}) {{Compounds}},  \textit{J.
  Chem. Phys.} \textbf{1967,} \textsl{47,} 2074--2082.

\bibitem{fan1989}
Fan,~Y.-P.;\ \ Du,~G.-Y.;\ \ Zhang,~W.-X.  The Molecular and
  Crystalline-Structure of
  [{{Ba}}(18-Crown-6)({{H2O}})({{OH}})]{{Fe}}({{NCS}})4,  \textit{Acta Chim.
  Sinica} \textbf{1989,} \textsl{47,} 276--278.

\bibitem{renovitch1969}
Renovitch,~G.~A.;\ \ Baker,~W.~A.  Complexes of Iron with O-Phenylenediamine,
  \textit{J. Chem. Soc. A} \textbf{1969,}  75--78.

\bibitem{maxcy2000}
Maxcy,~K.~R.;\ \ Smith,~R.;\ \ Willett,~R.~D.;\ \ Vij,~A.
  {{Di}}$\-$chloro$\-$bis(1,2-Phenyl$\-$enedi$\-$amine)$\-$nickel({{II}}),
  \textit{Acta Cryst. C} \textbf{2000,} \textsl{56,} e454.

\bibitem{coucouvanis1981}
Coucouvanis,~D.;\ \ Swenson,~D.;\ \ Baenziger,~N.~C.;\ \ Murphy,~C.;\ \
  Holah,~D.~G.;\ \ Sfarnas,~N.;\ \ Simopoulos,~A.;\ \ Kostikas,~A.  Tetrahedral
  Complexes Containing the {{Fe}}({{II}}){{S4}} Core. {{The}} Syntheses,
  Ground-State Electronic Structures and Crystal and Molecular Structures of
  the Bis(Tetraphenylphosphonium) Tetrakis (Thiophenolato)Ferrate ({{II}}) and
  Bis (Tetrapheny Lphosphonium) Bis(Dithiosquarato)Ferrate ({{II}}) Complexes.
  {{An}} Analog for the Active Site in Reducd Rubredoxins ({{Rdred}}),
  \textit{J. Am. Chem. Soc.} \textbf{1981,} \textsl{103,} 3350--3362.

\bibitem{burstyn1988}
Burstyn,~J.~N.;\ \ Roe,~J.~A.;\ \ Miksztal,~A.~R.;\ \ Shaevitz,~B.~A.;\ \
  Lang,~G.;\ \ Valentine,~J.~S.  Magnetic and Spectroscopic Characterization of
  an Iron Porphyrin Peroxide Complex. {{Peroxoferrioctaethylporphyrin}}(1-),
  \textit{J. Am. Chem. Soc.} \textbf{1988,} \textsl{110,} 1382--1388.

\bibitem{strauss1985}
Strauss,~S.~H.;\ \ Silver,~M.~E.;\ \ Long,~K.~M.;\ \ Thompson,~R.~G.;\ \
  Hudgens,~R.~A.;\ \ Spartalian,~K.;\ \ Ibers,~J.~A.  Comparison of the
  Molecular and Electronic Structures of
  (2,3,7,8,12,13,17,18-Octaethylporphyrinato)Iron({{II}}) and
  (Trans-7,8-Dihydro-2,3,7,8,12,13,17,18-Octaethylporphyrinato)Iron({{II}}),
  \textit{J. Am. Chem. Soc.} \textbf{1985,} \textsl{107,} 4207--4215.

\bibitem{daida2004}
Daida,~E.~J.;\ \ Peters,~J.~C.  Considering {{FeII}}/{{IV Redox Processes}} as
  {{Mechanistically Relevant}} to the {{Catalytic Hydrogenation}} of
  {{Olefins}} by [{{PhBPiPr3}}]{{Fe}}-{{Hx Species}},  \textit{Inorg. Chem.}
  \textbf{2004,} \textsl{43,} 7474--7485.

\bibitem{armstrong1984}
Armstrong,~W.~H.;\ \ Spool,~A.;\ \ Papaefthymiou,~G.~C.;\ \ Frankel,~R.~B.;\ \
  Lippard,~S.~J.  Assembly and Characterization of an Accurate Model for the
  Diiron Center in Hemerythrin,  \textit{J. Am. Chem. Soc.} \textbf{1984,}
  \textsl{106,} 3653--3667.

\bibitem{hartman1987}
Hartman,~J. A.~R.;\ \ Rardin,~R.~L.;\ \ Chaudhuri,~P.;\ \ Pohl,~K.;\ \
  Wieghardt,~K.;\ \ Nuber,~B.;\ \ Weiss,~J.;\ \ Papaefthymiou,~G.~C.;\ \
  Frankel,~R.~B.;\ \ Lippard,~S.~J.  Synthesis and Characterization of
  (.Mu.-Hydroxo)Bis(.Mu.-Acetato)Diiron({{II}}) and
  (.Mu.-Oxo)Bis(.Mu.-Acetato)Diiron({{III}})
  1,4,7-Trimethyl-1,4,7-Triazacyclononane Complexes as Models for Binuclear
  Iron Centers in Biology; Properties of the Mixed Valence
  Diiron({{II}},{{III}}) Species,  \textit{J. Am. Chem. Soc.} \textbf{1987,}
  \textsl{109,} 7387--7396.

\bibitem{li2002}
Li,~M.;\ \ Bonnet,~D.;\ \ Bill,~E.;\ \ Neese,~F.;\ \ Weyherm{\"u}ller,~T.;\ \
  Blum,~N.;\ \ Sellmann,~D.;\ \ Wieghardt,~K.  Tuning the {{Electronic
  Structure}} of {{Octahedral Iron Complexes}} [{{FeL}}({{X}})]
  ({{L}}\,=\,1-{{Alkyl}}-4,7-Bis(4-Tert-Butyl-2-Mercaptobenzyl)-1,4,7-Triazacyclononane,
  {{X}} = {{Cl}}, {{CH3O}}, {{CN}}, {{NO}}). {{The S}} = 1/2
  $\rightleftharpoons$ {{S}} = 3/2 {{Spin Equilibrium}} of [{{FeLPr}}({{NO}})],
   \textit{Inorg. Chem.} \textbf{2002,} \textsl{41,} 3444--3456.

\bibitem{berry1983}
Berry,~K.~J.;\ \ Clark,~P.~E.;\ \ Murray,~K.~S.;\ \ Raston,~C.~L.;\ \
  White,~A.~H.  Structure, Magnetism, and {{Moessbauer}} Spectrum of the
  Five-Coordinate Complex
  Chlorobis({{N}}-Methylbenzothiohydroxamato)Iron({{III}}),  \textit{Inorg.
  Chem.} \textbf{1983,} \textsl{22,} 3928--3934.

\bibitem{mankad2007}
Mankad,~N.;\ \ Whited,~M.;\ \ Peters,~J.  Terminal {{FeI}}--{{N2}} and
  {{FeII}}$\cdot\cdot\cdot${{H}}--{{C Interactions Supported}} by
  {{Tris}}(Phosphino)Silyl {{Ligands}},  \textit{Angew. Chem. Int. Ed.}
  \textbf{2007,} \textsl{46,} 5768--5771.

\bibitem{meyer1999}
Meyer,~K.;\ \ Bill,~E.;\ \ Mienert,~B.;\ \ Weyherm{\"u}ller,~T.;\ \
  Wieghardt,~K.  Photolysis of Cis- and Trans-[{{FeIII}}(Cyclam)({{N3}})2]+
  {{Complexes}}:\, {{Spectroscopic Characterization}} of a
  {{Nitridoiron}}({{V}}) {{Species}},  \textit{J. Am. Chem. Soc.}
  \textbf{1999,} \textsl{121,} 4859--4876.

\bibitem{silvernail2006}
Silvernail,~N.~J.;\ \ Noll,~B.~C.;\ \ Schulz,~C.~E.;\ \ Scheidt,~W.~R.
  Coordination of {{Diatomic Ligands}} to {{Heme}}:\, {{Simply CO}},
  \textit{Inorg. Chem.} \textbf{2006,} \textsl{45,} 7050--7052.

\bibitem{maelia1992}
Maelia,~L.~E.;\ \ Millar,~M.;\ \ Koch,~S.~A.  General Synthesis of
  Iron({{III}}) Tetrathiolate Complexes. {{Structural}} and Spectroscopic
  Models for the [{{Fe}}({{Cys}}-{{S}})4] Center in Oxidized Rubredoxin,
  \textit{Inorg. Chem.} \textbf{1992,} \textsl{31,} 4594--4600.

\bibitem{vogel1994}
Vogel,~E.;\ \ Will,~S.;\ \ Tilling,~A.~S.;\ \ Neumann,~L.;\ \ Lex,~J.;\ \
  Bill,~E.;\ \ Trautwein,~A.~X.;\ \ Wieghardt,~K.  Metallocorroles with
  {{Formally Tetravalent Iron}},  \textit{Angew. Chem. Int. Ed.} \textbf{1994,}
  \textsl{33,} 731--735.

\bibitem{harrop2008}
Harrop,~T.~C.;\ \ Tonzetich,~Z.~J.;\ \ Reisner,~E.;\ \ Lippard,~S.~J.
  Reactions of {{Synthetic}} [{{2Fe}}-{{2S}}] and [{{4Fe}}-{{4S}}] {{Clusters}}
  with {{Nitric Oxide}} and {{Nitrosothiols}},  \textit{J. Am. Chem. Soc.}
  \textbf{2008,} \textsl{130,} 15602--15610.

\bibitem{sellmann1991}
Sellmann,~D.;\ \ Geck,~M.;\ \ Knoch,~F.;\ \ Ritter,~G.;\ \ Dengler,~J.
  Transition-Metal Complexes with Sulfur Ligands. 57. {{Stabilization}} of
  High-Valent Iron({{IV}}) Centers and Vacant Coordination Sites by Sulfur
  .Pi.-Donation: Syntheses, x-Ray Structures, and Properties of
  [{{Fe}}("{{S2}}")2({{PMe3}})n] (n = 1, 2) and
  ({{NMe4}})[{{Fe}}("{{S2}}")2({{PMe3}})2].Cntdot.{{CH3OH}} ("{{S2}}"2- =
  1,2-Benzenedithiolate(2-)),  \textit{J. Am. Chem. Soc.} \textbf{1991,}
  \textsl{113,} 3819--3828.

\bibitem{rohde2003}
Rohde,~J.-U.;\ \ In,~J.-H.;\ \ Lim,~M.~H.;\ \ Brennessel,~W.~W.;\ \
  Bukowski,~M.~R.;\ \ Stubna,~A.;\ \ M{\"u}nck,~E.;\ \ Nam,~W.;\ \ Que,~L.
  Crystallographic and {{Spectroscopic Characterization}} of a {{Nonheme
  Fe}}({{IV}})={{O Complex}},  \textit{Science} \textbf{2003,} \textsl{299,}
  1037--1039.

\bibitem{havlin1998}
Havlin,~R.~H.;\ \ Godbout,~N.;\ \ Salzmann,~R.;\ \ Wojdelski,~M.;\ \
  Arnold,~W.;\ \ Schulz,~C.~E.;\ \ Oldfield,~E.  An {{Experimental}} and
  {{Density Functional Theoretical Investigation}} of {{Iron}}-57
  {{M{\"o}ssbauer Quadrupole Splittings}} in {{Organometallic}} and
  {{Heme}}-{{Model Compounds}}:\, {{Applications}} to {{Carbonmonoxy}}-{{Heme
  Protein Structure}},  \textit{J. Am. Chem. Soc.} \textbf{1998,} \textsl{120,}
  3144--3151.

\bibitem{kukolich1993}
Kukolich,~S.~G.;\ \ Roehrig,~M.~A.;\ \ Wallace,~D.~W.;\ \ Henderson,~G.~L.
  Microwave Measurements of the Rotational Spectrum and Structure of
  Butadieneiron Tricarbonyl,  \textit{J. Am. Chem. Soc.} \textbf{1993,}
  \textsl{115,} 2021--2027.

\bibitem{obrist2009}
Obrist,~B.~V.;\ \ Chen,~D.;\ \ Ahrens,~A.;\ \ Sch{\"u}nemann,~V.;\ \
  Scopelliti,~R.;\ \ Hu,~X.  An {{Iron Carbonyl Pyridonate Complex Related}} to
  the {{Active Site}} of the [{{Fe}}]-{{Hydrogenase}} ({{Hmd}}),
  \textit{Inorg. Chem.} \textbf{2009,} \textsl{48,} 3514--3516.

\bibitem{chen2011}
Chen,~D.;\ \ Ahrens-Botzong,~A.;\ \ Sch{\"u}nemann,~V.;\ \ Scopelliti,~R.;\ \
  Hu,~X.  Synthesis and {{Characterization}} of a {{Series}} of {{Model
  Complexes}} of the {{Active Site}} of [{{Fe}}]-{{Hydrogenase}} ({{Hmd}}),
  \textit{Inorg. Chem.} \textbf{2011,} \textsl{50,} 5249--5257.

\bibitem{chen2010}
Chen,~D.;\ \ Scopelliti,~R.;\ \ Hu,~X.  Synthesis and {{Reactivity}} of {{Iron
  Acyl Complexes Modeling}} the {{Active Site}} of [{{Fe}}]-{{Hydrogenase}},
  \textit{J. Am. Chem. Soc.} \textbf{2010,} \textsl{132,} 928--929.

\bibitem{kostka1993}
Kostka,~K.~L.;\ \ Fox,~B.~G.;\ \ Hendrich,~M.~P.;\ \ Collins,~T.~J.;\ \
  Rickard,~C. E.~F.;\ \ Wright,~L.~J.;\ \ Munck,~E.  High-Valent Transition
  Metal Chemistry. {{Moessbauer}} and {{EPR}} Studies of High-Spin ({{S}} = 2)
  Iron({{IV}}) and Intermediate-Spin ({{S}} = 3/2) Iron({{III}}) Complexes with
  a Macrocyclic Tetraamido-{{N}} Ligand,  \textit{J. Am. Chem. Soc.}
  \textbf{1993,} \textsl{115,} 6746--6757.

\bibitem{greatrex1969}
Greatrex,~R.;\ \ Greenwood,~N.~N.  M{\"o}ssbauer Spectra, Structure, and
  Bonding in Iron Carbonyl Derivatives,  \textit{Discuss. Faraday Soc.}
  \textbf{1969,} \textsl{47,} 126--135.

\bibitem{rasmussen2006}
Rasmussen, \textit{{Gaussian Processes for Machine Learning};} {MIT University
  Press Group Ltd}: Cambridge, MA, USA, 2006.

\bibitem{reiher2001}
Reiher,~M.;\ \ Salomon,~O.;\ \ Hess,~B.~A.  Reparameterization of Hybrid
  Functionals Based on Energy Differences of States of Different Multiplicity,
  \textit{Theor Chem Acc} \textbf{2001,} \textsl{107,} 48--55.

\bibitem{reiher2002}
Reiher,~M.  Theoretical {{Study}} of the {{Fe}}(Phen)2({{NCS}})2
  {{Spin}}-{{Crossover Complex}} with {{Reparametrized Density Functionals}},
  \textit{Inorg. Chem.} \textbf{2002,} \textsl{41,} 6928--6935.

\bibitem{salomon2002}
Salomon,~O.;\ \ Reiher,~M.;\ \ Hess,~B.~A.  Assertion and Validation of the
  Performance of the {{B3LYP}}$\star$ Functional for the First Transition Metal
  Row and the {{G2}} Test Set,  \textit{The Journal of Chemical Physics}
  \textbf{2002,} \textsl{117,} 4729--4737.

\bibitem{ioannidis2015}
Ioannidis,~E.~I.;\ \ Kulik,~H.~J.  Towards Quantifying the Role of Exact
  Exchange in Predictions of Transition Metal Complex Properties,  \textit{The
  Journal of Chemical Physics} \textbf{2015,} \textsl{143,} 034104.

\bibitem{gani2016}
Gani,~T. Z.~H.;\ \ Kulik,~H.~J.  Where {{Does}} the {{Density Localize}}?
  {{Convergent Behavior}} for {{Global Hybrids}}, {{Range Separation}}, and
  {{DFT}}+{{U}},  \textit{J. Chem. Theory Comput.} \textbf{2016,} \textsl{12,}
  5931--5945.

\bibitem{janet2017}
Janet,~J.~P.;\ \ Kulik,~H.~J.  Predicting {{Electronic Structure Properties}}
  of {{Transition Metal Complexes}} with {{Neural Networks}},   \textbf{2017,}
  arXiv:1702.05771.

\bibitem{ioannidis2017}
Ioannidis,~E.~I.;\ \ Kulik,~H.~J.  Ligand-{{Field}}-{{Dependent Behavior}} of
  {{Meta}}-{{GGA Exchange}} in {{Transition}}-{{Metal Complex Spin}}-{{State
  Ordering}},  \textit{J. Phys. Chem. A} \textbf{2017,} \textsl{121,} 874--884.

\bibitem{schwabl2007}
Schwabl,~F. \textit{{Quantum Mechanics};} {Springer}: Berlin, Germany; New
  York, NY, USA, 4th ed. 2007 ed.; 2007.

\bibitem{toth2010}
T{\'o}th,~P.~V.  Boundary Conditions for Many-Electron Systems,
  \textbf{2010,}  arXiv:1010.2700.

\bibitem{neugebauer2002}
Neugebauer,~J.;\ \ Reiher,~M.;\ \ Hinze,~J.  Analysis of the Asymptotic and
  Short-Range Behavior of Quasilocal {{Hartree}}-{{Fock}} and
  {{Dirac}}-{{Fock}}-{{Coulomb}} Electron-Electron Interaction Potentials,
  \textit{Phys. Rev. A} \textbf{2002,} \textsl{65,} 032518.

\bibitem{Filatov2007}
Filatov,~M.  On the Calculation of {{M{\"o}ssbauer}} Isomer Shift,  \textit{J.
  Chem. Phys.} \textbf{2007,} \textsl{127,} 084101.

\bibitem{kurian2008}
Kurian,~R.;\ \ Filatov,~M.  {{DFT Approach}} to the {{Calculation}} of
  {{M{\"o}ssbauer Isomer Shifts}},  \textit{J. Chem. Theory Comput.}
  \textbf{2008,} \textsl{4,} 278--285.

\bibitem{eaton2009}
Eaton,~J.~W.;\ \ Bateman,~D.;\ \ Hauberg,~S. \textit{{{GNU Octave}} Version
  3.0.1 Manual: A High-Level Interactive Language for Numerical Computations;}
  {CreateSpace Independent Publishing Platform}: 2009.

\bibitem{tao2003}
Tao,~J.;\ \ Perdew,~J.~P.;\ \ Staroverov,~V.~N.;\ \ Scuseria,~G.~E.  Climbing
  the {{Density Functional Ladder}}: {{Nonempirical Meta}}-{{Generalized
  Gradient Approximation Designed}} for {{Molecules}} and {{Solids}},
  \textit{Phys. Rev. Lett.} \textbf{2003,} \textsl{91,} 146401.

\bibitem{weigend2005}
Weigend,~F.;\ \ Ahlrichs,~R.  Balanced Basis Sets of Split Valence, Triple Zeta
  Valence and Quadruple Zeta Valence Quality for {{H}} to {{Rn}}: {{Design}}
  and Assessment of Accuracy,  \textit{Phys. Chem. Chem. Phys.} \textbf{2005,}
  \textsl{7,} 3297--3305.

\bibitem{weigend2006}
Weigend,~F.  Accurate {{Coulomb}}-Fitting Basis Sets for {{H}} to {{Rn}},
  \textit{Phys. Chem. Chem. Phys.} \textbf{2006,} \textsl{8,} 1057--1065.

\bibitem{klamt1993}
Klamt,~A.;\ \ Sch{\"u}{\"u}rmann,~G.  {{COSMO}}: A New Approach to Dielectric
  Screening in Solvents with Explicit Expressions for the Screening Energy and
  Its Gradient,  \textit{J. Chem. Soc. Perkin Trans. 2} \textbf{1993,}
  \textsl{0,} 799--805.

\bibitem{peterson2003}
Peterson,~K.~A.;\ \ Figgen,~D.;\ \ Goll,~E.;\ \ Stoll,~H.;\ \ Dolg,~M.
  Systematically Convergent Basis Sets with Relativistic Pseudopotentials.
  {{II}}. {{Small}}-Core Pseudopotentials and Correlation Consistent Basis Sets
  for the Post-d Group 16\textendash{}18 Elements,  \textit{J. Chem. Phys.}
  \textbf{2003,} \textsl{119,} 11113--11123.

\bibitem{ahlrichs1989}
Ahlrichs,~R.;\ \ B{\"a}r,~M.;\ \ H{\"a}ser,~M.;\ \ Horn,~H.;\ \ K{\"o}lmel,~C.
  Electronic Structure Calculations on Workstation Computers: {{The}} Program
  System Turbomole,  \textit{Chem. Phys. Lett.} \textbf{1989,} \textsl{162,}
  165--169.

\bibitem{neese2012}
Neese,~F.  The {{ORCA}} Program System,  \textit{WIREs Comput. Mol. Sci.}
  \textbf{2012,} \textsl{2,} 73--78.

\bibitem{perdew1992}
Perdew,~J.~P.;\ \ Wang,~Y.  Accurate and Simple Analytic Representation of the
  Electron-Gas Correlation Energy,  \textit{Phys. Rev. B} \textbf{1992,}
  \textsl{45,} 13244--13249.

\bibitem{perdew1986}
Perdew,~J.~P.  Density-Functional Approximation for the Correlation Energy of
  the Inhomogeneous Electron Gas,  \textit{Phys. Rev. B} \textbf{1986,}
  \textsl{33,} 8822--8824.

\bibitem{becke1988}
Becke,~A.~D.  Density-Functional Exchange-Energy Approximation with Correct
  Asymptotic Behavior,  \textit{Phys. Rev. A} \textbf{1988,} \textsl{38,}
  3098--3100.

\bibitem{lee1988}
Lee,~C.;\ \ Yang,~W.;\ \ Parr,~R.~G.  Development of the {{Colle}}-{{Salvetti}}
  Correlation-Energy Formula into a Functional of the Electron Density,
  \textit{Phys. Rev. B} \textbf{1988,} \textsl{37,} 785--789.

\bibitem{miehlich1989}
Miehlich,~B.;\ \ Savin,~A.;\ \ Stoll,~H.;\ \ Preuss,~H.  Results Obtained with
  the Correlation Energy Density Functionals of Becke and {{Lee}}, {{Yang}} and
  {{Parr}},  \textit{Chem. Phys. Lett.} \textbf{1989,} \textsl{157,} 200--206.

\bibitem{perdew1992a}
Perdew,~J.~P.;\ \ Chevary,~J.~A.;\ \ Vosko,~S.~H.;\ \ Jackson,~K.~A.;\ \
  Pederson,~M.~R.;\ \ Singh,~D.~J.;\ \ Fiolhais,~C.  Atoms, Molecules, Solids,
  and Surfaces: {{Applications}} of the Generalized Gradient Approximation for
  Exchange and Correlation,  \textit{Phys. Rev. B} \textbf{1992,} \textsl{46,}
  6671--6687.

\bibitem{perdew1996}
Perdew,~J.~P.;\ \ Burke,~K.;\ \ Ernzerhof,~M.  Generalized {{Gradient
  Approximation Made Simple}},  \textit{Phys. Rev. Lett.} \textbf{1996,}
  \textsl{77,} 3865--3868.

\bibitem{zhao2006a}
Zhao,~Y.;\ \ Truhlar,~D.~G.  A New Local Density Functional for Main-Group
  Thermochemistry, Transition Metal Bonding, Thermochemical Kinetics, and
  Noncovalent Interactions,  \textit{J. Chem. Phys.} \textbf{2006,}
  \textsl{125,} 194101.

\bibitem{becke1993}
Becke,~A.~D.  Density-functional Thermochemistry. {{III}}. {{The}} Role of
  Exact Exchange,  \textit{J. Chem. Phys.} \textbf{1993,} \textsl{98,}
  5648--5652.

\bibitem{adamo1999}
Adamo,~C.;\ \ Barone,~V.  Toward Reliable Density Functional Methods without
  Adjustable Parameters: {{The PBE0}} Model,  \textit{J. Chem. Phys.}
  \textbf{1999,} \textsl{110,} 6158--6170.

\bibitem{zhao2008}
Zhao,~Y.;\ \ Truhlar,~D.~G.  The {{M06}} Suite of Density Functionals for Main
  Group Thermochemistry, Thermochemical Kinetics, Noncovalent Interactions,
  Excited States, and Transition Elements: Two New Functionals and Systematic
  Testing of Four {{M06}}-Class Functionals and 12 Other Functionals,
  \textit{Theor. Chem. Acc.} \textbf{2008,} \textsl{120,} 215--241.

\bibitem{staroverov2003}
Staroverov,~V.~N.;\ \ Scuseria,~G.~E.;\ \ Tao,~J.;\ \ Perdew,~J.~P.
  Comparative Assessment of a New Nonempirical Density Functional:
  {{Molecules}} and Hydrogen-Bonded Complexes,  \textit{J. Chem. Phys.}
  \textbf{2003,} \textsl{119,} 12129--12137.

\end{thebibliography}
\providecommand{\refin}[1]{\\ \textbf{Referenced in:} #1}

\end{document}